%% file: ms_apj.tex
\documentclass[twocolumn]{emulateapj}

\usepackage{lscape}

\slugcomment{Accepted for publication in ApJ, 05-Dec-2005}%

\shorttitle{Stacking at 24\,$\mu$m}
\shortauthors{Zheng et al.}

\begin{document}

\title{ Detecting Faint Galaxies by Stacking at 24\,$\micron$}


\author{X. Z. \ Zheng\altaffilmark{1}, Eric F. Bell\altaffilmark{1}, Hans-Walter Rix\altaffilmark{1}, 
Casey Papovich \altaffilmark{2}, Emeric Le Floc'h \altaffilmark{2,3}, G. H. Rieke\altaffilmark{2},
and P. G. P\'erez-Gonz\'alez\altaffilmark{2}}
\altaffiltext{1} {Max-Planck Institut f\"ur Astronomie, K\"onigstuhl 17, 
D-69117 Heidelberg, Germany}
\altaffiltext{2} {Steward Observatory, University of Arizona, 933 N
Cherry Ave, Tucson, AZ 85721, USA}
\altaffiltext{3} {GEPI, Observatoire de Paris-Meudon, 92195 Meudon, France}



\email{zheng@mpia.de}

\begin{abstract}
We stack {\it Spitzer} 24\,$\mu$m images for $\sim$ 7000 galaxies with 
$0.1 \le z < 1$ in the Chandra Deep Field South to probe the thermal 
dust emission in low-luminosity galaxies over this redshift range. 
Through stacking, we can detect mean 24\,$\mu$m
fluxes that are more than an order of magnitude below the individual
detection limit.  We find that 
the correlations for low and moderate luminosity galaxies 
between the average $L_{\rm IR}/L_{\rm UV}$ and 
rest-frame $B$-band luminosity, and between the star formation rate (SFR) 
and $L_{\rm IR}/L_{\rm UV}$, are similar to those in the local Universe. 
This verifies that oft-used assumption in 
deep UV/optical surveys that the dust obscuration--SFR 
relation for galaxies with SFR $\leq$ 20\,M$_\odot$\,yr$^{-1}$
varies little with epoch. We have used this relation 
to derive the cosmic IR luminosity density from $z = 1$ to $z = 0.1$. 
The results also demonstrate directly that little of the bolometric 
luminosity of the galaxy population arises from the faint end of the 
luminosity function,
indicating a relatively flat faint-end slope of the IR luminosity 
function with a power law index of $1.2\pm0.3$.
\end{abstract}



\section{Introduction}

The rapid decay of the cosmic mean star formation rate (SFR) density 
from $z = 1$ to the present epoch has been convincingly established 
over the last decade (see, e.g., Hopkins~\citeyear{Hopkins04} and 
references therein). 
The focus has now turned to characterizing the types of galaxies 
responsible for this decay. Rest-frame ultraviolet (UV) and 
optical emission-line studies indicate the decay since z $\sim$ 1 
is strongly influenced by the behavior of the relatively low-mass 
galaxies (e.g., Brinchmann \& Ellis et al. ~\citeyear{Brinchmann00}; 
Juneau et al. \citeyear{Juneau05}; Bundy et al. \citeyear{Bundy05}; 
Bauer et al. \citeyear{Bauer05}; Wolf et al. \citeyear{Wolf05}).  
On the other hand, surveys in the thermal infrared (IR) have found 
many galaxies of intermediate and high mass with intense, 
deeply obscured star formation (Franceschini et al. 
\citeyear{Franceschini03}; Zheng et al. \citeyear{Zheng04}; 
Bell et al. \citeyear{Bell05}; P\'erez-Gonz\'alez et al. 
\citeyear{Perez05}; LeFloc'h et al. \citeyear{LeFloc'h05}). 
However, the role of obscured star formation in lower-mass galaxies 
remains unknown (although see Heavens et al. \citeyear{Heavens04} 
for a powerful and complementary approach).

Among the many observational SFR estimators (see Kennicutt 
\citeyear{Kennicutt98a} for a review), at high redshifts UV radiation 
is the most easily measured proxy for the SFR.  
It provides the basis for much of our understanding of the evolution of 
the cosmic SFR density (e.g., Madau et al. \citeyear{Madau96}; 
Steidel et al. \citeyear{Steidel99}; Sullivan et al. \citeyear{Sullivan00}; 
Hopkins et al. \citeyear{Hopkins01}; Schiminovich et al. 
\citeyear{Schiminovich05}). Yet, young stars are usually 
born in dust-rich environments; the dust absorbs the vast majority of 
the UV light and re-radiates this energy in the thermal IR. 
Therefore, to obtain a complete census of the bolometric luminosity 
from young stars, observations in both the UV and IR are required 
(see, e.g., Gordon et al. \citeyear{Gordon00}). 
However, the available infrared facilities have lacked 
the sensitivity and imaging resolution to explore obscured 
star formation in low-luminosity, relatively low-mass galaxies, 
except for nearby examples.  Thus, indirect estimates have had to 
be used for the IR outputs of such galaxies even at moderate redshift, 
using tools such as the empirical calibration of UV color to 
extinction (e.g., Meurer et al. \citeyear{Meurer99}; 
Calzetti \& Heckman \citeyear{Calzetti99}), or the locally-observed 
trend between $L_{\rm IR}/L_{\rm UV}$ and luminosity 
(showing that more luminous galaxies tend to have higher extinction; 
e.g., Wang \& Heckman \citeyear{Wang96}; Bell \citeyear{Bell03}; 
Buat et al. \citeyear{Buat05}). 
These approaches have been applied to observations of 
high-redshift galaxies to derive the SFR per unit comoving 
volume (e.g., Adelberger \& Steidel \citeyear{Adelberger00}; 
Hopkins \citeyear{Hopkins01}).

\begin{figure*} \centering
\includegraphics[width=0.35\textwidth,clip]{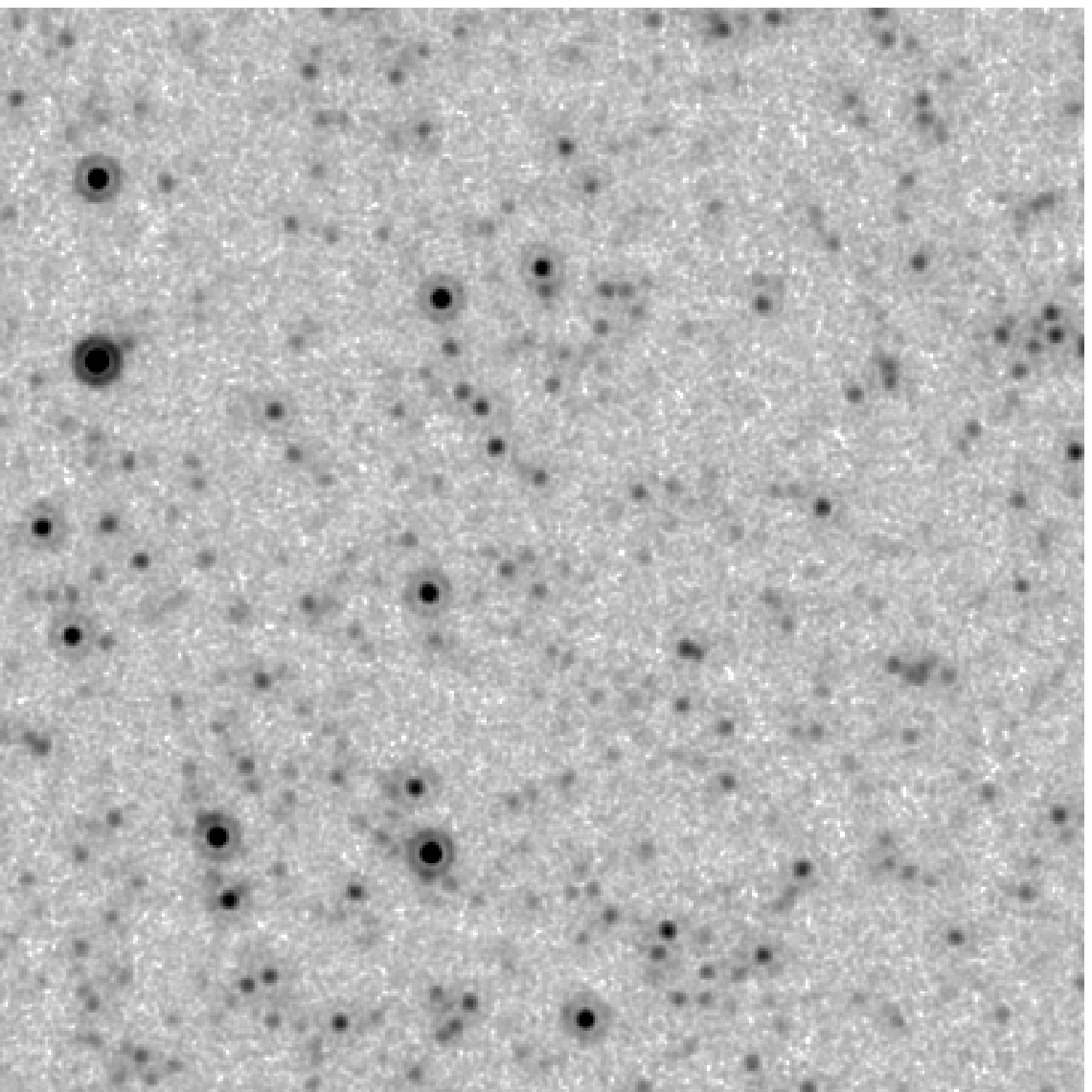}\hskip 6mm
\includegraphics[width=0.35\textwidth,clip]{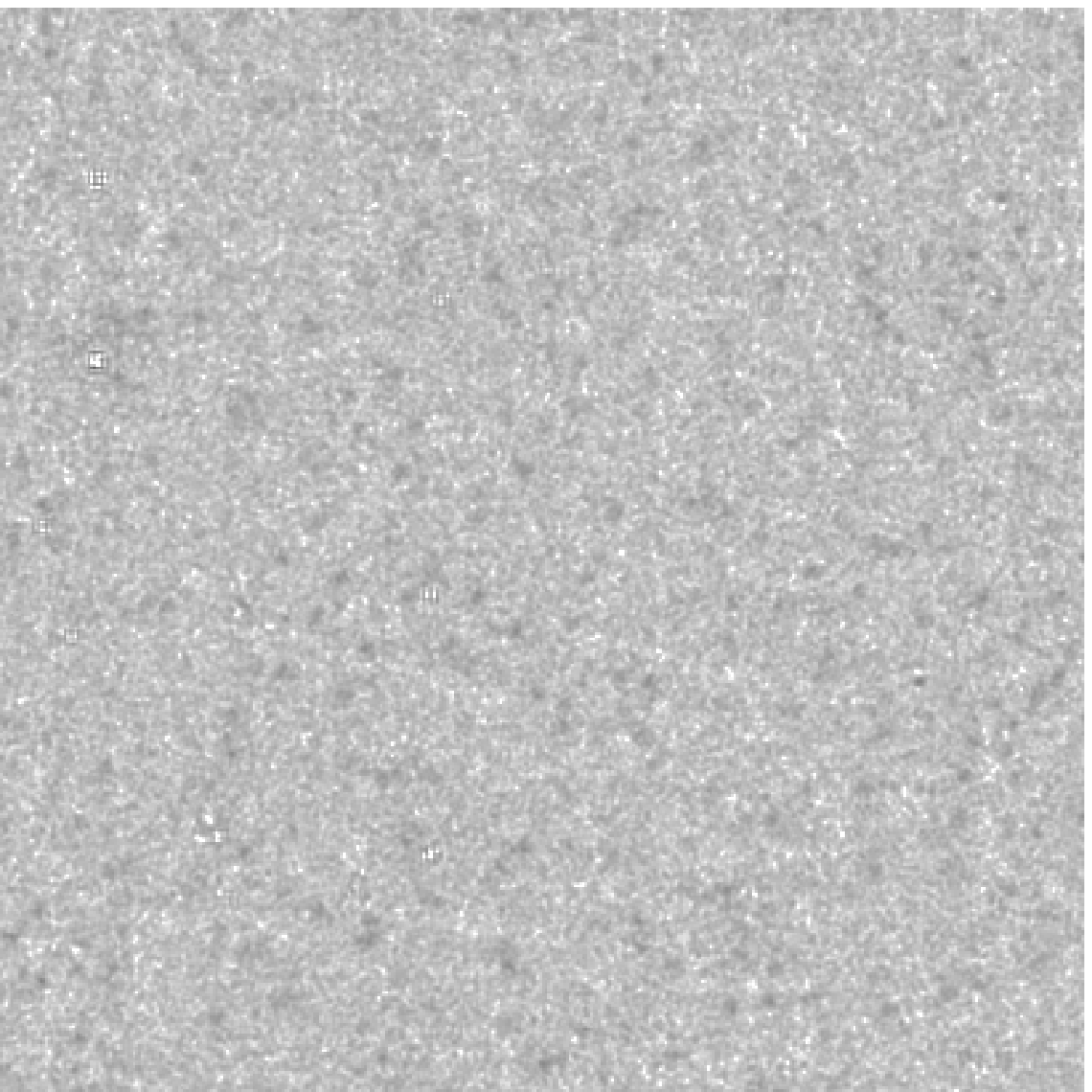} 
\caption{Illustration of Spitzer/MIPS 24\,$\mu$m images before
  and after PSF subtraction of individually-detected sources.  {\it Left}: an
  image section (10$\arcmin \times 10\arcmin$) in the CDFS; {\it Right}: the
  corresponding residual image after removing all detected sources brighter
  than 83\,$\mu$Jy (at the 80\% completeness).  Two images are shown in the
  same greylevel scale.  Sources seen in the residual images are fainter than
  83\,$\mu$Jy.  } \label{exampleimage}
\end{figure*}

The goal of this paper is to measure directly the average IR outputs 
of low-luminosity galaxies at intermediate redshift. 
We use a deep 24\,$\mu$m map of the Chandra Deep Field South (CDFS) 
from the Multiband Imaging Photometer on 
{\it Spitzer} (MIPS: Rieke et al. \citeyear{Rieke04}). 
The 24\,$\mu$m map is limited by both photon and confusion noise 
(Dole et al. \citeyear{Dole04}).
We find that stacking on the position of known intermediate redshift 
galaxies substantially reduces both sources of noise, 
allowing secure detection of average flux densities substantially 
below the conventional confusion limit. We have stacked 
24\,$\mu$m images of several thousand intermediate- and 
low-luminosity galaxies at $0.1 \le z < 1$.  
We combine the resulting detections with the COMBO-17 data set 
to investigate the effects of dust obscuration in galaxies 
with $m_R < 24$ mag out to $z \sim$ 1. 
We find no evidence for evolution in the relationship between 
obscuration and the SFR over this range of redshift. 
Our results suggest that the IR luminosity function 
has a relatively flat slope (power law index of $1.2\pm 0.3$) 
toward low luminosities, at least out to $z \sim$ 1. 

The paper is organized as follows. Section 2 describes the data 
and the selection of the galaxy sample. Section 3 addresses 
the methods for stacking the images and describes tests of 
their validity. Section 4 shows the results. 
Section 5 uses these results to refine estimates of 
the cosmic SFR density from z $\sim$ 1 to the present epoch. 
Our work is summarized in Section 6. 
Throughout the paper we adopt a cosmology with 
$H_0$\,=\,70\,km\,s$^{-1}$\,Mpc$^{-1}$, 
$\Omega_{\rm M}$\,=\,0.3 and $\Omega_{\Lambda}$\,=\,0.7.

\section{The data and galaxy sample}

In this paper, we combine ground-based optical data from 
the COMBO-17 photometric redshift survey
with 24\,$\mu$m data from MIPS on board the Spitzer Space Telescope.
As part of COMBO-17, the extended CDFS (centered on $\alpha_{2000}$\,=\,03$^h32^m25^s$, $\delta_{2000}$\,=\,$-27\degr48\arcmin50\arcsec$)
was imaged in 5 broad and 12 medium-band optical filters from 
350 to 930\,nm (Wolf et al. \citeyear{Wolf03}). 
The CDFS catalog is available for public use (Wolf et al. \citeyear{Wolf04}).
The catalog presents astrometry and photometry for 
63501 objects in an area of 30$\arcmin.5\,\times\,30\arcmin$.  
Of these, photometric redshifts and 
rest-frame luminosities in the standard Johnson $U,B,V$ bands and a 
synthetic UV band centered at 280\,nm\footnote{The synthetic 280\,nm 
passband has a square bandpass and 40\,nm FWHM.  Rest-frame luminosities
in the 280\,nm bandpass are extrapolated using the best-fit
galaxy templates for galaxies with $z \la 0.3$.  Rest-frame luminosities
in $U$- and $B$-band are always interpolated in the redshift
range of interest, whereas $V$-band luminosities are extrapolations for 
galaxies with $z \ga 0.7$.}
are measured for $\sim$9\,000 classified galaxies with aperture magnitudes
$m_R < 24$ mag and 
$z < 1.1$.  Redshift uncertainties have been tested against large 
spectroscopic datasets and are $\delta z/(1+z)\sim 0.02$ 
at the median galaxy magnitude of $R \sim 22$; astrometric
uncertainties are $\sim 0.1\arcsec$ (Wolf et al. \citeyear{Wolf04}).

MIPS observations of the CDFS at 24\,$\mu$m were performed 
under MIPS Guaranteed Time Observations (GTOs).
A rectangular field of 
$90\arcmin \times 30\arcmin$ was observed in slow scan-map mode,
with a total integration time of $\sim 1380$\,seconds per pixel. 
The  24\,$\mu$m image reduction was done with 
the MIPS Data Analysis Tool (Gordon et al. \citeyear{Gordon}). 
A final mosaic image was produced with resolution $1\arcsec .25$ per pixel 
and a Point-Spread Function (PSF) with  
Full Width at Half Maximum (FWHM) $\simeq$\,6$\arcsec$. 
We used Sextractor (Bertin \& Arnouts et al. \citeyear{Bertin96}) 
for source detection.  Because the typical size of an intermediate 
redshift galaxy is $\ll 6\arcsec$, we chose to fit all galaxies 
as unresolved point sources. 
The few clearly extended sources were individually analyzed.
An empirical PSF was built by stacking 18 
bright point sources; this PSF was subsequently used 
for PSF fitting within 10$\arcsec$ of all other objects\footnote{Aperture
corrections from this empirical PSF are similar to 
the updated PSFs in the ``Spitzer Space Telescope 
Multiband Imaging Photometer for Spitzer (MIPS) Data
Handbook'' (version 2.3) to within the errors.}. 
The ALLSTAR routine in IRAF's DAOPHOT was used to 
simultaneously fit multiple sources to the entire 24\,$\mu$m image
mosaic.  Local background values were estimated to be the mode
value of an annulus with an inner radius of 17$\arcsec$ (5\,FWHM) 
and width 20$\arcsec$ centered on a given target\footnote{The use
of median or mean values for the sky would make little difference
for uncrowded sources; the mode is adopted as it is substantially
more outlier-resistant than either mean or median in the case
of sky annuli contaminated by sources.}.
The brightest sources were fit first and then removed. 
The fainter sources hidden by the brightest sources 
were then detected and progressively included into ALLSTAR fitting. 
The flux of each detected object is derived from the PSF 
fits.  We cut the 24\,$\mu$m catalog at the 80\% completeness
level of 83\,$\mu$Jy (``5$\sigma$'', see 
Papovich et al. \citeyear{Papovich04} for the
description of the completeness limit; the Papovich 
et al. catalog is very similar to that described here to
within the errors). 

The MIPS and COMBO-17 areas on the sky do not completely overlap.  
For the present analysis, we 
choose the subsample of galaxies contained
within both the MIPS and the COMBO-17 areas, avoiding
the edges in the MIPS data.  This sample covers
an area of $\sim$800\,arcmin$^2$ and contains 9785 objects with $m_R < 24$
(of which 7892 are galaxies in the redshift range 0.1\,$\leq\,z\,<$\,1). 
Before the core analysis of this paper,
we constructed a catalog of 24\,$\mu$m detected sources that could
be matched to COMBO-17 sources with redshifts.
We cross-correlated the two catalogs with a 
tolerance of 1$\arcsec$.25, or 3$\sigma$ positional uncertainty, 
as estimated from bright stars and compact sources. 
Of the 9785 COMBO-17 objects at $m_R < 24$, 
1725 objects are MIPS 24\,$\mu$m resolved sources. Of these, 
1352 objects are selected in our sample as galaxies 
in the redshift range 0.1\,$\leq\,z\,<$\,1.

\begin{figure} \centering
\includegraphics[width=0.45\textwidth,clip]{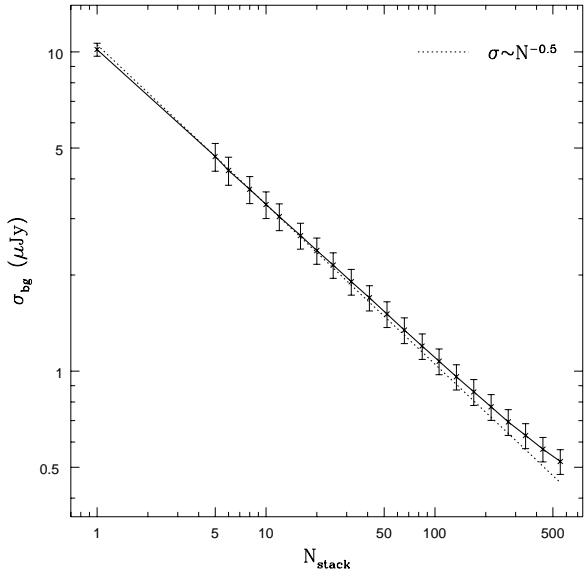}
\caption{Reduction of the background RMS with increasing number 
  of stacked images.  Here $\sigma_{\rm bg}$ refers to the RMS of the flux
  within an aperture of radius 5$\arcsec$. For comparison, the dotted line
  shows the inverse square root of the stack number, normalized to the
  measured $\sigma_{\rm bg}$ at N$_{\rm stack}$\,=\,10.  }\label{stacksig}
\end{figure}

\begin{figure*} \centering
\includegraphics[width=0.75\textwidth,clip]{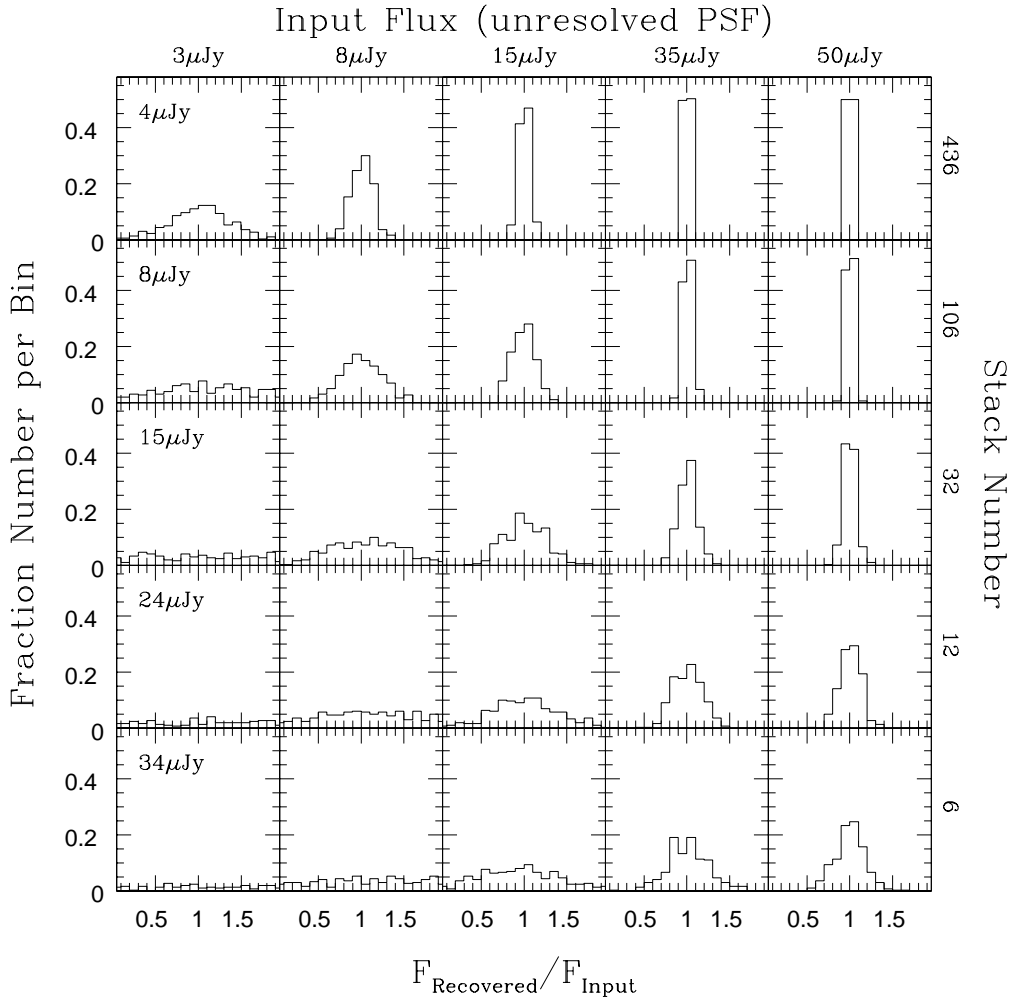}
\caption{Recovery of simulated sources 
  with different input flux levels as a function of stack number.  The
  intervals between stack numbers are identical in logarithm space.  Assuming
  Poisson noise, the expected ``5$\sigma$'' limits for corresponding stack
  numbers are given in the left hand panels. The detection limit 83\,$\mu$Jy
  corresponds to stack number $N_{\rm stack}$\,=\,1.  As the simulated sources
  in one stack set are identical, median stack and mean stack give almost the
  same recovery.  Here median stack values are adopted.  }\label{stacksimu}
\end{figure*}

\begin{figure*} \centering
\includegraphics[width=0.9\textwidth,clip]{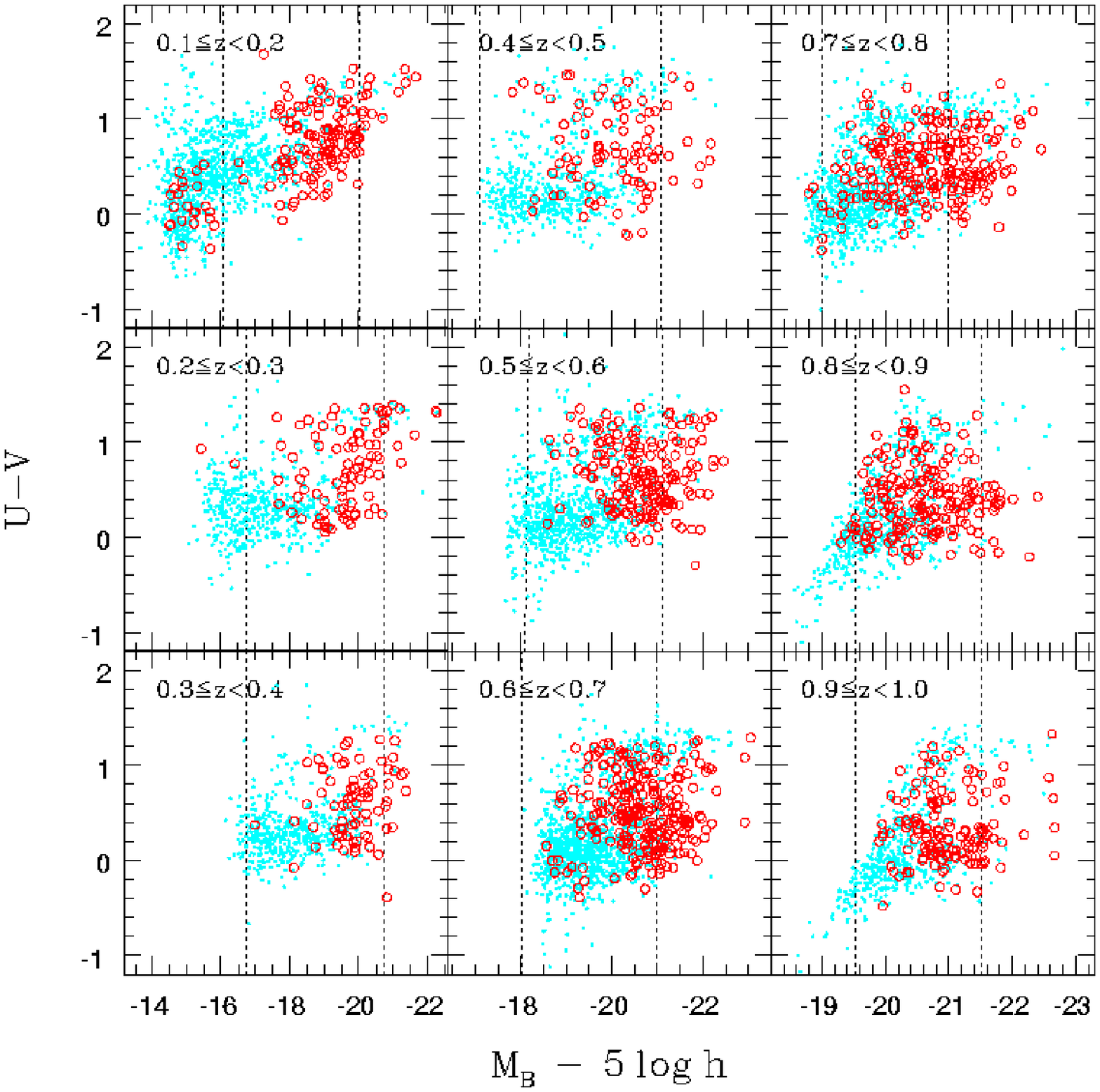} 
\caption{COMBO-17 galaxy sample in the CDFS: each panel shows 
the rest-frame color $U-V$ as a function of 
$B$ band absolute magnitude for objects in 9 different 
redshift slices, increasing in redshift from top to bottom and left
to right.  In each panel, the right-hand side dotted line shows 
$M^\ast$ and the dotted line on the left shows the limit of the faintest
bin considered in this work. 
The $M^\ast_B$ adopted from Ilbert et al. (\citeyear{Ilbert}) is $-$20.0, $-$20.7, $-$21.1, $-$21.0 and $-$21.5 for redshift bins 0.1\,$\leq\,z\,<$\,0.2, 0.2\,$\leq\,z\,<$\,0.4, 0.4\,$\leq\,z\,<$\,0.6, 0.6\,$\leq\,z\,<$\,0.8 and 0.8\,$\leq\,z\,<$\,1, respectively.  
Sources individually detected at
24\,$\mu$m are shown as open circles; undetected 
sources are denoted by dots.  The population of very faint 
24\,$\mu$m individually-detected sources at $z \le 0.3$ consists of 
much more luminous galaxies at high redshift which were mistakenly
assigned COMBO-17 photometric redshifts $z \le 0.3$ (see text for more 
discussion).  } \label{colormag}
\end{figure*}

\section{MIPS 24\,$\mu$m image stacking}

In this section we describe the specifics of deriving the mean 24\,$\mu$m
fluxes for galaxy sub-sets through stacking. Basically, there are three steps:
first, the removal of all individually-detected sources in the image;
second, the stacking of the residual image postage stamps through averaging
or medianing; finally, the addition of the individual 24\,$\mu$m fluxes
for all MIPS sources that coincide with the optical positions of the galaxy
sub-set at hand.

\subsection{Stacking: method and testing}

Deep Spitzer surveys at 24\,$\mu$m, including 
these data, are limited by a combination of photon noise and 
confusion noise (Dole et al. \citeyear{Dole04}).
The confusion noise at a given position on the sky
is related to the unresolved extragalactic population;
increasing exposure time will not reduce this source of noise
and therefore it sets a limit to one's ability to individually detect
galaxies in Spitzer 24\,$\mu$m observations.
However, the confusion noise does vary across the sky.
If one is interested in estimates of the average flux
for classes of objects (e.g. galaxies at a given redshift) with
known coordinates, one can improve the detection 
threshold by stacking images centered on the object coordinates.
The degree to which the confusion noise is thereby
reduced depends on the degree to which the (confusing) sources
are randomly distributed with respect to the population of interest. 
In the case of deep 24{\micron} observations, the clustering
is relatively weak (i.e., most close pairs of galaxies are
projections rather than physical associations) and the background
is already largely resolved 
(i.e., more than 50\% of the total extragalactic 24{\micron}
flux is in individually-detected sources; Dole et al. \citeyear{Dole04}; 
Papovich et al. \citeyear{Papovich04}), making these data almost
ideal for a stacking analysis.    
Hence, the combination of many images can
substantially reduce the background noise and allow
detection of the average flux of a given population,
well below the canonical detection limit.

How well this works in practice and how best to stack,
we explore with a number of tests.
As is clear from Fig.\ \ref{exampleimage}, a substantial portion
of 24\,$\mu$m images have significant contributions from bright sources.
In order to avoid that the ``background'' in stacked images
is contaminated by small
numbers of bright sources, we first identified and PSF-subtracted
all sources that were individually detected above the
83\,$\mu$Jy level.  The right-hand panel of Fig.\ \ref{exampleimage}
shows a residual image constructed in this way.
We tested the efficacy of stacking these cleaned images in two ways.  
First, we measured the 
RMS of the background in a randomly placed 5$\arcsec$ aperture 
(the same aperture used for the galaxy photometry) 
as a function of number of images stacked.  
Postage stamp images with a size of $1\arcmin .68\,\times\,1\arcmin .68$ 
were randomly extracted from the residual image and stacked.
We quantify the background noise by modified, outlier-resistant 
measure of the variance, defined as the half width of the range 
centered at the median value and including 68\% of the sample values.
Fig.~\ref{stacksig} shows the decrease in $\sigma_{\rm bg}$ with 
increasing stack number: the effective background noise decreased as
the inverse square root of the stack number.
This behavior is as expected for ideal, uncorrelated
random noise. The ideal behavior of the noise is a
consequence of the flat number count spectrum at low flux
levels at 24\,$\mu$m, and it has the benefit that a standard
error analysis is useful on the stacked images.
Only below $\sigma_{\rm bg} \sim$ 0.4, which corresponds to 
a ``5$\sigma$'' limit of $\sim$3 to 4\,$\mu$Jy, does the
effective background noise become significantly larger than that
predicted by the inverse square law extrapolation 
(see also Fig.~\ref{stacksimu}).
Second, we tested the accuracy of recovering
artificial sources from image stacks, by randomly placing 
large numbers of individually faint, identical sources into the
source subtracted image.
Fig.~\ref{stacksimu} shows the results: stacking 
is able to recover progressively fainter signals with larger stack numbers. 
Roughly speaking, for data of this quality 
($\sim 1380$\,seconds per pixel slow scan map in a low background field), 
a 35\,$\mu$Jy signal can be recovered in a $\sim$10-image 
stack with a flux accuracy of $\pm$50\% at the $3\sigma$ detection level,
and a 8\,$\mu$Jy signal can be recovered with a $\sim$100-image stack.
In our stacking, postage stamp images are aligned to a few tenths 
of one pixel.  Alignment to the centroid pixel, i.e., to an
accuracy of one pixel, causes an additional 5\% error on the final recovered
flux.

\begin{figure*} \centering
\includegraphics[width=0.3\textwidth,clip]{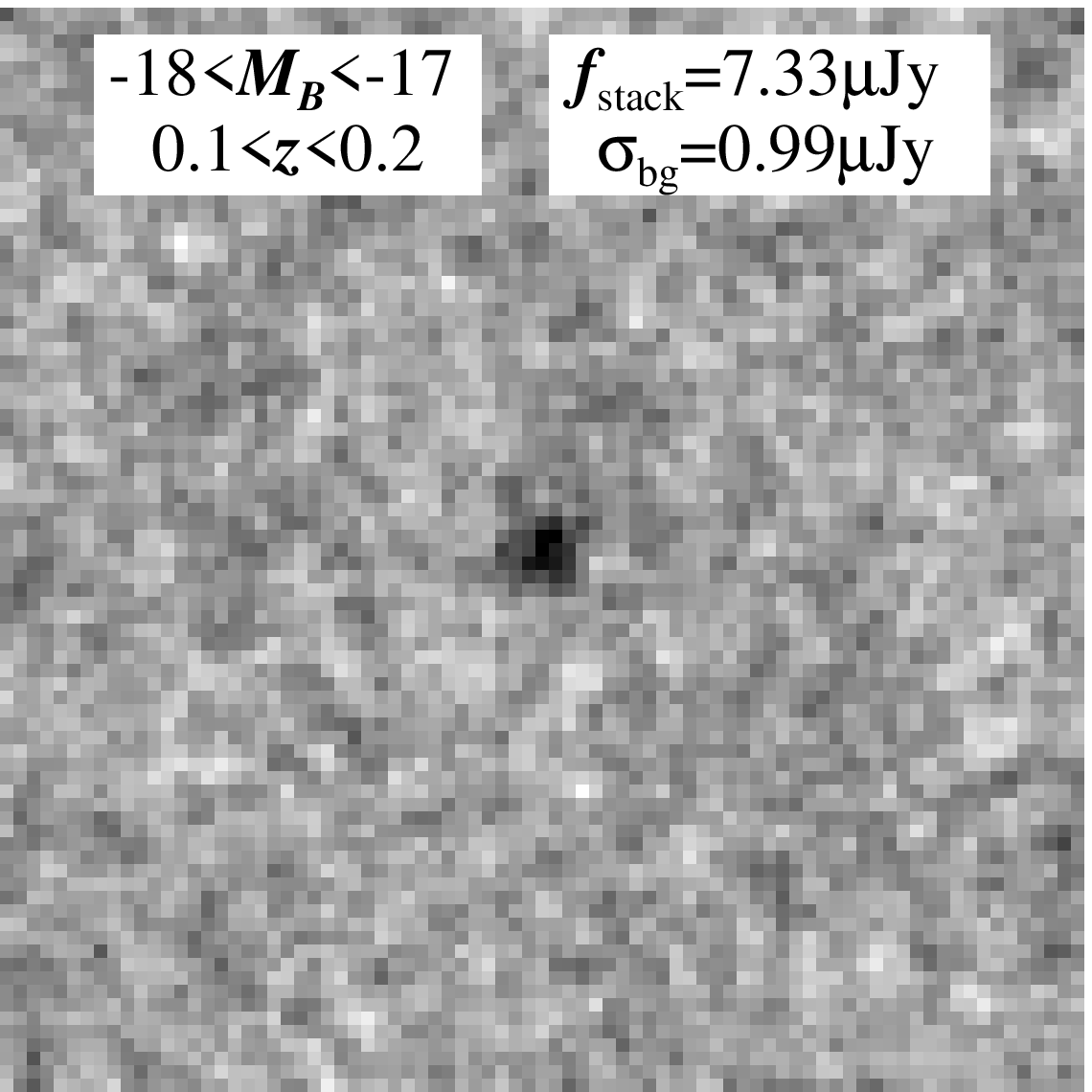}\hskip 1mm
\includegraphics[width=0.3\textwidth,clip]{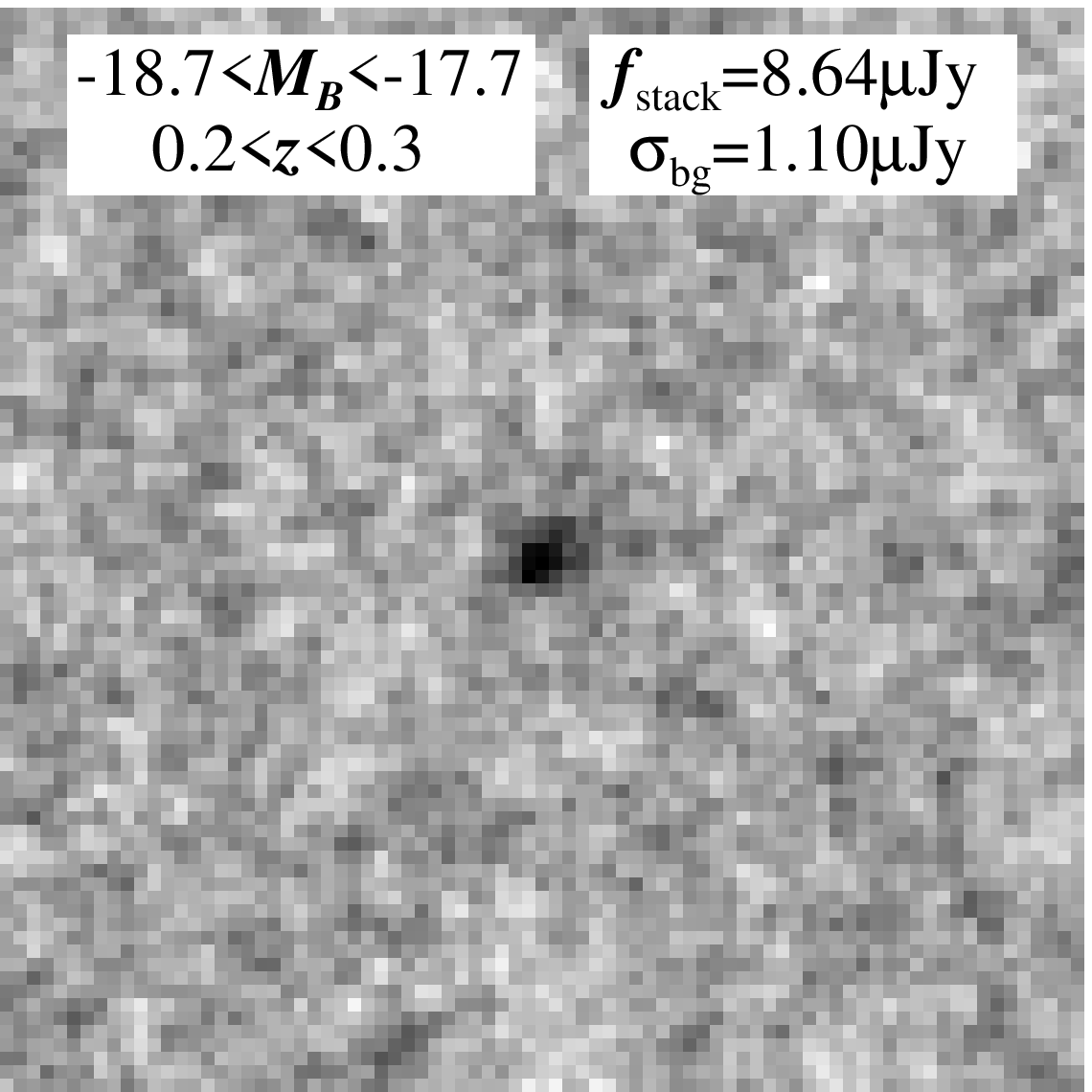}
\vskip 1mm
\includegraphics[width=0.3\textwidth,clip]{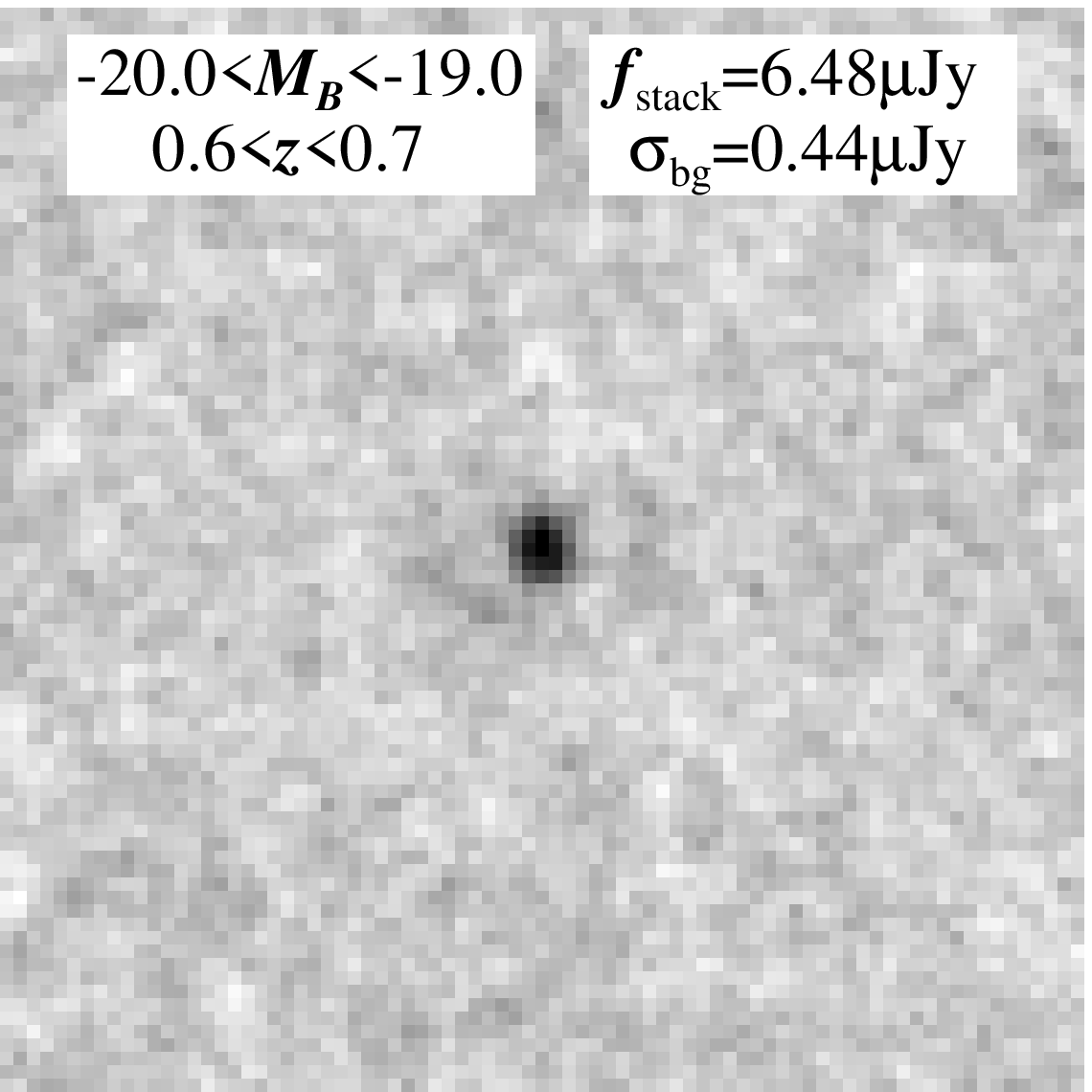}\hskip 1mm
\includegraphics[width=0.3\textwidth,clip]{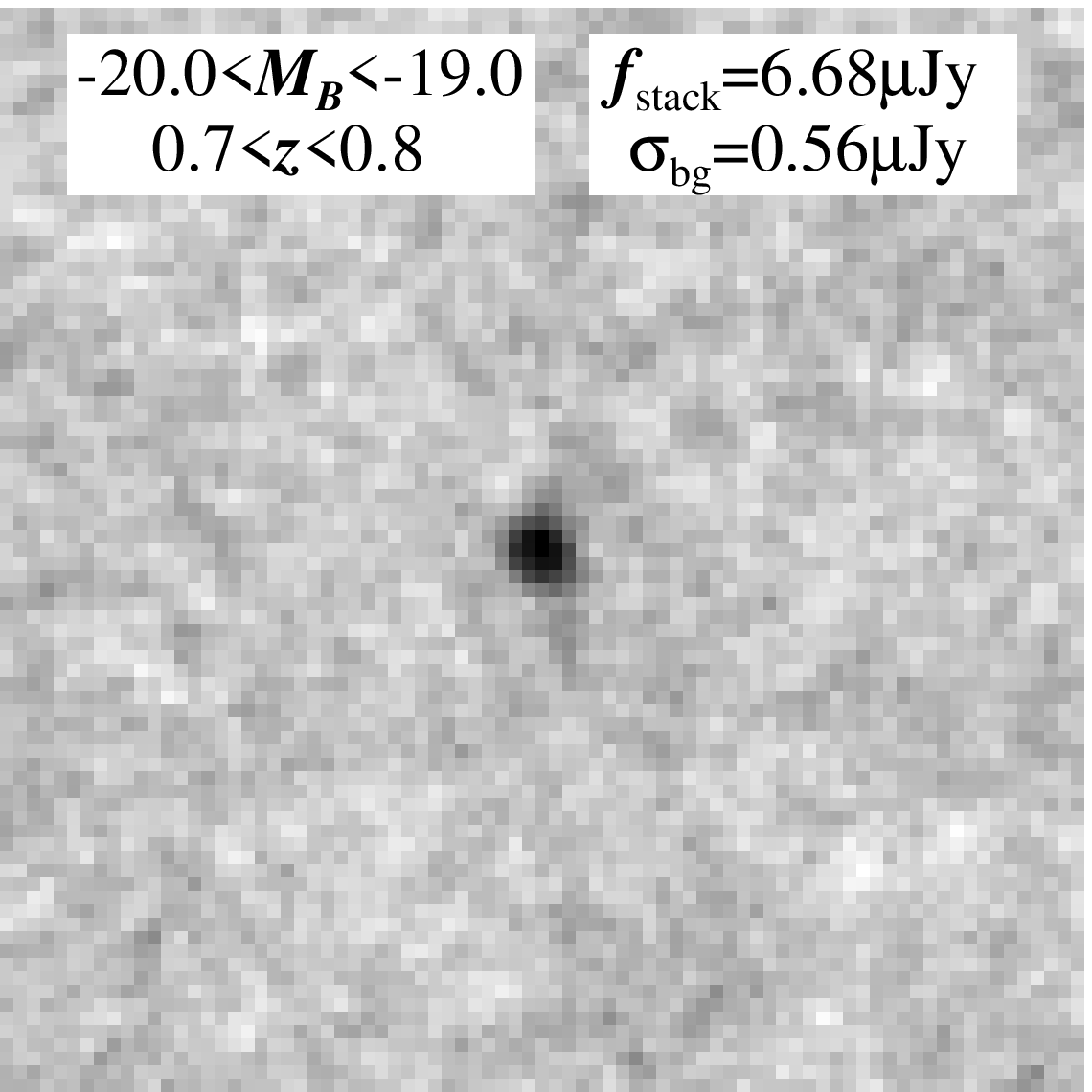}
\caption{Examples of median-stacked 24\,$\mu$m images. 
The magnitude bin, redshift slice, integrated flux of the central target within an aperture of radius 5$\arcsec$, $f_{\rm stack}$ and background fluctuation $\sigma_{\rm bg}$ on the same aperture are labeled. Both $f_{\rm stack}$ and $\sigma_{\rm bg}$ are given in units of $\mu$Jy. No aperture correction is implemented. The size of each image is $1\arcmin .68\,\times\,1\arcmin .68$. In bottom-left image, the second Airy ring of the 24\,$\mu$m PSF is visible.}\label{examplestack}
\end{figure*}

\begin{figure*} \centering
\includegraphics[width=0.8\textwidth,clip]{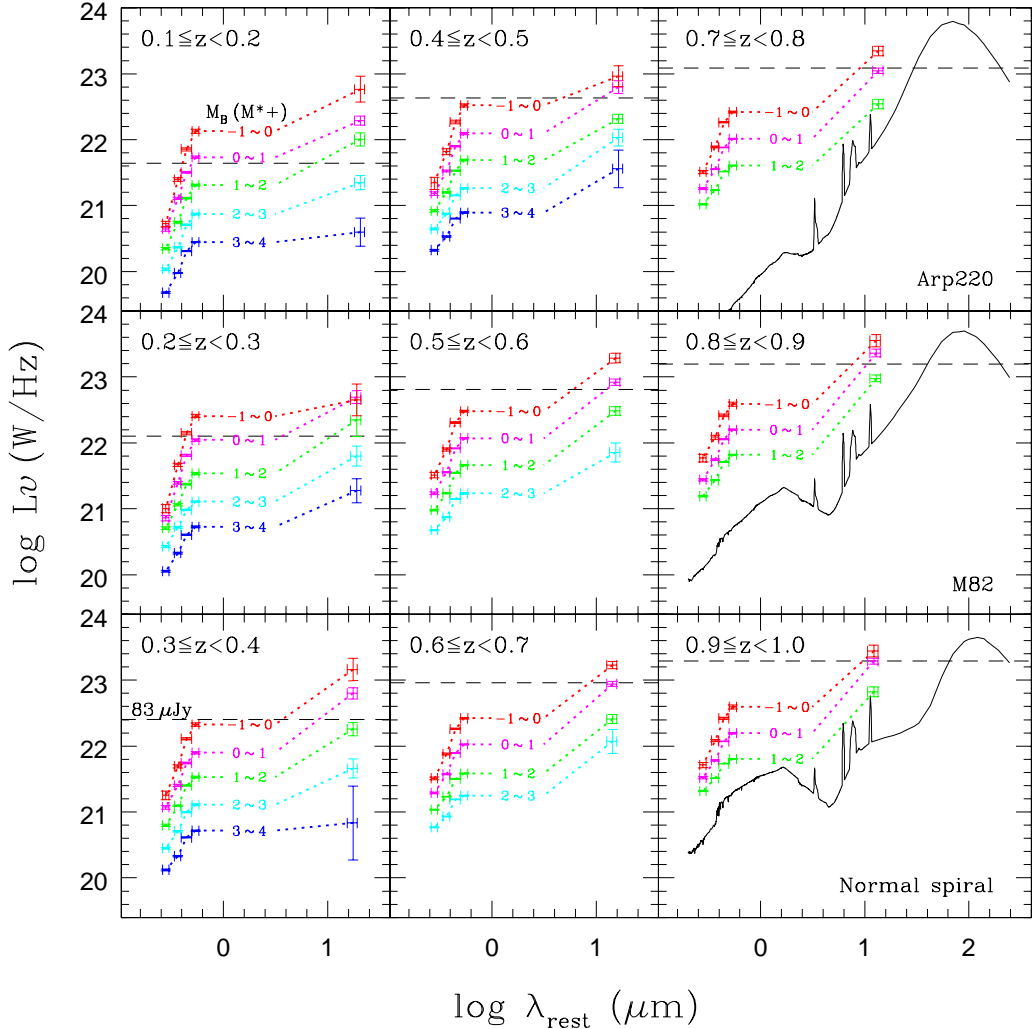}
\caption{Average luminosities at rest-frame 2800\AA, 
Johnson $U$, $B$, $V$-bands, and at observed-frame 24\,$\mu$m for 
different magnitude bins (relative to an evolving
$M^\ast_B$) and redshift slices. The label ``$-1\sim\,$0'' refers to the
magnitude bin $M^{\ast}-1 < M_{\rm B}\leq M^\ast+0$ and so forth.
Three arbitrarily shifted SED templates 
are presented for comparison, taken from 
Devriendt et al. (\citeyear{Devriendt99}). 
Horizontal dashed lines show the 83\,$\mu$Jy detection limit.}\label{sedplot}
\end{figure*}

\begin{figure} \centering
\includegraphics[width=0.45\textwidth,clip]{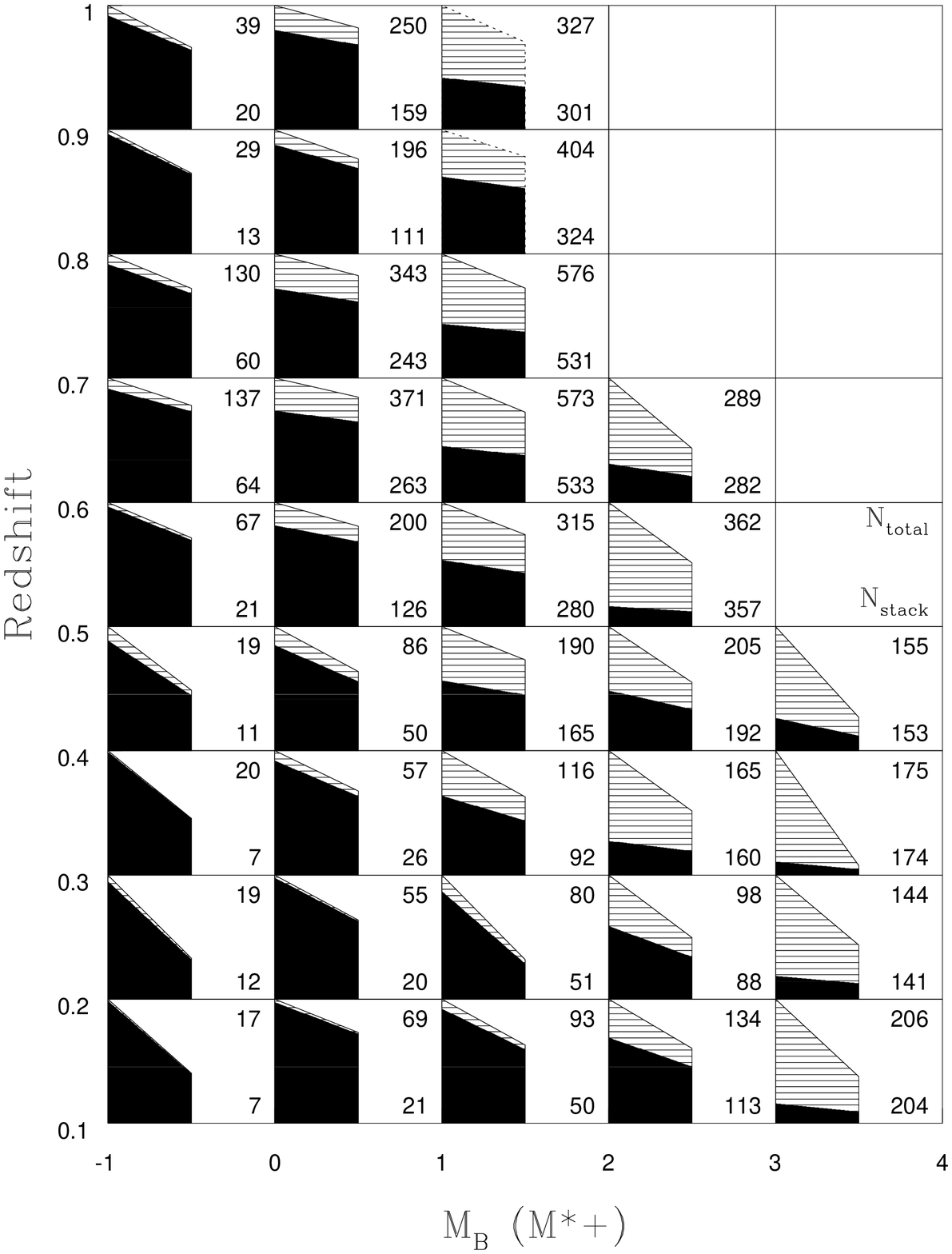}
\caption{The relative contribution from individually-detected sources (heavily
  shaded region) at 24\,$\mu$m compared to the contribution from undetected
  sources (lightly shaded region) for different magnitude bins in different
  redshift slices. In each panel, the left-hand side of shaded region shows
  the results for mean stacking scaled to unit and the right-hand 
  side shows those for median stacking linearly normalized to mean-stack
  values, incorporating bootstrap error (the left-hand side shows the 
  total mean-stack flux$+$1$\sigma$ bootstrapping error, and the right-hand 
  side shows the total median-stack flux$-$1$\sigma$); the heavily shaded 
  region shows the contribution from individually-detected 
  sources (from the left-hand side to the right-hand side, the height of 
  heavily shaded region relative to overall shaded region shows minimal 
  estimate of the contribution to maximal estimate).
  The total number of galaxies and number of individually-undetected
  galaxies stacked are shown in the upper and lower-right respectively.
  Dotted lines denote those magnitude bins having bias against red galaxies
  due to the $R$-band selection effect.  
}\label{stackfrac}
\end{figure}

\subsection{Stacking of intermediate redshift galaxies} 

To determine the characteristic thermal IR (24\,$\mu$m) flux for sub-sets of
galaxies to $z\,\sim$\,1, a sample was drawn from the COMBO-17 
redshift survey, where it overlaps with the MIPS 24\,$\mu$m
image.  We divided the total sample of 7892 galaxies into 9 
equally-spaced redshift slices from 0.1\,$\leq\,z\,<$\,1 and 
within each redshift slice, into several rest-frame 
$B$-band luminosity bins, scaled at each redshift to the characteristic 
absolute magnitude ($M^\ast_B$) of the luminosity function 
derived by the VIMOS survey (Ilbert et al. \citeyear{Ilbert}). 
Each luminosity bin covers one magnitude;
adoption of a smaller magnitude bin would not change our results.
We first illustrate which galaxies are individually detected through
Fig.~\ref{colormag}, which  shows the rest-frame color $U-V$ 
of the sample galaxies versus the absolute $B$ band magnitude.
Open circles denote galaxies individually detected above 83\,$\mu$Jy 
by MIPS; dots denote galaxies without individual detections.
The dotted lines on the right-hand side of each panel show
the value of $M^\ast_B$ in each redshift bin, while the dotted
line on the left-hand side denotes the faintest magnitude of galaxies
stacked in this paper.
In the two lowest redshift slices, 
0.1\,$\leq\,z\,<$\,0.2 and 0.2\,$\leq\,z\,<$\,0.3, 
the objects at the faint end have an abnormal color distribution, and 
have an unusually high incidence of detections at 24\,$\mu$m.
Inspection of VVDS redshifts (Le Fevre et al. \citeyear{LeFevre04}) and GEMS
morphologies (Rix et al. \citeyear{Rix04}) of galaxies in the low-luminosity,
low-$z$ bin argues 
strongly for a significant contamination by intrinsically more
luminous galaxies at higher redshift.
For very faint objects ($m_R \sim$ 23.5 mag) with a featureless
blue continuum, it is difficult to determine their redshift by either 
photometry or spectroscopy. Such objects are likely to be low-$z$ blue dwarf 
galaxies or luminous star-forming galaxies at high-$z$ ($>\sim$1.1;
where the 4000\AA\ break is redshifted out of COMBO-17 broad $I$-band coverage). 
It is supported by the phenomenon that the high incidence of detections 
at 24\,$\mu$m occurs only
in the low-luminosity and low-$z$ bins. Contamination by 
high-$z$ objects to the low-$z$ objects is difficult to quantify. 
Accordingly, we excluded these faint objects from our investigation.
It is worth noting that owing to COMBO-17's $R$-band selection, 
the faintest bin of the sample becomes incomplete at 
all redshifts $z \ga 0.6$; future work with a near IR-selected
galaxy population can remedy this incompleteness.

For each bin in magnitude and redshift, $1\arcmin .68\,\times\,1\arcmin .68$ 
24\,$\mu$m postage stamps from the PSF-subtracted image centered on 
the positions of the individually-undetected objects 
(see Fig.\ \ref{exampleimage}) were cut out and then stacked.
The size of the stamps is chosen to have sufficient area to properly 
estimate background.
Two stacks were constructed for each sub-sample: a stack in which 
each pixel is the mean, or alternatively the median
of all contributing pixels. 
The mean stack has the advantage that it includes the 
flux contributed from each galaxy, but has the disadvantage that it 
is more susceptible to residual flux from nearby sources. 
The median stack will underestimate the total flux somewhat, as the 
contributions from bright, nearly-detected sources will not be 
incorporated properly, but has the advantage that it is more robust
to contamination from nearby sources.  Furthermore, the background
values for the mean stack will be contaminated by sources, whereas
the median will be more robust.  Therefore, in the following
we adopt two measures of the stack flux:
the integrated 5$\arcsec$ aperture flux of the median stack (using 
the median stack background value for sky subtraction), 
a lower limit to the flux; and, the
integrated central flux within an aperture of radius 5$\arcsec$
of the mean-stacked image, using 
the background estimate from the median-stacked image as the background, 
giving an upper limit to the flux.  Finally, 
the empirically-derived correction of a
factor of 1.881 was used to aperture correct the estimates
of stack flux to total.
Fig.~\ref{examplestack} shows 4 examples of median-stacked images.
All median-stacked and mean-stacked images are electronically provided
in Appendix. Stack fluxes and their uncertainties are listed in 
Table~\ref{tableflux}, along with total object numbers and stack numbers
in each bin.
Note that these mean flux estimates do not contain the
flux contribution from the individual detections from each 
galaxy sub-sample.

\section{Results}

\subsection{Average SEDs and the contribution of individually-undetected sources}

Average luminosities at {\it rest-frame} 2800\AA, Johnson $U$, $B$, 
and $V$-bands and at {\it observed-frame} 24\,$\mu$m were calculated 
for each (redshift, luminosity) bin of galaxies.  
The 24\,$\mu$m luminosity includes contributions from 
both individually-detected and individually-undetected sources. 
For the latter, the mean 24\,$\mu$m intensity was estimated by stacking. 
Bootstrapping was used to estimate the uncertainty in 
total luminosity, i.e. randomly extracting the same number of sources 
from the parent set, either individually-detected sources or 
individually-undetected sources, then summing the luminosity from each subset.
%
Table~\ref{table} lists average monochromatic luminosities ($L_\nu$) 
in the rest-frame 2800\,\AA, standard Johnson $U, B, V$, and observed 
MIPS 24\,$\mu$m bands for all magnitude bins in our investigation. 
Total UV luminosities derived from the $UV 2800$\,{\AA}
luminosities are given in unit of solar luminosity. 
Total IR luminosities and SFRs are also tabulated 
(see Sect.~\ref{secIR} \& \ref{secSFR} for details).
Fig.~\ref{sedplot} shows the average 24\,$\mu$m luminosities in
different $B$-band magnitude bins over the redshift range 0.1\,$\leq\,z\,<$\,1;
luminosities in some of the optical bands (from Wolf et al. \citeyear{Wolf03})
are shown to delineate the overall SED;
the vertical errorbar is 1\,``$\sigma$'', which shows the 68\% confidence
region from bootstrapping. 
The horizonal bars show the band widths at each wavelength. 
Brighter galaxies in rest-frame $B$-band show higher mean 
24\,$\mu$m luminosities in all bands and redshifts than fainter galaxies. 
To place our measurements 
in context, three Spectral Energy Distribution (SED) templates 
from Devriendt et al.~(\citeyear{Devriendt99}) are arbitrarily scaled and 
shown alongside the observed SEDs. The three templates represent 
galaxies in three relevant modes of star formation: 
normal spiral galaxies, starbursts (M82) 
and massive starbursts (Arp\,220). 
The importance of stacking in determining the average IR
properties of distant galaxy populations can easily be discerned:
the mean 24\,$\mu$m fluxes of the faintest galaxies are
an order of magnitude below the canonical ``5$\sigma$'' detection 
limit for individual detections for galaxies with $z \la 0.7$, 
and are a factor of 5 below the individual detection limit for 
faint systems with 0.7\,$<\,z\,<$\,1.
Since the average 24\,$\mu$m luminosity accounts for contributions
from individually-detected sources, the mean 24\,$\mu$m intensity of 
individually-undetected sources derived from stacking 
is indeed even lower. 

Fig.~\ref{stackfrac} shows the 24\,$\mu$m flux fractions
derived from individual detections and from stacking. 
For the brightest galaxies ($M^\ast-1\,<\,M_B\,\leq\,M^\ast$), 
almost all of the flux arises from individual 24\,$\mu$m detections
across the redshift range 0.1\,$\leq\,z\,<$\,1.  
The galaxies that are not individually
detected have low average 24\,$\mu$m luminosities; this is
expected given that a substantial fraction of the most luminous
galaxies in rest-frame $B$-band are ``red sequence'' galaxies which 
contain little recent star formation  (Bell et al. \citeyear{Bell04}). 
For sub-$L^\ast$ galaxies, the majority of galaxies in each bin are not 
detected individually. 
However, a large fraction of the total luminosity in each 
redshift and luminosity bin comes from the individually-undetected sources, 
especially for samples with $z>$\,0.4. 
For the faintest bin in each redshift slice, 
the contribution from individually-detected sources is typically
small.

COMBO-17 survey presents optically identified type I AGNs (Wolf et al. 
\citeyear{Wolf04}). Such objects are not included in our sample.
However, optically obscured type II AGNs are not eliminated from our sample. 
Type II AGNs could potentially contribute to 24\,$\mu$m flux. 
To estimate the contamination from the type II AGNs, we cross-correlated 
our sample with the extended CDFS X-ray point source catalogs from 1\,Ms 
CDFS and 250\,ks $Chandra$ observations (Lehmer et al. \citeyear{Lehmer}). 
We identified 123 galaxies in our sample as X-ray detected 
type II AGNs. We found that the average 24\,$\mu$m luminosities 
decrease approximately 5\% for galaxies at $M_B<M^\ast$+2 by removing 
the 123 objects. Low luminosity galaxies are almost not affected.
For X-ray undetected type II AGNs (to 250 ks exposure) with optical
counterparts at $m_R<$24, radio and mid-IR investigations
(Donley et al. \citeyear{Donley}; Mart\'inez-Sansigre et al. 
\citeyear{Martinez}) suggest that they are not more than X-ray detected
ones. Then type II AGN's contribution to the average 24\,$\mu$m 
luminosity is most likely up to $\sim$10\% in our analysis. 
This effect is negligible compared with uncertainties in estimating 
average IR luminosities (typically $\sim$0.3 dex; see Table~\ref{table}). 
On the other hand, it is still unclear whether IR emission in
obscured AGNs is mostly related to AGN activity or star formation
in their host galaxies. We conclude that obscured AGNs have insignificant 
effect on our conclusions.

\subsection{The relationship between IR/UV and rest-frame $B$-band luminosity}
\label{secIR}

\begin{figure} \centering
\includegraphics[width=0.4\textwidth,clip]{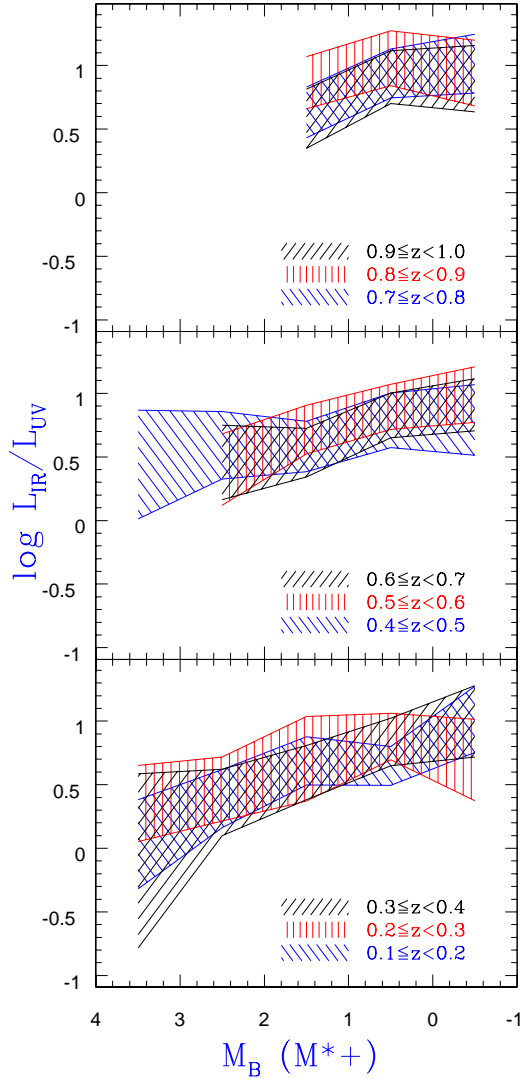}
\caption{Relationship between the mean $L_{\rm IR}/L_{\rm UV}$ ratio 
and rest-frame $B$-band luminosity, scaled to $M^\ast_B$, covering 
0.1\,$\leq\,z\,<$\,1. The $L_{\rm IR}/L_{\rm UV}$ ratio increases with
luminosity, i.e., brighter objects show a higher $L_{\rm IR}/L_{\rm UV}$ 
ratio.
}\label{IRUVMB}
\end{figure}

\begin{figure*} \centering
\includegraphics[width=0.65\textwidth,clip]{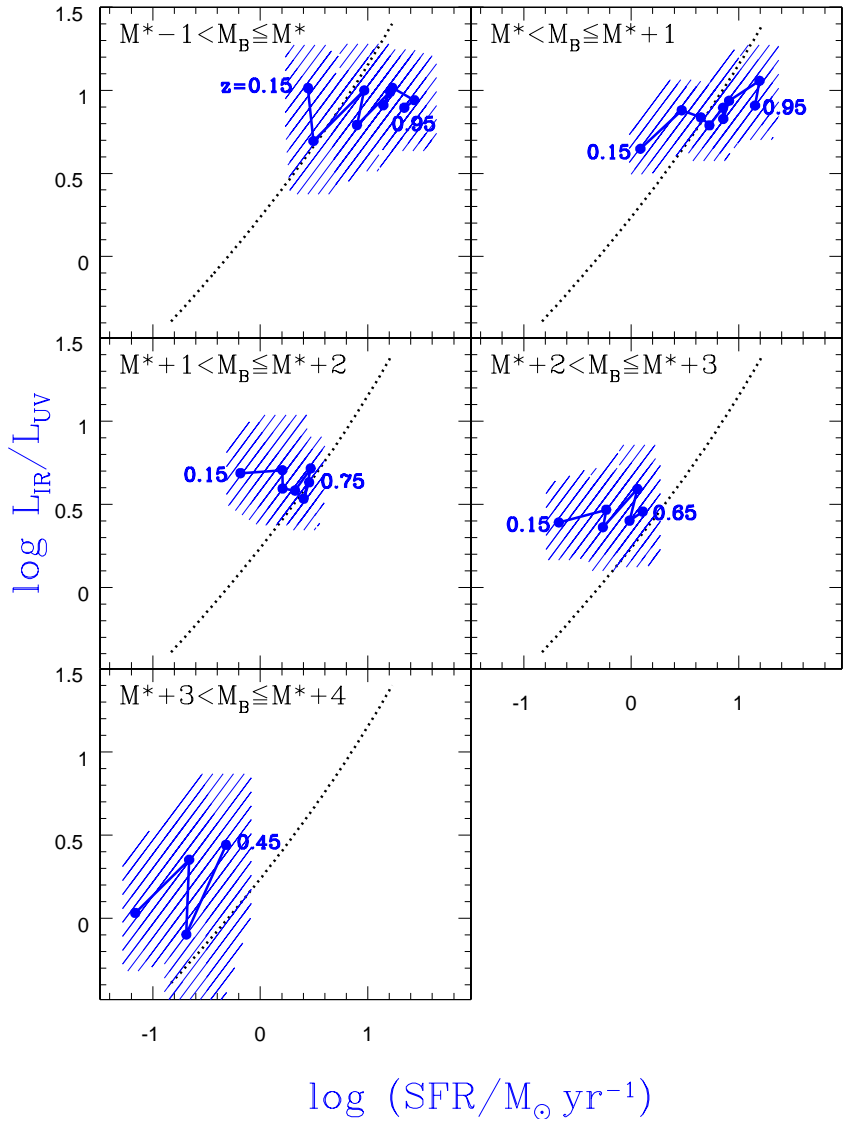}
\caption{The mean $L_{\rm IR}/L_{\rm UV}$ ratio as a function of 
SFR for galaxies in magnitude bins from 
$M^\ast-1<M_B\leq M^\ast$ to $M^\ast+3<M_B\leq M^\ast+4$.
The heavy lines show increasing redshift, from left to right, with the
minimum and maximum values of $z$ at the corresponding ends of the lines.
The local relation, recently re-derived using GALEX and IRAS data, 
is shown with dotted lines (Martin et al. \citeyear{Martin05}).}
\label{IRUVSFR}
\end{figure*}

Local star-forming galaxies exhibit a clear
correlation between the $L_{\rm IR}/L_{\rm UV}$ ratio 
and rest-frame $B$-band luminosity
(e.g. Wang \& Heckman~\citeyear{Wang96}; Bell~\citeyear{Bell03}), 
in the sense that lower-luminosity galaxies typically possess a 
lower  $L_{\rm IR}/L_{\rm UV}$ ratio albeit with large scatter.  
While we cannot test
the validity of this relation galaxy-by-galaxy at intermediate
redshift using the current datasets, we can investigate its
validity of this relation in an average sense using stacking. 

Fig.~\ref{IRUVMB} shows the $L_{\rm IR}/L_{\rm UV}$ ratio 
as a function of rest-frame $B$-band luminosity at 
0.1\,$\leq\,z\,<$\,1. 
COMBO-17's observed-frame $R$-band selection allows access
to galaxies as faint as 0.04\,$L^\ast$\ (i.e., $M^\ast +3.5)$ up to 
$z\,\sim$0.45, 0.1\,$L^\ast$\ (i.e., $M^\ast +2.5)$ up to 
$z\,\sim$0.65, and 0.25\,$L^\ast$\ (i.e., $M^\ast +1.5)$ up to $z\,\sim$0.95.
Owing to Spitzer's 
lower sensitivity and larger PSF at wavelengths beyond 24\,$\mu$m, 
the vast majority of galaxies are 
undetected at longer wavelengths. Therefore the total IR flux must 
be estimated from the 24\,$\mu$m flux by a rather large extrapolation,  
using suites of local templates\footnote{We will tackle the 
technically-challenging stacking of longer-wavelength Spitzer
data in an upcoming work.}.  Several sets of luminosity-dependent templates
are used to estimate the total IR luminosity over 
$\lambda_{\rm rest}$\,=\,$8\,-1000\,\mu$m and its uncertainty (See Le Floc'h et
al. \citeyear{LeFloc'h05} for more details and an extensive discussion 
of observational tests of 24\,$\mu$m-to-total-IR corrections; 
see also Chary \& Elbaz \citeyear{Chary01}, Papovich \& Bell \citeyear{Papovich02} and Dale et al. \citeyear{Dale05} for more discussions).
In Fig.\ \ref{IRUVMB} the shaded region shows the lower and upper limits 
for the $L_{\rm IR}/L_{\rm UV}$ ratio mainly due to the uncertainty of $L_{\rm 
IR}$. Here the lower limit refers to median-stack total flux\,$-\,1\sigma$
bootstrapping error\,$-\,1\sigma$ IR luminosity error and the upper limit to
mean-stack total flux\,+\,1$\sigma$ bootstrapping error\,+\,1$\sigma$ IR
luminosity error. 
The $L_{\rm IR}/L_{\rm UV}$ ratio is correlated with 
rest-frame $B$-band luminosity at all redshifts up to at least $z \sim 0.8$. 
Brighter galaxies on average show a higher $L_{\rm IR}/L_{\rm UV}$ ratio, 
consistent with the sense of the local relation. 
As COMBO-17's observed-frame $R$-band selection may exclude
red galaxies in the faintest bin at $z \ga 0.6$,
it is conceivable in these cases that the $L_{\rm IR}/L_{\rm UV}$ ratio is 
underestimated (and especially 
at $z \ga 0.8$) due to a deficiency of red galaxies.

\subsection{The relationship between $L_{\rm IR}/L_{\rm UV}$ and SFR at $z < 1$}\label{secSFR}

Hopkins et al. (\citeyear{Hopkins01}) and 
Adelberger \& Steidel (\citeyear{Adelberger00}) explore the 
relationship between UV-extinction and SFR, arguing for a
universal correlation between extinction and SFR.  
In this section, we explore this issue using 
our stacked dataset.

For the data at hand, we estimate the SFR from a combination of UV 
(the directly-observed emission from young stars) and IR luminosities
(dominated by reprocessed UV light from young stars; e.g., 
Gordon et al. \citeyear{Gordon00}).  The IR luminosity in turn
is estimated from the 24\,$\mu$m luminosity.
Following Bell et al. (\citeyear{Bell05}), we estimate 
the ``total'' UV luminosity, from 1216\,\AA\ to 3000\,\AA, from COMBO-17's 
2800\AA\ monochromatic luminosity using 
$L_{UV}\,=\,1.5\,\nu L_{\nu,2800}$.  This conversion assumes 
a stellar population with a constant SFR for 100\,Myr and a Kroupa IMF. 
The same stellar population is then used to calibrate the total SFR:
\begin{equation}
{\rm SFR}/(M_\sun\,{\rm yr}^{-1})\,=\,9.8\,\times\,10^{-11}\,\times\,(L_{\rm IR}+2.2\,L_{\rm UV}),
\end{equation}
where $L_{\rm IR}$ is the total IR luminosity in units of solar luminosity 
(see  Bell et al. \citeyear{Bell05} for further description 
of the assumptions behind this SFR calibration).

Fig.~\ref{IRUVSFR} shows the $L_{\rm IR}/L_{\rm UV}$ ratio as a function of
SFR split into 5 bins of optical luminosity.  The local relationship, recently
re-derived by Martin et al.~(\citeyear{Martin05}) using GALEX and IRAS data,
is also shown\footnote{ Martin et al. derive UV luminosities using $\nu
  l_{\nu}$ at 1500\AA, which yields values comparable to our definition of
  total UV for constant SFR populations.  That work uses $L_{\rm IR} = (\nu
  L_{\nu})_{60\,\mu{\rm m}}$; these values are roughly a factor of two lower
  than our definition of IR luminosity (see, e.g., Bell~\citeyear{Bell03}).
  We account for these differences in definition when comparing with the
  Martin et al. relation. }.  From the nearby universe back to $z\,\sim$\,1,
both $L_{\rm IR}/L_{\rm UV}$ and SFR increase at a given optical luminosity
relative to $L^\ast_B$.  At all redshifts and luminosities, galaxies follow
approximately the local trend for $L_{\rm IR}/L_{\rm UV}$ and SFR to within
the (considerable) uncertainties.  In particular, as one considers fainter and
fainter optical luminosity (from comparison of the different panels), galaxies
have lower SFRs and $L_{\rm IR}/L_{\rm UV}$ values, again in agreement with
locally-observed galaxies.  Thus, to within the systematic uncertainties, we
see no evidence against the proposition that $0.1<z<0.8$ star-forming galaxies
obey on average the local $L_{\rm IR}/L_{\rm UV}$--SFR correlation.  Beyond $z
\sim 0.8$, the incompleteness of the sample makes such inferences
inconclusive.  This relationship has been studied on a galaxy-by-galaxy basis
at $z \sim 0.7$ by Bell et al. (2005) for individually-detected IR-luminous
galaxies.  These distant star-forming galaxies were found to approximately
follow the local $L_{\rm IR}/L_{\rm UV}$--SFR correlation, again to within the
considerable uncertainties and with very large ($>0.5$\,dex) scatter.  We
stress that such scatter invalidates attempts at implementing the relationship
to individual very luminous infrared galaxies with SFR
$>$20\,M$_\odot$\,yr$^{-1}$.

\section{Cosmic IR luminosity density}

\begin{figure} \centering
\includegraphics[width=0.4\textwidth,clip]{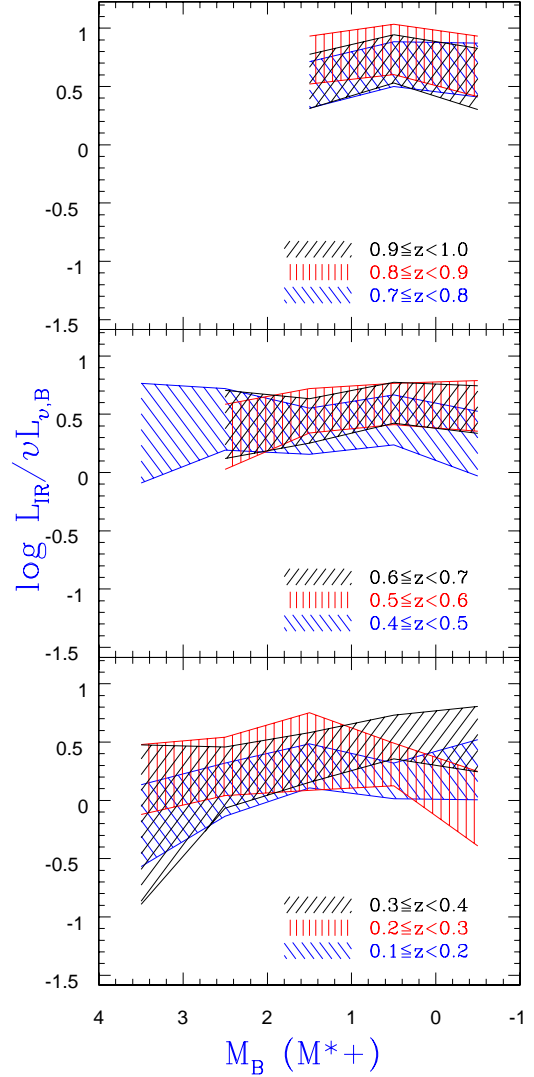} 
\caption{The relationship between mean $L_{\rm IR}/L_{\rm B}$ ratio 
and rest-frame $B$-band luminosity, scaled to $M^\ast_B$, covering 
0.1\,$\leq\,z\,<$\,1. The $L_{\rm IR}/L_{\rm B}$ ratio is roughly 
independent of luminosity to first order.} \label{IRMBMB}
\end{figure}

\begin{figure*} \centering
\includegraphics[width=0.8\textwidth,clip]{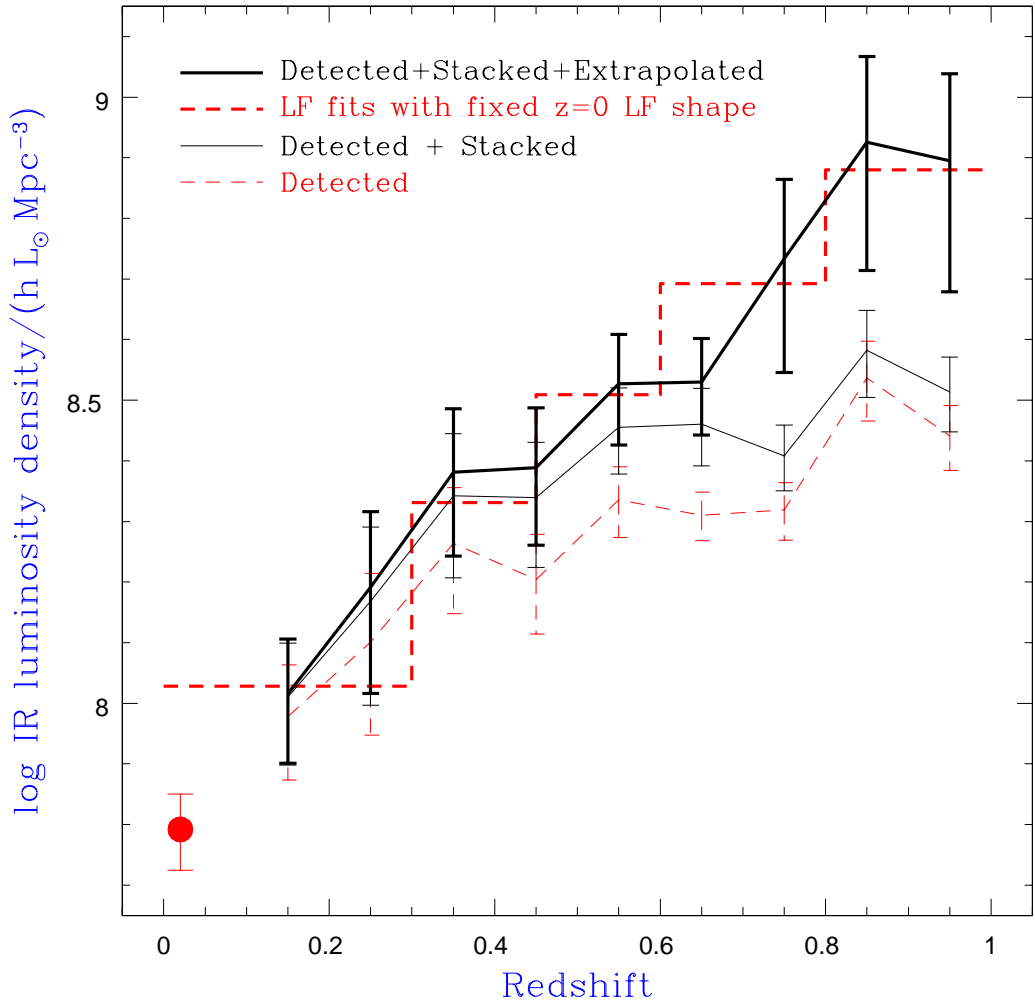}
\caption{IR luminosity density as a function of redshift. 
  The thin solid line is the IR luminosity density attributed to galaxies
  whose population-averaged $L_{\rm IR}/L_{\rm B}$ ratio is known from
  Fig.~\ref{IRMBMB}. By extrapolating the relation between $L_{\rm IR}/L_{\rm
    B}$ and $L_{\rm B}$ to the faint end, the total IR luminosity density
  derived from the $B$-band luminosity function is obtained and shown as the
  thick solid line. The thin dashed line shows the contribution from galaxies
  individually detected at 24\,$\mu$m ($>83\,\mu$Jy). The thick dashed line
  shows the IR luminosity density given by the infrared luminosity function
  (Le Floc'h et al. \citeyear{LeFloc'h05}).  The dot shows the local IR
  luminosity density (Takeuchi et al. \citeyear{Takeuchi03}).  The thick dashed
  line and thick solid line are well matched.} \label{irdensity}
\end{figure*}

The cosmic IR luminosity density is essential to estimate of the cosmic star
formation rate (e.g., Le Floc'h et al. \citeyear{LeFloc'h05}). 
It can be derived from the infrared luminosity function. Due to limited
detection depth in the infrared, the faint end slope of the infrared luminosity
function is poorly constrained, resulting in large uncertainty in estimating
the cosmic IR luminosity density 
(P\'erez-Gonz\'alez et al. \citeyear{Perez05}).
Our stacked dataset probes flux levels below the
individually-resolved flux in deep 24\,$\mu$m images. 
Hence, our results can provide new constraints on 
the cosmic IR luminosity density, which can be then used to constrain
the slope of the faint end of the infrared luminosity 
function. 

We derived average IR luminosity for galaxies of known rest-frame 
$B$-band luminosity. By Comparing with $B$-band luminosity function,
we are able to estimate the total infrared luminosity density.
Similar to Fig.~\ref{IRUVMB}, we show the relation between
$L_{\rm IR}/\nu L_{\nu,\rm B}$ ratio and the $B-$band magnitude in
Fig.~\ref{IRMBMB}. We fitted the behavior in this figure and used
the fits to extrapolate to fainter luminosities. The fits are given
by the equations
\begin{equation}
\log\,L_{\rm IR}/\nu L_{\nu,\rm B} = -0.14 \times M_{\rm B} + (0.6\pm 0.3)
\end{equation}
and
\begin{equation}
\log\,L_{\rm IR}/ \nu L_{\nu,\rm B} = -0.03 \times M_{\rm B} + (0.5\pm 0.3)
\end{equation}
for 0.1$\leq z <$0.4 and 0.4$\leq z <$0.7, respectively. Here the $B$-band
magnitude is scaled to $M^{\ast}_{\rm B}$. For redshift bins
0.7$\leq z <$1, we adopted the same slope as found for 0.4$\leq z <$0.7 
but left the normalization free, to derive:
\begin{equation}
\log\,L_{\rm IR}/ \nu L_{\nu,\rm B} = -0.03 \times M_{\rm B} + (0.8\pm 0.3).
\end{equation}
The fits indicate that galaxies at the faint end are similar to bright ones 
in the $L_{\rm IR}/ \nu L_{\nu,\rm B}$ ratio, consistent with the finding
of no correlation between the $L_{\rm IR}/ \nu L_{\nu,\rm B}$ ratio and 
$B$-band luminosity by Wang \& Heckman (\citeyear{Wang96}).
Together with the rest-frame $B$-band luminosity function of 
Ilbert et al. (\citeyear{Ilbert}), we calculated 
the total IR luminosity density. In Fig.~\ref{irdensity} 
we show the total IR luminosity density as a function 
of redshift. Due to the observed-frame $R$-band selection, 
an increasing fraction of the total IR luminosity comes from 
the extrapolated part of the LF with increasing redshift. 
Since the $B$-band luminosity function is well determined 
in several deep redshift surveys (e.g. Faber et al. \citeyear{Faber05}),
our results should not underestimate the cosmic infrared luminosity density.
Similar results were obtained with the $B$-band luminosity 
function of the DEEP2 survey (Willmer et al. \citeyear{Willmer05}).
Fig.~\ref{irdensity} exhibits that the cosmic luminosity density 
increases by a factor of 9$\pm$3 from $z$\,=\,0.1 to $z$\,=\,1. 

We compare our result with the total IR luminosity density given by
the infrared luminosity function of Le Floc'h et al. (\citeyear{LeFloc'h05}) in
Fig.~\ref{irdensity}. The two estimates of the total IR luminosity density are
identical to within the errors over the redshift range 0.1$\leq z < $1. 
This is important, as both estimates were derived in very different ways.
Le Floc'h et al. fit the evolving 
infrared luminosity function of individually-detected
galaxies using the same functional form as determined in the local 
Universe, incorporating luminosity and density evolution. 
In contrast, we directly detect the average flux of faint galaxies not
individually detected by MIPS, and then use the trend in $L_{\rm IR}/L_{\rm B}$
as a function of $L_{\rm B}$ to estimate the IR luminosity residing in 
galaxies faintwards of our optical completeness limit (it should 
be noted that the shape of the optical luminosity function to deeper
limits has been tested directly by the VVDS and e.g., in the HDF N and S, 
thus our extrapolation could be considered to be relatively well-posed).
This comparison would tend to support Le Floc'h et al.'s 
claim that the faint end slope of IR 
luminosity function is unlikely to be very steep. 
We have tested this directly by re-fitting the IR luminosity functions of
Le Floc'h et al. (\citeyear{LeFloc'h05}), using our estimated 
total IR luminosity (with our derived error bars) as a constraint.
Such an analysis gives a power law slope of 1.2$\pm$0.3 (to be compared to
the local value of 1.23; Takeuchi et al. \citeyear{Takeuchi03}; see also 
P\'erez-Gonz\'alez et al. \citeyear{Perez05} for a power law index of 1 to 1.3
up to $z$\,=\,1 and beyond).

\section{Discussion and Conclusion}

Estimating the importance of dust-obscured star formation for 
intermediate-redshift normal, or even low-luminosity galaxies on 
a galaxy-by-galaxy basis is impossible through direct thermal IR 
observations with current technology.  Using 24\,$\mu$m images
from Spitzer, we show that one can determine mean thermal IR fluxes
and hence 10-fold fainter average obscured star formation rates
by stacking 24\,$\mu$m images centered on the optical positions of 
known intermediate-redshift galaxies.  We use a sample
of galaxies from the COMBO-17 photometric redshift survey
of the Extended Chandra Deep Field South, which provides
astrometry, photometric redshift and rest-frame
2800\AA, $U$, $B$, and $V$-band absolute magnitudes for 
thousands of $z \la 1$ galaxies.  We stack MIPS 24\,$\mu$m
images for subsamples of galaxies in redshift slices and 
rest-frame $B$-band luminosity bins, allowing 
detection of average 24\,$\mu$m fluxes down to $<$\,10\,$\mu$Jy, 
an order of magnitude
deeper than those accessible on a galaxy-by-galaxy basis.

The mean total IR luminosity of these galaxy subsets 
is estimated from the observed 24\,$\mu$m luminosity,
taking into account the uncertainty of IR SED shapes.
Analogous mean UV luminosities are derived, and 
average SFRs are estimated from the UV and IR luminosities of each subsample.
We use these data to examine the correlations 
among optical luminosity, dust obscuration, SFR and redshift.

We find that the correlation between dust obscuration, i.e. 
the ratio of $L_{\rm IR}/L_{\rm UV}$, and rest-frame $B$-band 
luminosity seen in local star-forming galaxies 
holds over all redshifts $z \la 1$, with brighter 
galaxies showing a higher $L_{\rm IR}/L_{\rm UV}$ ratio. 
Our averaged 24\,$\mu$m detections show directly that even 
in low luminosity galaxies (to 0.05\,$L^\ast$) the majority of
the bolometric luminosity from young stars is re-radiated in the thermal
IR. Nonetheless, the decrease of $L_{\rm IR}/L_{\rm UV}$ ratio with
decreasing $L_{\rm B}$ implies that globally star formation in faint objects
is lower than the estimate one would derive the level of dust obscuration
typical of normal galaxies.

We explore the correlation between average 
$L_{\rm IR}/L_{\rm UV}$ ratios and SFRs
for the different optically-selected subsamples.
Different subsamples populate different parts
of the $L_{\rm IR}/L_{\rm UV}$--SFR plane; however,
our data indicate that this correlation does not evolve much 
between $z = 1$ and the present day.

In closing, we briefly consider some of the factors determining the degree of
dust obscuration indicated by $L_{\rm IR}/L_{\rm UV}$ and roughly 
parameterized as an optical depth, $\tau$. 
The optical depth of a galaxy $\tau \propto \Sigma_{\rm gas}\,Z\,\alpha$, 
where $\Sigma_{\rm gas}$ is gas density, $Z$ is metallicity and $\alpha$ 
is a geometric term to account for the gas and dust distribution relative 
to massive stars.  Star formation intensity is strongly correlated 
with gas density, e.g., the Schmidt law (Kennicutt~\citeyear{Kennicutt98b}). 
Since our results are drawn from a large number of galaxies, 
geometric effects may cancel out, to first order. 
Then the correlation of $L_{\rm IR}/L_{\rm UV}$, or  $\tau$ with SFR (i.e.,
more intense star-forming environments show larger dust obscuration) 
may suggest that gas density drives 
$L_{\rm IR}/L_{\rm UV}$ correlation to a much greater extent than 
$Z$, at least over the magnitude and redshift ranges we 
consider in this work.

Independent from the infrared luminosity function, we estimate the cosmic
infrared luminosity density from the rest-frame $B$-band luminosity function.
An increase by a factor of $9\pm$3 is found for the comoving infrared 
luminosity density from $z$\,=\,0.1 to $z$\,=\,1,  consistent with
Le Floc'h et al. (\citeyear{LeFloc'h05}). Based on our estimate of 
the comic infrared luminosity density, it is suggested that the 
infrared luminosity function of Le Floc'h et al. (\citeyear{LeFloc'h05}) is
well determined at intermediate redshifts, supporting their claim 
that faint end slope of the luminosity function is relatively flat.
Our result suggests a power law index of $1.2\pm 0.3$.

\begin{acknowledgements}

We thank John Peacock for helpful discussions. We thank the referee for
helpful comments, which improved this manuscript. Support for this work was
provided by NASA through contract number 960785 issued by JPL/Caltech.
E.\ F.\ B.\ was supported by the European
Community's Human Potential Program under contract
HPRN-CT-2002-00316, SISCO.
This publication made use of NASA's Astrophysics Data System
Bibliographic Services.  

\end{acknowledgements}



\clearpage

\begin{deluxetable}{ccrrcccc}
\centering
\tabletypesize{\scriptsize}
\tablewidth{0pt}
\tablecaption{Total object numbers, stack object numbers, 24$\micron$ stack
  fluxes and background RMS for 37 subsamples defined by redshift and luminosity.
\label{tableflux}}
\tablehead{
 & & & & \multicolumn{2}{c}{Median-stack} & \multicolumn{2}{c}{Mean-stack} \\
 $z$ & $<M_{B}>$ & N$_{total}$ & N$_{stack}$ & $f_{stack}$$^\mathrm{a}$ & $\sigma_{bg}$$^\mathrm{a}$ & $f_{stack}$$^\mathrm{a}$ & $\sigma_{bg}$$^\mathrm{a}$ \\ 
 & & & & (Jy) &  (Jy) & (Jy) & (Jy) \\ 
}
\startdata
 0.15 & $-16.5$  & 206 & 204 &    3.37 &  0.76 &    2.82 &  0.87  \\
      & $-17.5$  & 134 & 113 &    7.33 &  0.99 &    7.13 &  1.08  \\
      & $-18.5$  &  93 &  50 &   11.90 &  1.59 &   12.82 &  1.69  \\
      & $-19.5$  &  69 &  21 &   12.00 &  2.02 &   14.26 &  2.72  \\
      & $-20.5$  &  17 &   7 &   22.80 &  3.65 &   18.73 &  3.82  \\
 0.25 & $-17.2$  & 144 & 141 &    4.92 &  0.97 &    5.30 &  1.02  \\
      & $-18.2$  &  98 &  88 &    8.64 &  1.10 &    9.41 &  1.31  \\
      & $-19.2$  &  80 &  51 &   15.86 &  1.37 &   16.83 &  1.76  \\
      & $-20.2$  &  55 &  20 &    8.56 &  1.89 &    8.91 &  1.99  \\
      & $-21.2$  &  19 &  12 &   10.91 &  3.22 &   11.70 &  3.42  \\
 0.35 & $-17.2$  & 175 & 174 &    1.02 &  0.82 &    0.77 &  0.96  \\
      & $-18.2$  & 165 & 160 &    5.47 &  0.90 &    5.37 &  0.98  \\
      & $-19.2$  & 116 &  92 &   12.89 &  1.05 &   13.49 &  1.05  \\
      & $-20.2$  &  57 &  26 &   17.42 &  2.02 &   17.03 &  2.23  \\
      & $-21.2$  &  20 &   7 &    6.34 &  3.92 &    4.55 &  3.95  \\
 0.45 & $-17.6$  & 155 & 153 &    2.65 &  0.79 &    2.48 &  0.73  \\
      & $-18.6$  & 205 & 192 &    5.00 &  0.68 &    5.34 &  0.80  \\
      & $-19.6$  & 190 & 165 &    9.90 &  0.75 &    9.55 &  0.83  \\
      & $-20.6$  &  86 &  50 &   15.56 &  1.39 &   15.32 &  1.34  \\
      & $-21.6$  &  19 &  11 &   16.90 &  2.71 &   16.93 &  3.44  \\
 0.55 & $-18.6$  & 362 & 357 &    3.71 &  0.48 &    3.69 &  0.61  \\
      & $-19.6$  & 315 & 280 &   10.02 &  0.54 &    9.85 &  0.59  \\
      & $-20.6$  & 200 & 126 &   14.53 &  0.95 &   15.37 &  1.04  \\
      & $-21.6$  &  67 &  21 &   12.20 &  1.96 &   11.78 &  1.99  \\
 0.65 & $-18.5$  & 289 & 282 &    3.20 &  0.67 &    3.61 &  0.77  \\
      & $-19.5$  & 573 & 533 &    6.48 &  0.44 &    6.54 &  0.50  \\
      & $-20.5$  & 371 & 263 &   14.55 &  0.65 &   14.46 &  0.66  \\
      & $-21.5$  & 137 &  64 &   11.98 &  1.27 &   13.24 &  1.29  \\
 0.75 & $-19.5$  & 576 & 531 &    6.68 &  0.56 &    6.98 &  0.54  \\
      & $-20.5$  & 343 & 243 &   14.78 &  0.67 &   14.80 &  0.77  \\
      & $-21.5$  & 130 &  60 &   10.94 &  1.39 &   12.52 &  1.60  \\
 0.85 & $-20.0$  & 404 & 324 &   10.97 &  0.53 &   11.83 &  0.61  \\
      & $-21.0$  & 196 & 111 &   12.06 &  0.92 &   12.40 &  1.02  \\
      & $-22.0$  &  29 &  13 &    5.00 &  2.38 &    4.61 &  2.55  \\
 0.95 & $-20.0$  & 327 & 301 &    8.44 &  0.62 &    8.74 &  0.76  \\
      & $-21.0$  & 250 & 159 &   12.25 &  0.86 &   12.22 &  1.00  \\
      & $-22.0$  &  39 &  20 &    6.49 &  2.67 &    6.75 &  2.75  \\
\enddata

\tablenotetext{a}{Within an aperture of radius 5$\arcsec$.}
\end{deluxetable}

\clearpage

\begin{landscape}

\begin{deluxetable}{cccccccrrrrr}
\centering
\tabletypesize{\tiny}
\tablewidth{0pt}
  \tablecaption{average luminosities and SFRs for galaxy subpopulations of
  given magnitude bins over the redshift range 0.1$\leq z<$1.
  \label{table}}

\tablehead{
 & & & & & & & & \multicolumn{2}{c}{Minimum$^\mathrm{a}$} &
 \multicolumn{2}{c}{Maximum$^\mathrm{b}$} \\
$z$ & $<M_B>$ & log\,$L_{2800\AA}$ & log\,$L_{\rm U}$ & log\,$L_{\rm B}$ & log\,$L_{\rm V}$ & log\,$L_{\rm 24\mu m,obs}$ & log\,$L_{\rm UV}$  & log\,$L_{\rm IR}$ & log\,SFR & log\,$L_{\rm IR}$ & log\,SFR \\
 & & (W/Hz) & (W/Hz) & (W/Hz) & (W/Hz) & (W/Hz) & (L$_\odot$) & (L$_\odot$) & (M$_\odot$\,yr$^{-1}$) & (L$_\odot$) & (M$_\odot$\,yr$^{-1}$) \\
}

\startdata

0.15 & $ -16.5$ & 19.681$^{+0.012}_{-0.013}$ & 19.977$^{+0.009}_{-0.009}$ & 20.308$^{+0.008}_{-0.008}$ & 20.448$^{+0.009}_{-0.009}$ & 20.65$^{+0.16}_{-0.26}$ &  8.30$^{+0.01}_{-0.01}$ &  8.17$^{+0.13}_{-0.19}$ & $-1.24^{+0.04}_{-0.04}$ &  8.56$^{+0.12}_{-0.17}$ & $-1.10^{+0.06}_{-0.07}$ \\
     & $ -17.5$ & 20.033$^{+0.018}_{-0.019}$ & 20.367$^{+0.012}_{-0.012}$ & 20.710$^{+0.011}_{-0.011}$ & 20.874$^{+0.013}_{-0.013}$ & 21.36$^{+0.10}_{-0.12}$ &  8.65$^{+0.02}_{-0.02}$ &  8.96$^{+0.11}_{-0.15}$ & $-0.73^{+0.06}_{-0.07}$ &  9.17$^{+0.10}_{-0.14}$ & $-0.62^{+0.06}_{-0.08}$ \\
     & $ -18.5$ & 20.347$^{+0.019}_{-0.020}$ & 20.747$^{+0.013}_{-0.013}$ & 21.115$^{+0.012}_{-0.013}$ & 21.315$^{+0.015}_{-0.016}$ & 22.02$^{+0.09}_{-0.11}$ &  8.97$^{+0.02}_{-0.02}$ &  9.58$^{+0.09}_{-0.12}$ & $-0.24^{+0.06}_{-0.07}$ &  9.76$^{+0.09}_{-0.11}$ & $-0.12^{+0.06}_{-0.08}$ \\
     & $ -19.5$ & 20.646$^{+0.030}_{-0.032}$ & 21.110$^{+0.017}_{-0.018}$ & 21.506$^{+0.014}_{-0.015}$ & 21.734$^{+0.017}_{-0.018}$ & 22.29$^{+0.06}_{-0.07}$ &  9.27$^{+0.03}_{-0.03}$ &  9.86$^{+0.08}_{-0.10}$ & $ 0.05^{+0.05}_{-0.06}$ &  9.99$^{+0.08}_{-0.10}$ & $ 0.13^{+0.06}_{-0.07}$ \\
     & $ -20.5$ & 20.725$^{+0.039}_{-0.043}$ & 21.395$^{+0.025}_{-0.027}$ & 21.854$^{+0.025}_{-0.026}$ & 22.128$^{+0.028}_{-0.030}$ & 22.81$^{+0.15}_{-0.24}$ &  9.34$^{+0.04}_{-0.04}$ & 10.19$^{+0.07}_{-0.09}$ & $ 0.30^{+0.05}_{-0.07}$ & 10.54$^{+0.07}_{-0.09}$ & $ 0.59^{+0.06}_{-0.08}$ \\
0.25 & $ -17.2$ & 20.054$^{+0.015}_{-0.016}$ & 20.330$^{+0.012}_{-0.012}$ & 20.605$^{+0.010}_{-0.010}$ & 20.728$^{+0.012}_{-0.012}$ & 21.31$^{+0.14}_{-0.22}$ &  8.67$^{+0.02}_{-0.02}$ &  8.87$^{+0.11}_{-0.15}$ & $-0.76^{+0.05}_{-0.06}$ &  9.22$^{+0.11}_{-0.14}$ & $-0.58^{+0.07}_{-0.08}$ \\
     & $ -18.2$ & 20.429$^{+0.017}_{-0.018}$ & 20.723$^{+0.013}_{-0.014}$ & 20.982$^{+0.011}_{-0.011}$ & 21.109$^{+0.012}_{-0.012}$ & 21.83$^{+0.13}_{-0.18}$ &  9.05$^{+0.02}_{-0.02}$ &  9.40$^{+0.10}_{-0.14}$ & $-0.31^{+0.05}_{-0.06}$ &  9.67$^{+0.09}_{-0.12}$ & $-0.15^{+0.06}_{-0.08}$ \\
     & $ -19.2$ & 20.706$^{+0.022}_{-0.023}$ & 21.061$^{+0.015}_{-0.015}$ & 21.371$^{+0.013}_{-0.014}$ & 21.537$^{+0.018}_{-0.019}$ & 22.42$^{+0.18}_{-0.32}$ &  9.33$^{+0.02}_{-0.02}$ &  9.81$^{+0.09}_{-0.12}$ & $ 0.04^{+0.05}_{-0.07}$ & 10.28$^{+0.08}_{-0.10}$ & $ 0.37^{+0.07}_{-0.08}$ \\
     & $ -20.2$ & 20.862$^{+0.040}_{-0.045}$ & 21.394$^{+0.018}_{-0.019}$ & 21.812$^{+0.017}_{-0.018}$ & 22.044$^{+0.020}_{-0.021}$ & 22.71$^{+0.09}_{-0.11}$ &  9.48$^{+0.04}_{-0.04}$ & 10.28$^{+0.08}_{-0.10}$ & $ 0.40^{+0.06}_{-0.07}$ & 10.46$^{+0.08}_{-0.10}$ & $ 0.54^{+0.07}_{-0.08}$ \\
     & $ -21.2$ & 21.005$^{+0.057}_{-0.066}$ & 21.667$^{+0.025}_{-0.027}$ & 22.149$^{+0.022}_{-0.023}$ & 22.407$^{+0.023}_{-0.024}$ & 22.72$^{+0.18}_{-0.31}$ &  9.62$^{+0.06}_{-0.07}$ & 10.11$^{+0.09}_{-0.11}$ & $ 0.33^{+0.05}_{-0.06}$ & 10.56$^{+0.08}_{-0.10}$ & $ 0.65^{+0.07}_{-0.08}$ \\
0.35 & $ -17.2$ & 20.124$^{+0.013}_{-0.013}$ & 20.331$^{+0.008}_{-0.009}$ & 20.613$^{+0.007}_{-0.007}$ & 20.715$^{+0.010}_{-0.011}$ & 21.13$^{+0.27}_{-0.85}$ &  8.74$^{+0.01}_{-0.01}$ &  8.16$^{+0.14}_{-0.20}$ & $-0.87^{+0.02}_{-0.02}$ &  9.21$^{+0.12}_{-0.16}$ & $-0.55^{+0.07}_{-0.08}$ \\
     & $ -18.2$ & 20.454$^{+0.013}_{-0.014}$ & 20.709$^{+0.010}_{-0.010}$ & 20.998$^{+0.009}_{-0.009}$ & 21.111$^{+0.011}_{-0.011}$ & 21.68$^{+0.12}_{-0.17}$ &  9.07$^{+0.01}_{-0.01}$ &  9.33$^{+0.11}_{-0.16}$ & $-0.33^{+0.05}_{-0.06}$ &  9.59$^{+0.10}_{-0.14}$ & $-0.20^{+0.06}_{-0.08}$ \\
     & $ -19.2$ & 20.797$^{+0.015}_{-0.016}$ & 21.093$^{+0.011}_{-0.011}$ & 21.402$^{+0.010}_{-0.011}$ & 21.529$^{+0.013}_{-0.013}$ & 22.27$^{+0.09}_{-0.11}$ &  9.42$^{+0.02}_{-0.02}$ &  9.93$^{+0.10}_{-0.13}$ & $ 0.15^{+0.06}_{-0.08}$ & 10.12$^{+0.10}_{-0.13}$ & $ 0.27^{+0.07}_{-0.09}$ \\
     & $ -20.2$ & 21.073$^{+0.027}_{-0.028}$ & 21.406$^{+0.016}_{-0.016}$ & 21.743$^{+0.013}_{-0.013}$ & 21.898$^{+0.016}_{-0.017}$ & 22.81$^{+0.08}_{-0.09}$ &  9.69$^{+0.03}_{-0.03}$ & 10.46$^{+0.09}_{-0.12}$ & $ 0.59^{+0.07}_{-0.08}$ & 10.62$^{+0.09}_{-0.12}$ & $ 0.71^{+0.08}_{-0.09}$ \\
     & $ -21.2$ & 21.257$^{+0.059}_{-0.069}$ & 21.693$^{+0.025}_{-0.026}$ & 22.110$^{+0.016}_{-0.017}$ & 22.329$^{+0.024}_{-0.025}$ & 23.20$^{+0.14}_{-0.20}$ &  9.88$^{+0.06}_{-0.07}$ & 10.72$^{+0.10}_{-0.12}$ & $ 0.83^{+0.07}_{-0.09}$ & 11.05$^{+0.10}_{-0.13}$ & $ 1.10^{+0.09}_{-0.12}$ \\
0.45 & $ -17.6$ & 20.325$^{+0.013}_{-0.014}$ & 20.532$^{+0.009}_{-0.009}$ & 20.806$^{+0.007}_{-0.007}$ & 20.895$^{+0.010}_{-0.011}$ & 21.65$^{+0.20}_{-0.37}$ &  8.94$^{+0.01}_{-0.01}$ &  9.16$^{+0.14}_{-0.20}$ & $-0.48^{+0.06}_{-0.08}$ &  9.69$^{+0.12}_{-0.16}$ & $-0.17^{+0.09}_{-0.11}$ \\
     & $ -18.6$ & 20.643$^{+0.013}_{-0.014}$ & 20.874$^{+0.009}_{-0.010}$ & 21.157$^{+0.008}_{-0.008}$ & 21.264$^{+0.010}_{-0.010}$ & 22.05$^{+0.11}_{-0.15}$ &  9.26$^{+0.01}_{-0.01}$ &  9.75$^{+0.12}_{-0.16}$ & $-0.02^{+0.07}_{-0.09}$ & 10.00$^{+0.12}_{-0.16}$ & $ 0.14^{+0.09}_{-0.11}$ \\
     & $ -19.6$ & 20.923$^{+0.018}_{-0.019}$ & 21.202$^{+0.010}_{-0.010}$ & 21.530$^{+0.008}_{-0.008}$ & 21.692$^{+0.013}_{-0.013}$ & 22.32$^{+0.06}_{-0.07}$ &  9.54$^{+0.02}_{-0.02}$ & 10.09$^{+0.12}_{-0.16}$ & $ 0.29^{+0.07}_{-0.09}$ & 10.21$^{+0.11}_{-0.15}$ & $ 0.37^{+0.08}_{-0.10}$ \\
     & $ -20.6$ & 21.184$^{+0.026}_{-0.027}$ & 21.528$^{+0.013}_{-0.013}$ & 21.901$^{+0.011}_{-0.011}$ & 22.094$^{+0.015}_{-0.015}$ & 22.81$^{+0.09}_{-0.11}$ &  9.80$^{+0.03}_{-0.03}$ & 10.51$^{+0.10}_{-0.14}$ & $ 0.66^{+0.07}_{-0.09}$ & 10.70$^{+0.11}_{-0.14}$ & $ 0.80^{+0.08}_{-0.11}$ \\
     & $ -21.6$ & 21.348$^{+0.085}_{-0.106}$ & 21.822$^{+0.042}_{-0.047}$ & 22.270$^{+0.028}_{-0.029}$ & 22.520$^{+0.022}_{-0.024}$ & 22.99$^{+0.13}_{-0.18}$ &  9.97$^{+0.08}_{-0.11}$ & 10.62$^{+0.10}_{-0.14}$ & $ 0.78^{+0.07}_{-0.09}$ & 10.93$^{+0.11}_{-0.15}$ & $ 1.01^{+0.09}_{-0.11}$ \\
0.55 & $ -18.6$ & 20.678$^{+0.009}_{-0.009}$ & 20.870$^{+0.007}_{-0.007}$ & 21.151$^{+0.006}_{-0.006}$ & 21.234$^{+0.009}_{-0.009}$ & 21.88$^{+0.12}_{-0.17}$ &  9.30$^{+0.01}_{-0.01}$ &  9.59$^{+0.12}_{-0.17}$ & $-0.09^{+0.06}_{-0.07}$ &  9.86$^{+0.11}_{-0.16}$ & $ 0.06^{+0.07}_{-0.09}$ \\
     & $ -19.6$ & 20.981$^{+0.011}_{-0.011}$ & 21.237$^{+0.008}_{-0.008}$ & 21.544$^{+0.006}_{-0.007}$ & 21.660$^{+0.008}_{-0.008}$ & 22.49$^{+0.06}_{-0.07}$ &  9.60$^{+0.01}_{-0.01}$ & 10.27$^{+0.11}_{-0.15}$ & $ 0.43^{+0.08}_{-0.10}$ & 10.40$^{+0.11}_{-0.14}$ & $ 0.52^{+0.08}_{-0.10}$ \\
     & $ -20.6$ & 21.235$^{+0.019}_{-0.020}$ & 21.559$^{+0.010}_{-0.011}$ & 21.920$^{+0.007}_{-0.007}$ & 22.071$^{+0.009}_{-0.010}$ & 22.92$^{+0.04}_{-0.05}$ &  9.85$^{+0.02}_{-0.02}$ & 10.72$^{+0.11}_{-0.15}$ & $ 0.82^{+0.09}_{-0.11}$ & 10.81$^{+0.11}_{-0.15}$ & $ 0.90^{+0.09}_{-0.12}$ \\
     & $ -21.6$ & 21.511$^{+0.038}_{-0.042}$ & 21.901$^{+0.021}_{-0.022}$ & 22.309$^{+0.013}_{-0.014}$ & 22.482$^{+0.015}_{-0.015}$ & 23.29$^{+0.07}_{-0.08}$ & 10.13$^{+0.04}_{-0.04}$ & 11.07$^{+0.12}_{-0.16}$ & $ 1.16^{+0.10}_{-0.13}$ & 11.22$^{+0.12}_{-0.17}$ & $ 1.28^{+0.10}_{-0.14}$ \\
0.65 & $ -18.5$ & 20.769$^{+0.008}_{-0.008}$ & 20.932$^{+0.005}_{-0.005}$ & 21.191$^{+0.005}_{-0.005}$ & 21.248$^{+0.007}_{-0.007}$ & 22.11$^{+0.14}_{-0.22}$ &  9.39$^{+0.01}_{-0.01}$ &  9.68$^{+0.10}_{-0.13}$ & $ 0.00^{+0.05}_{-0.06}$ & 10.04$^{+0.10}_{-0.13}$ & $ 0.20^{+0.07}_{-0.08}$ \\
     & $ -19.5$ & 21.034$^{+0.008}_{-0.008}$ & 21.233$^{+0.005}_{-0.005}$ & 21.503$^{+0.005}_{-0.005}$ & 21.585$^{+0.006}_{-0.006}$ & 22.42$^{+0.06}_{-0.08}$ &  9.65$^{+0.01}_{-0.01}$ & 10.13$^{+0.10}_{-0.13}$ & $ 0.36^{+0.06}_{-0.07}$ & 10.27$^{+0.10}_{-0.14}$ & $ 0.45^{+0.07}_{-0.09}$ \\
     & $ -20.5$ & 21.288$^{+0.013}_{-0.013}$ & 21.570$^{+0.007}_{-0.007}$ & 21.895$^{+0.006}_{-0.006}$ & 22.025$^{+0.008}_{-0.008}$ & 22.94$^{+0.03}_{-0.04}$ &  9.91$^{+0.01}_{-0.01}$ & 10.72$^{+0.12}_{-0.16}$ & $ 0.84^{+0.09}_{-0.11}$ & 10.79$^{+0.12}_{-0.16}$ & $ 0.89^{+0.09}_{-0.12}$ \\
     & $ -21.5$ & 21.512$^{+0.021}_{-0.023}$ & 21.882$^{+0.012}_{-0.012}$ & 22.261$^{+0.010}_{-0.010}$ & 22.422$^{+0.013}_{-0.013}$ & 23.23$^{+0.05}_{-0.06}$ & 10.13$^{+0.02}_{-0.02}$ & 11.01$^{+0.12}_{-0.17}$ & $ 1.11^{+0.10}_{-0.13}$ & 11.12$^{+0.13}_{-0.18}$ & $ 1.20^{+0.10}_{-0.14}$ \\
0.75 & $ -19.5$ & 21.021$^{+0.009}_{-0.009}$ & 21.235$^{+0.005}_{-0.005}$ & 21.517$^{+0.005}_{-0.005}$ & 21.607$^{+0.007}_{-0.007}$ & 22.55$^{+0.06}_{-0.08}$ &  9.64$^{+0.01}_{-0.01}$ & 10.21$^{+0.11}_{-0.14}$ & $ 0.40^{+0.07}_{-0.08}$ & 10.36$^{+0.11}_{-0.15}$ & $ 0.51^{+0.08}_{-0.10}$ \\
     & $ -20.5$ & 21.257$^{+0.014}_{-0.014}$ & 21.556$^{+0.007}_{-0.007}$ & 21.882$^{+0.006}_{-0.006}$ & 22.016$^{+0.008}_{-0.008}$ & 23.05$^{+0.04}_{-0.04}$ &  9.88$^{+0.01}_{-0.01}$ & 10.79$^{+0.12}_{-0.17}$ & $ 0.88^{+0.10}_{-0.13}$ & 10.88$^{+0.12}_{-0.17}$ & $ 0.96^{+0.10}_{-0.14}$ \\
     & $ -21.5$ & 21.511$^{+0.023}_{-0.024}$ & 21.890$^{+0.012}_{-0.012}$ & 22.263$^{+0.010}_{-0.010}$ & 22.422$^{+0.012}_{-0.013}$ & 23.35$^{+0.07}_{-0.08}$ & 10.13$^{+0.02}_{-0.02}$ & 11.09$^{+0.13}_{-0.18}$ & $ 1.18^{+0.10}_{-0.14}$ & 11.25$^{+0.13}_{-0.18}$ & $ 1.31^{+0.11}_{-0.15}$ \\
0.85 & $ -20.0$ & 21.193$^{+0.011}_{-0.011}$ & 21.433$^{+0.006}_{-0.007}$ & 21.710$^{+0.006}_{-0.006}$ & 21.820$^{+0.009}_{-0.010}$ & 22.98$^{+0.05}_{-0.06}$ &  9.81$^{+0.01}_{-0.01}$ & 10.64$^{+0.12}_{-0.17}$ & $ 0.75^{+0.09}_{-0.12}$ & 10.76$^{+0.12}_{-0.17}$ & $ 0.85^{+0.10}_{-0.13}$ \\
     & $ -21.0$ & 21.442$^{+0.017}_{-0.018}$ & 21.745$^{+0.011}_{-0.011}$ & 22.059$^{+0.008}_{-0.009}$ & 22.200$^{+0.010}_{-0.010}$ & 23.37$^{+0.05}_{-0.06}$ & 10.06$^{+0.02}_{-0.02}$ & 11.07$^{+0.12}_{-0.17}$ & $ 1.15^{+0.10}_{-0.14}$ & 11.21$^{+0.12}_{-0.18}$ & $ 1.26^{+0.11}_{-0.15}$ \\
     & $ -22.0$ & 21.773$^{+0.056}_{-0.064}$ & 22.086$^{+0.032}_{-0.034}$ & 22.419$^{+0.020}_{-0.021}$ & 22.589$^{+0.028}_{-0.029}$ & 23.56$^{+0.08}_{-0.10}$ & 10.39$^{+0.06}_{-0.06}$ & 11.25$^{+0.13}_{-0.18}$ & $ 1.36^{+0.10}_{-0.13}$ & 11.46$^{+0.13}_{-0.18}$ & $ 1.53^{+0.11}_{-0.15}$ \\
0.95 & $ -20.0$ & 21.318$^{+0.009}_{-0.009}$ & 21.513$^{+0.005}_{-0.005}$ & 21.734$^{+0.005}_{-0.005}$ & 21.808$^{+0.007}_{-0.007}$ & 22.83$^{+0.07}_{-0.08}$ &  9.94$^{+0.01}_{-0.01}$ & 10.45$^{+0.12}_{-0.17}$ & $ 0.67^{+0.08}_{-0.09}$ & 10.62$^{+0.13}_{-0.18}$ & $ 0.78^{+0.09}_{-0.12}$ \\
     & $ -21.0$ & 21.520$^{+0.016}_{-0.016}$ & 21.784$^{+0.010}_{-0.010}$ & 22.071$^{+0.006}_{-0.007}$ & 22.197$^{+0.010}_{-0.011}$ & 23.29$^{+0.04}_{-0.05}$ & 10.14$^{+0.02}_{-0.02}$ & 11.03$^{+0.13}_{-0.19}$ & $ 1.13^{+0.10}_{-0.14}$ & 11.13$^{+0.13}_{-0.18}$ & $ 1.21^{+0.11}_{-0.14}$ \\
     & $ -22.0$ & 21.713$^{+0.038}_{-0.041}$ & 22.084$^{+0.019}_{-0.020}$ & 22.422$^{+0.018}_{-0.019}$ & 22.594$^{+0.030}_{-0.032}$ & 23.45$^{+0.08}_{-0.10}$ & 10.33$^{+0.04}_{-0.04}$ & 11.15$^{+0.13}_{-0.18}$ & $ 1.27^{+0.10}_{-0.13}$ & 11.36$^{+0.13}_{-0.19}$ & $ 1.43^{+0.11}_{-0.15}$ \\

\enddata
\tablenotetext{a}{Minimal estimate of IR luminosity and SFR from
  median-stack total flux $-$ 1\,$\sigma$ bootstrapping error.}
\tablenotetext{b}{Maximal estimate of IR luminosity and SFR from
  mean-stack total flux + 1\,$\sigma$ bootstrapping error.}
\end{deluxetable}
\clearpage
\end{landscape}

\clearpage


\begin{center}{\bf Appendix: MIPS 24\,$\micron$ stack images}\end{center}

Here we provide MIPS 24\,$\micron$ median-stack and mean-stack images of 37
subsets of galaxies in our sample. The images are available in the 
electronic edition of the Journal.

\include{appendix}

\end{document}

%% file: appendix.tex
\begin{figure*}[] 
\includegraphics[width=0.24\textwidth,clip]{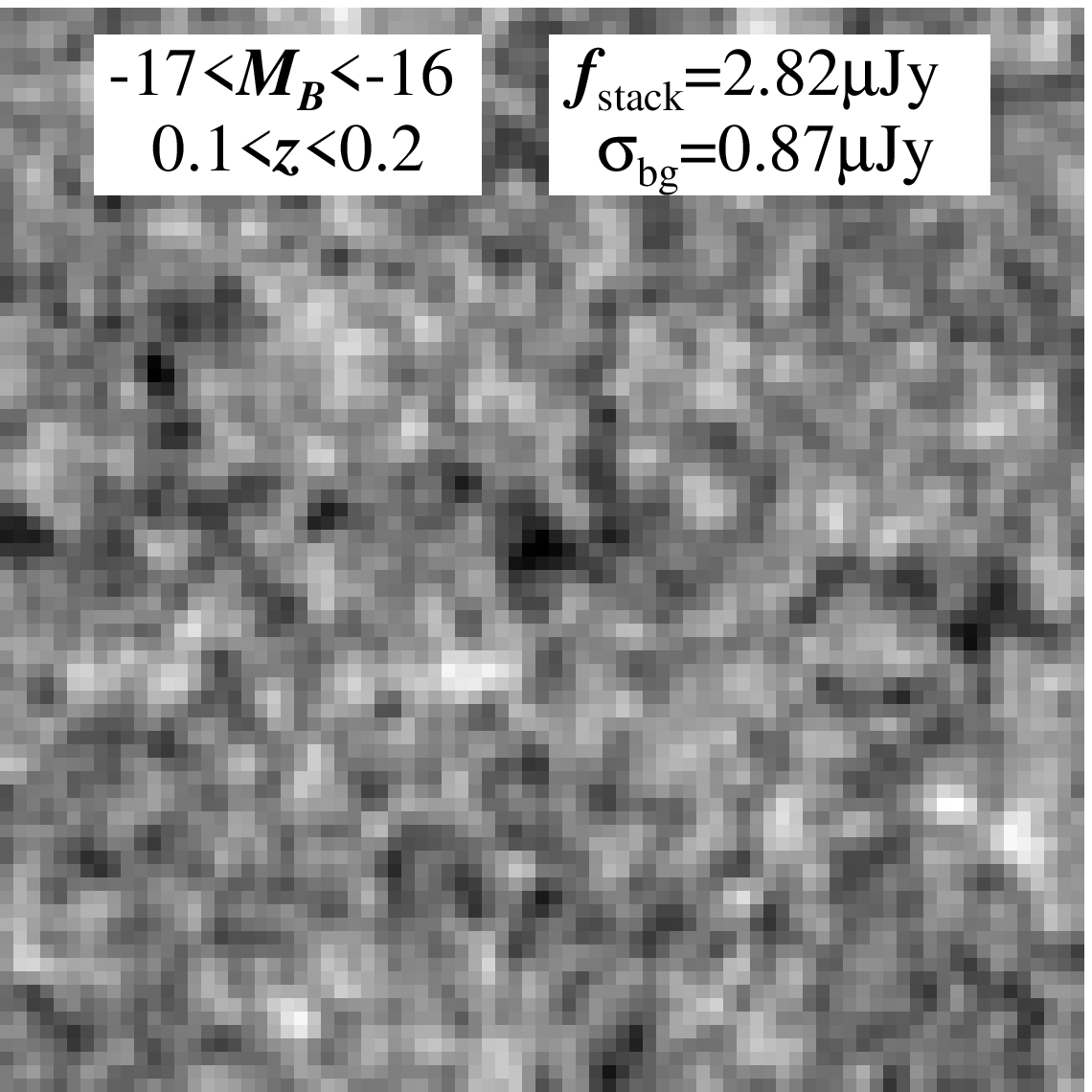}  \includegraphics[width=0.24\textwidth,clip]{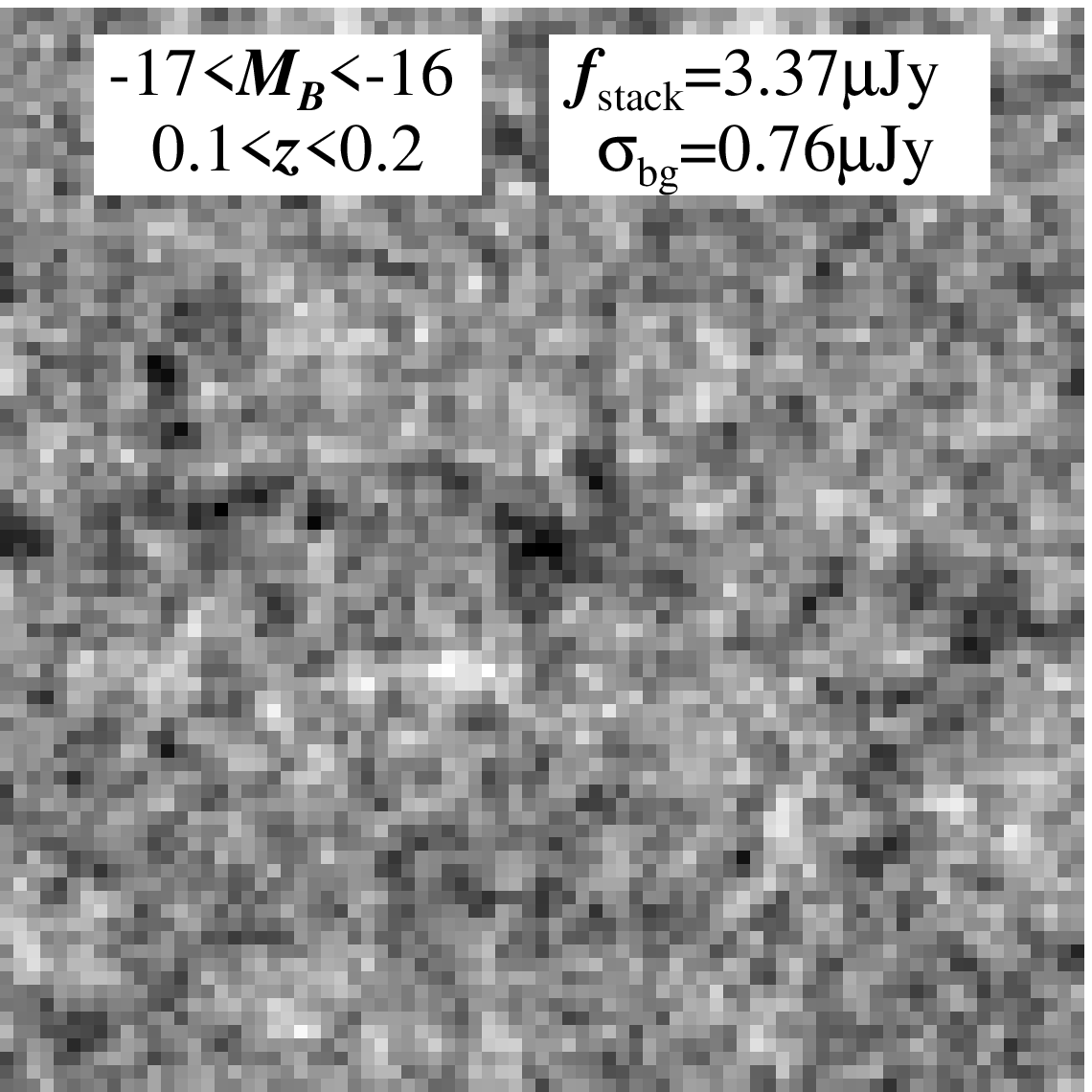} \hfill
\includegraphics[width=0.24\textwidth,clip]{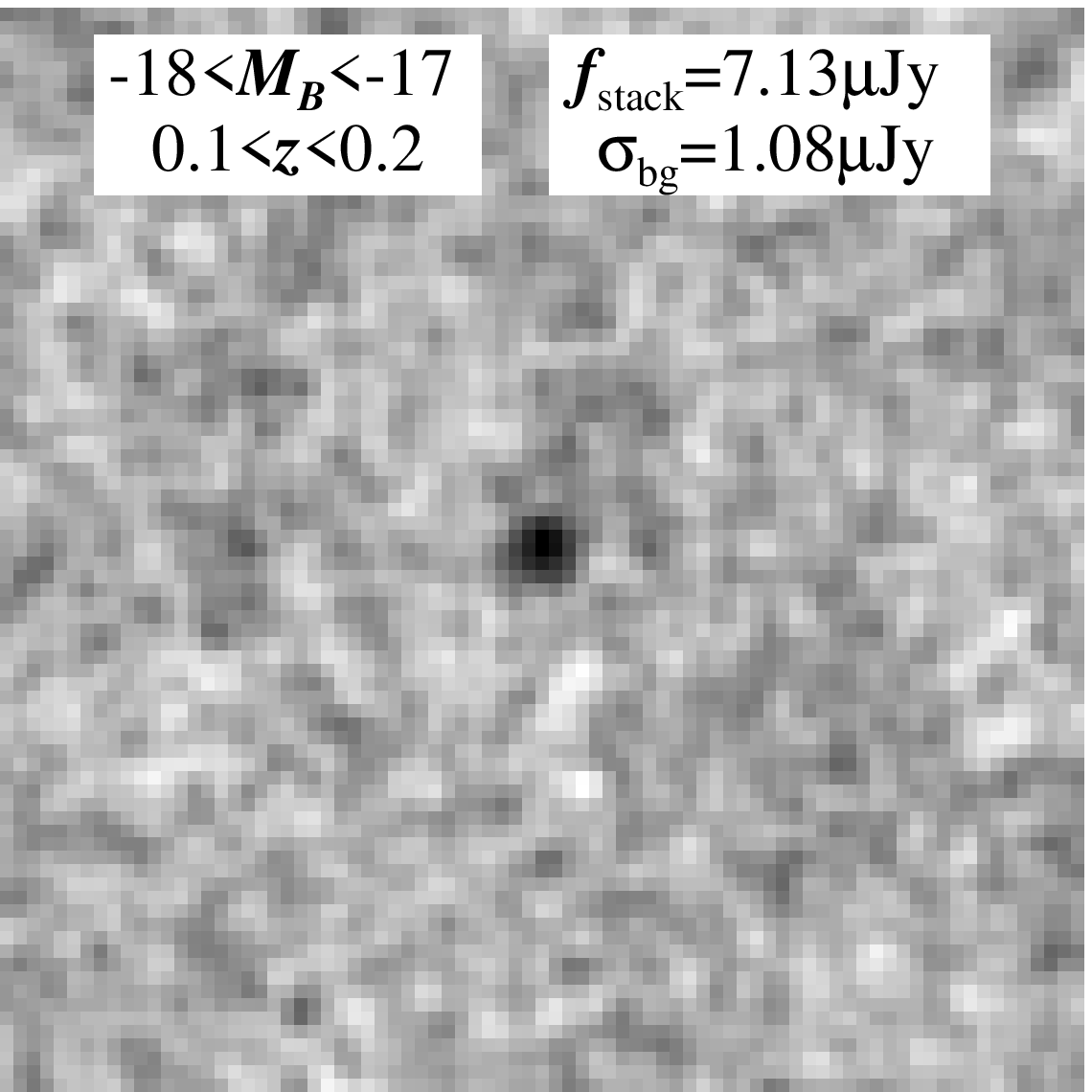}  \includegraphics[width=0.24\textwidth,clip]{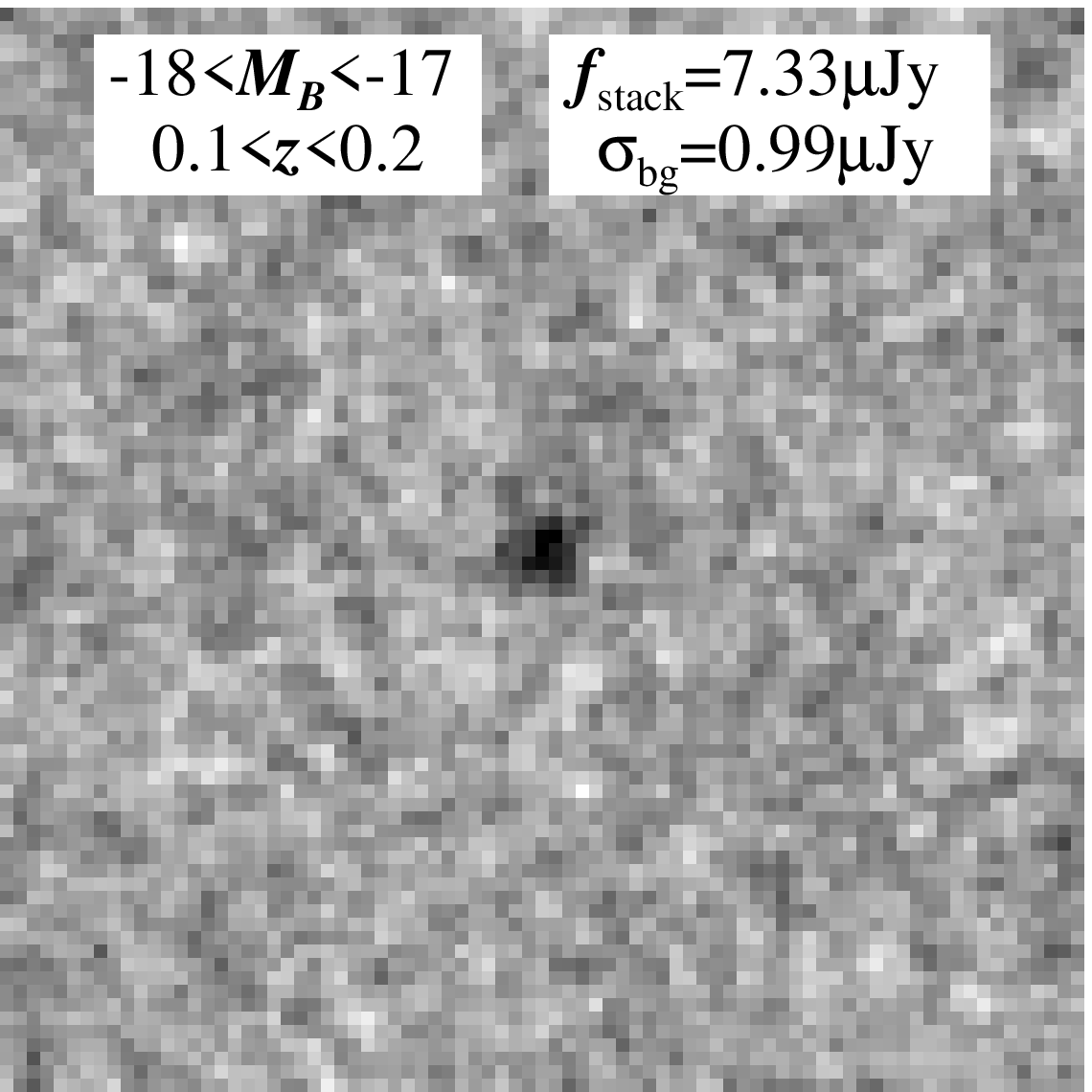}
\vskip 3mm
\includegraphics[width=0.24\textwidth,clip]{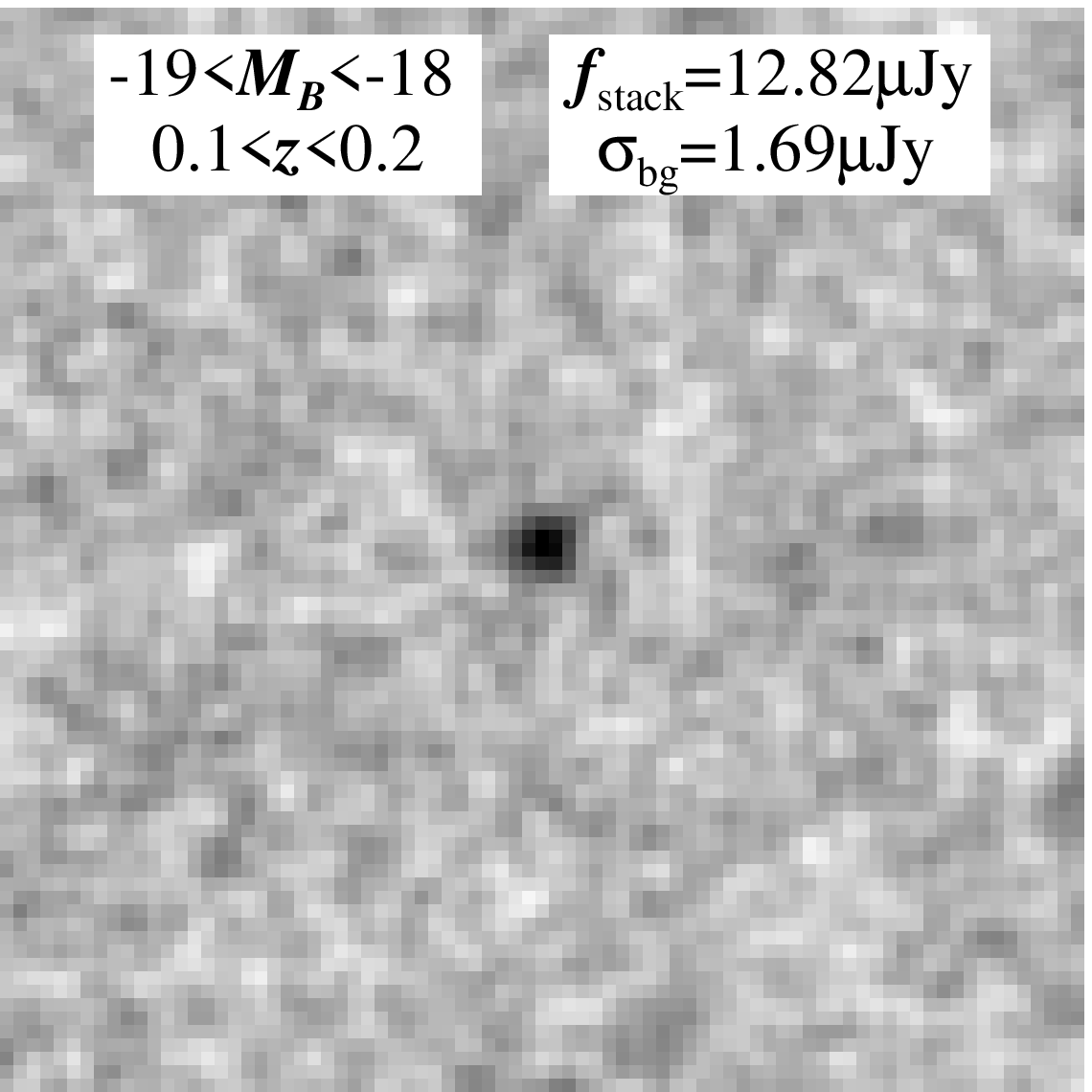}  \includegraphics[width=0.24\textwidth,clip]{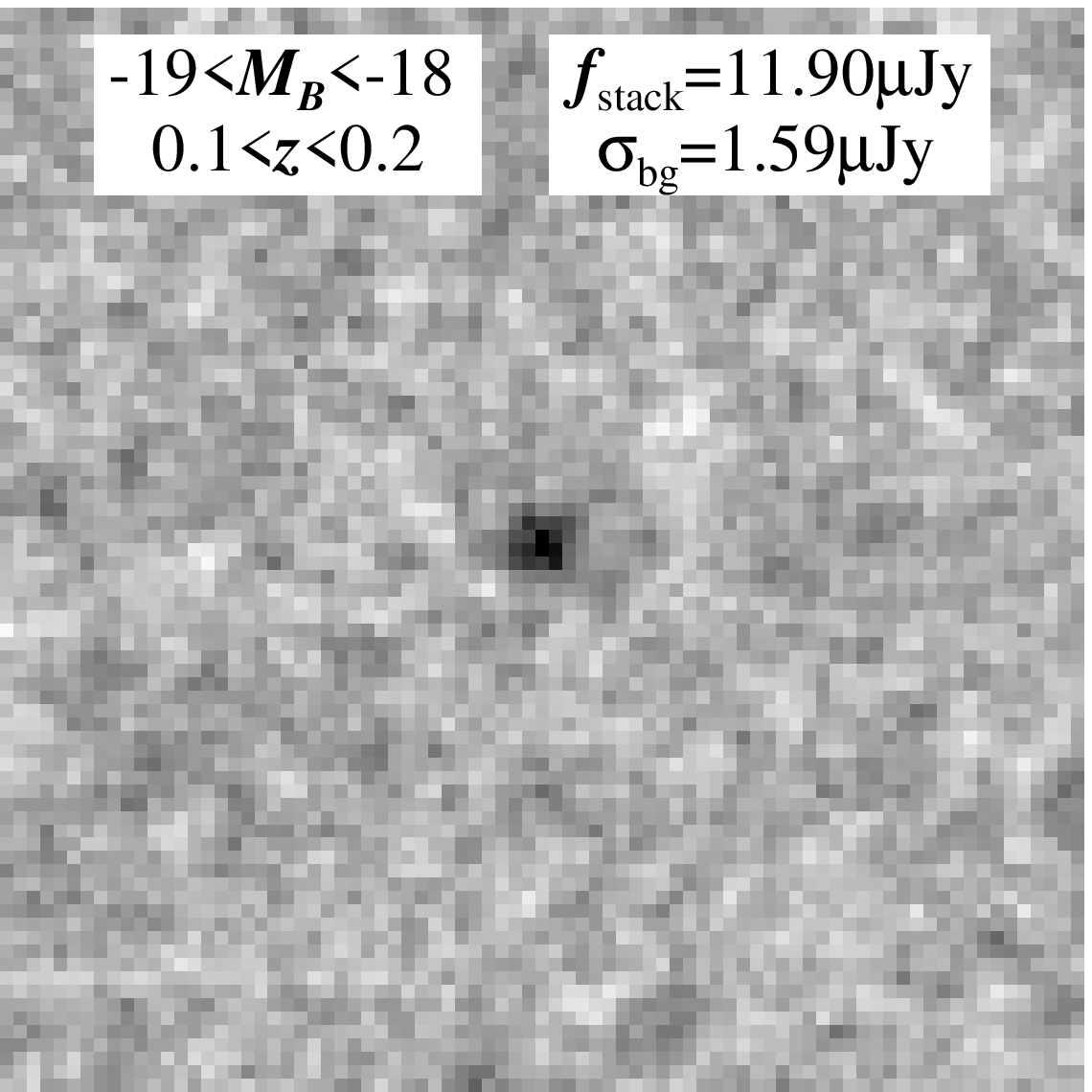}\hfill
\includegraphics[width=0.24\textwidth,clip]{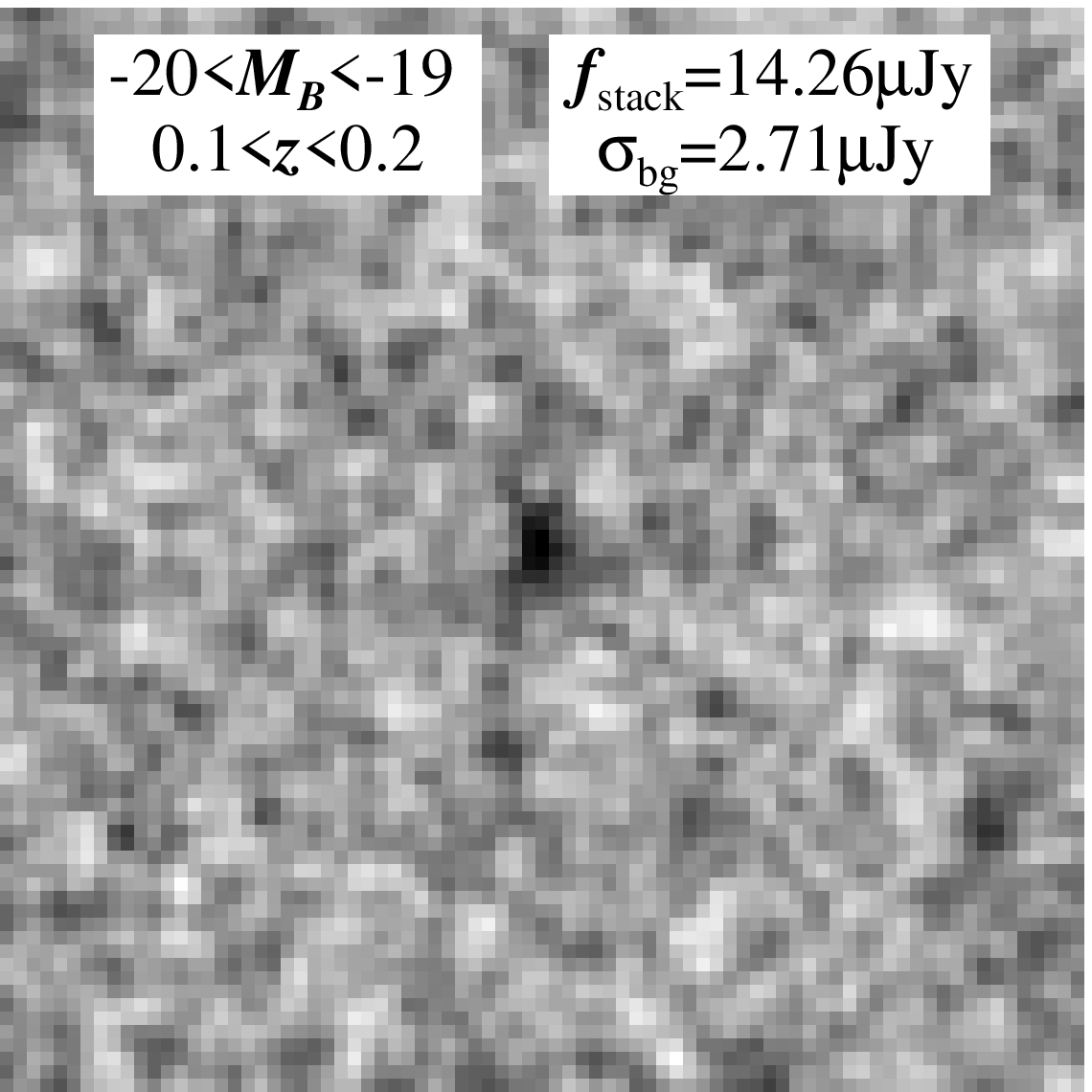}  \includegraphics[width=0.24\textwidth,clip]{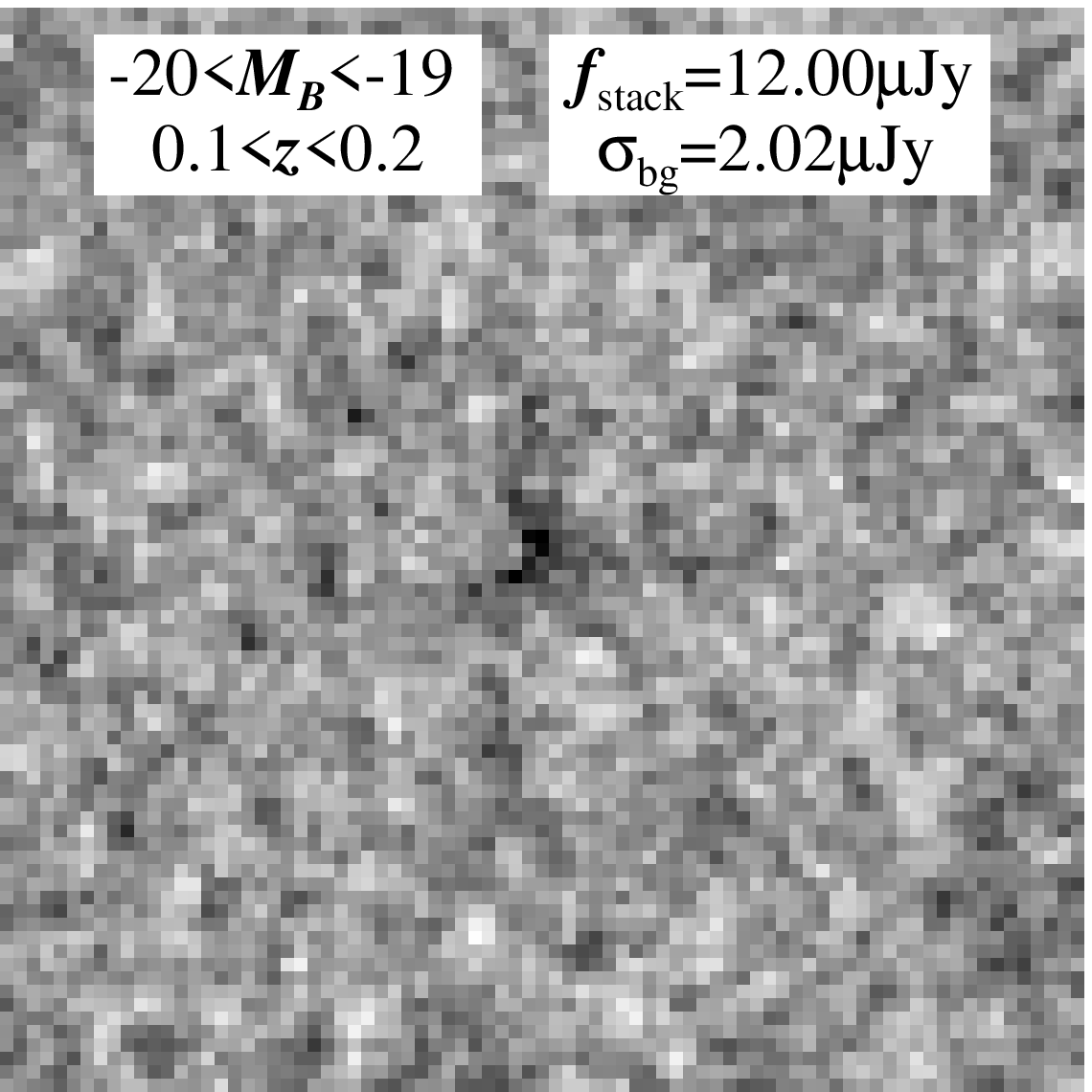}
\vskip 3mm
\includegraphics[width=0.24\textwidth,clip]{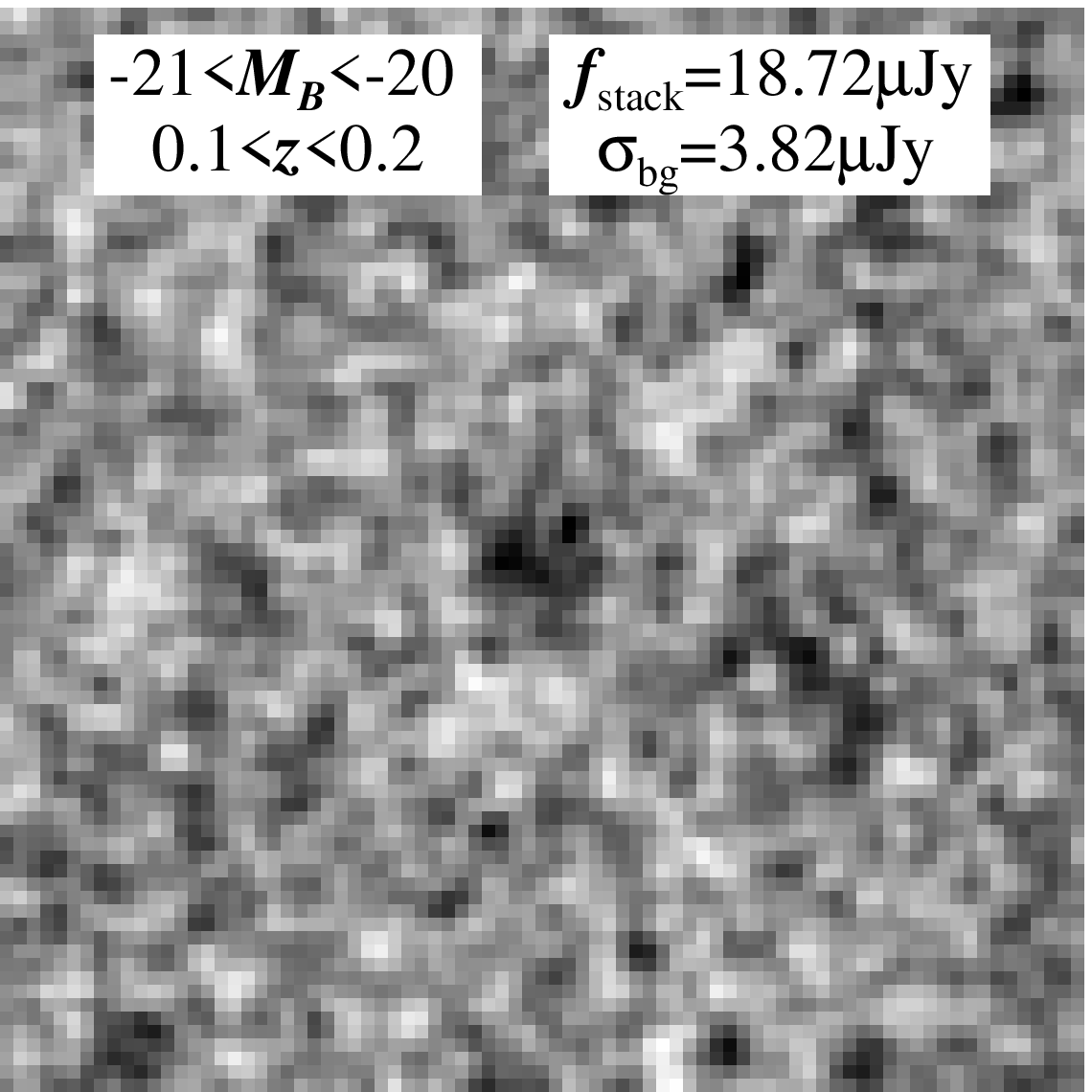}  \includegraphics[width=0.24\textwidth,clip]{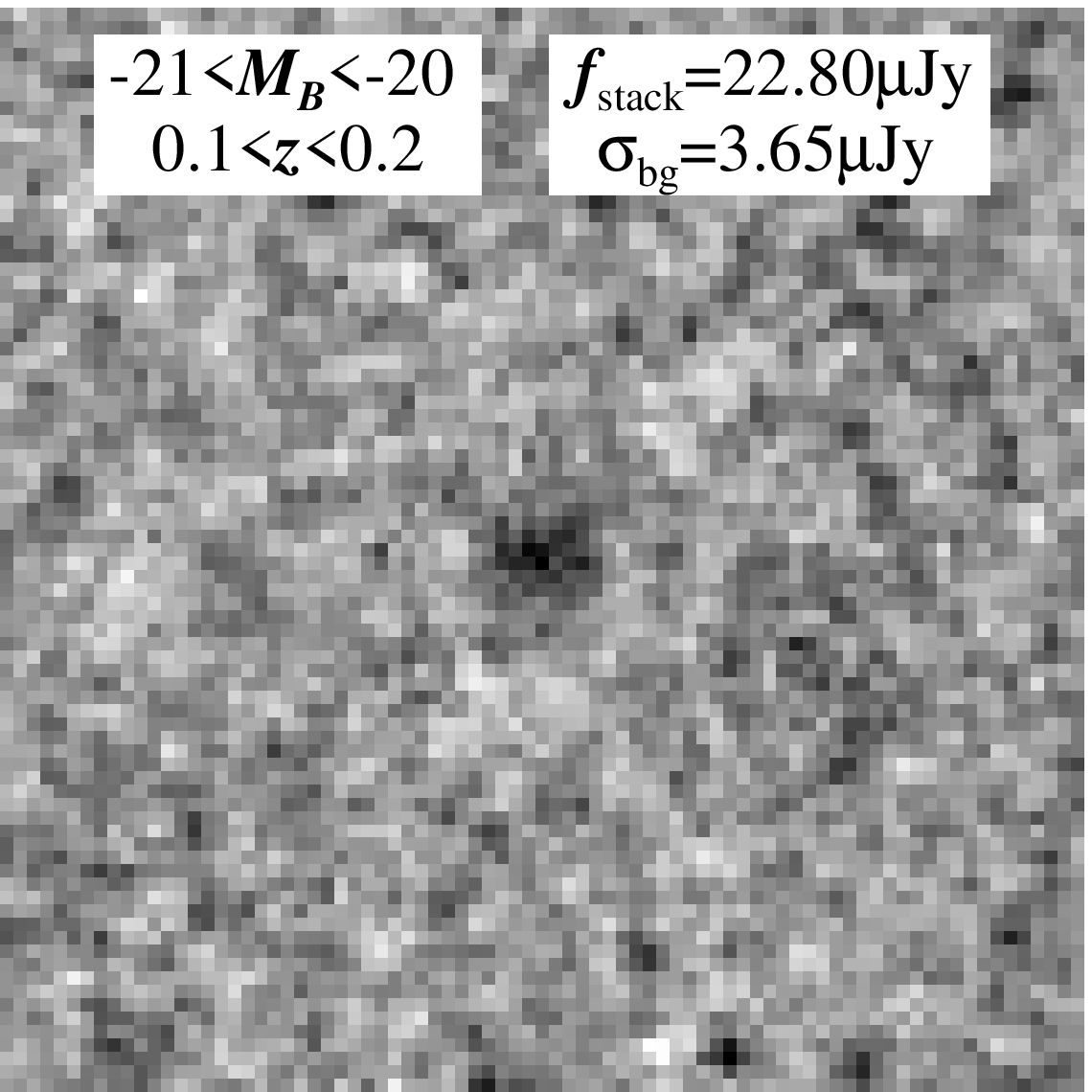}\hfill
\includegraphics[width=0.24\textwidth,clip]{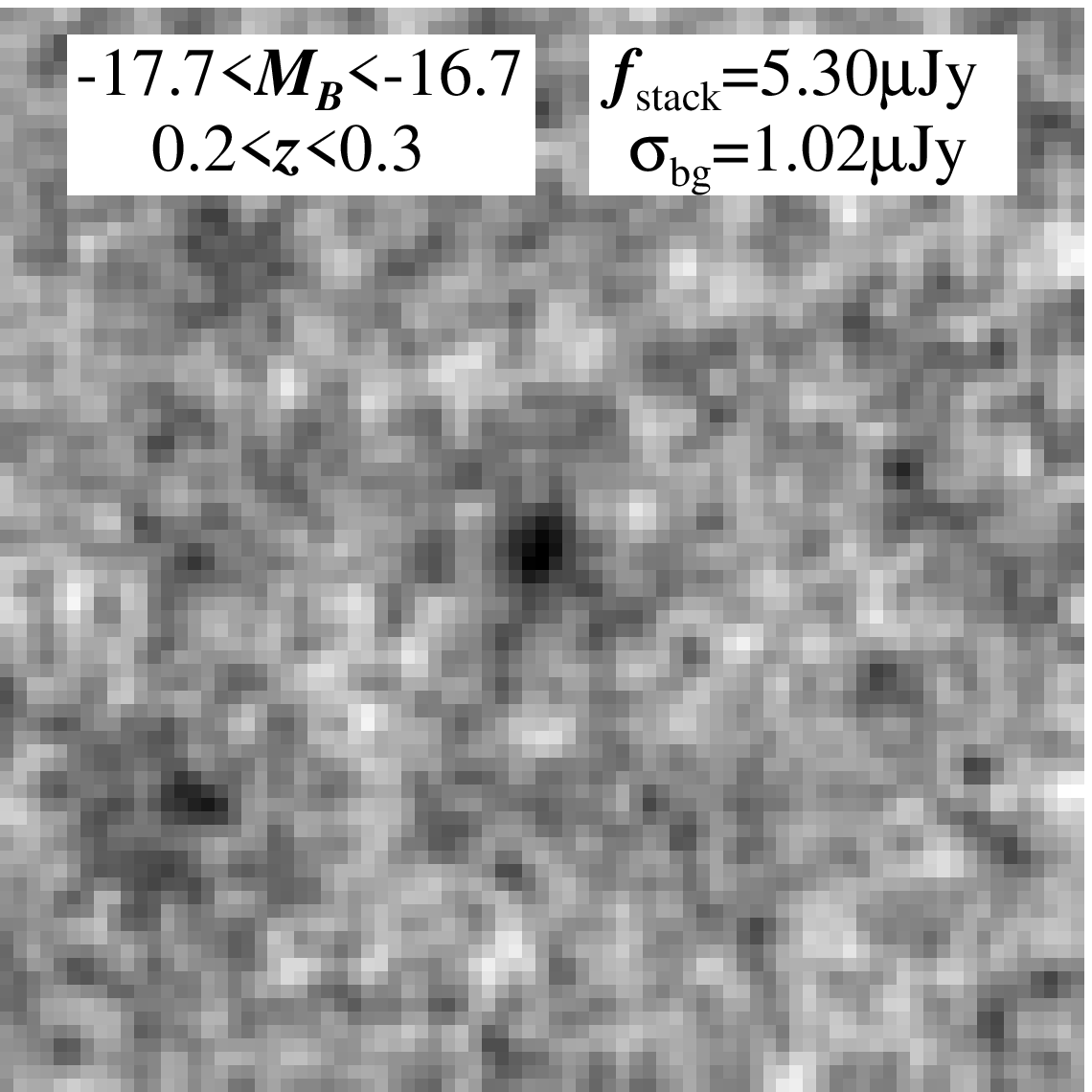}  \includegraphics[width=0.24\textwidth,clip]{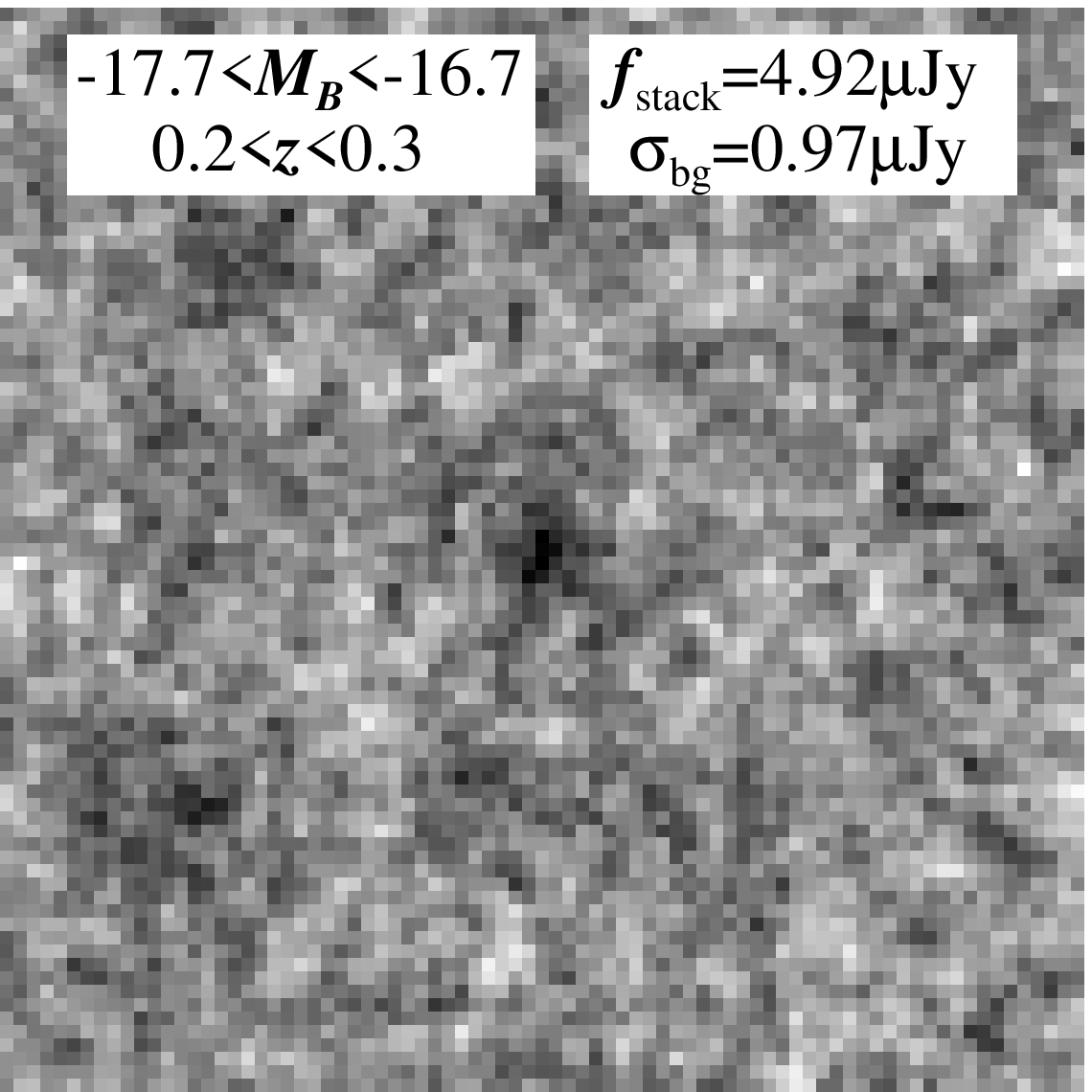}
\vskip 3mm
\includegraphics[width=0.24\textwidth,clip]{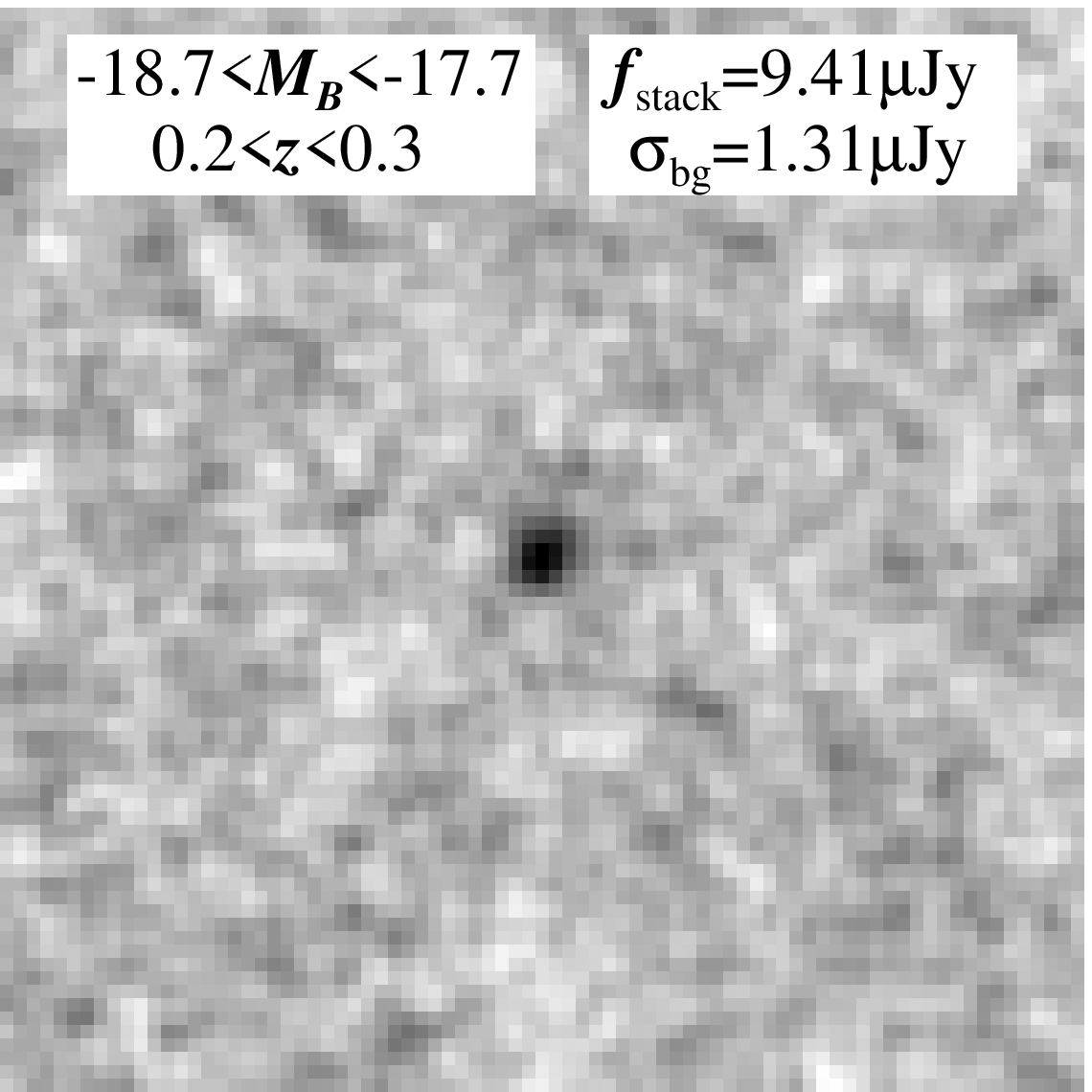}  \includegraphics[width=0.24\textwidth,clip]{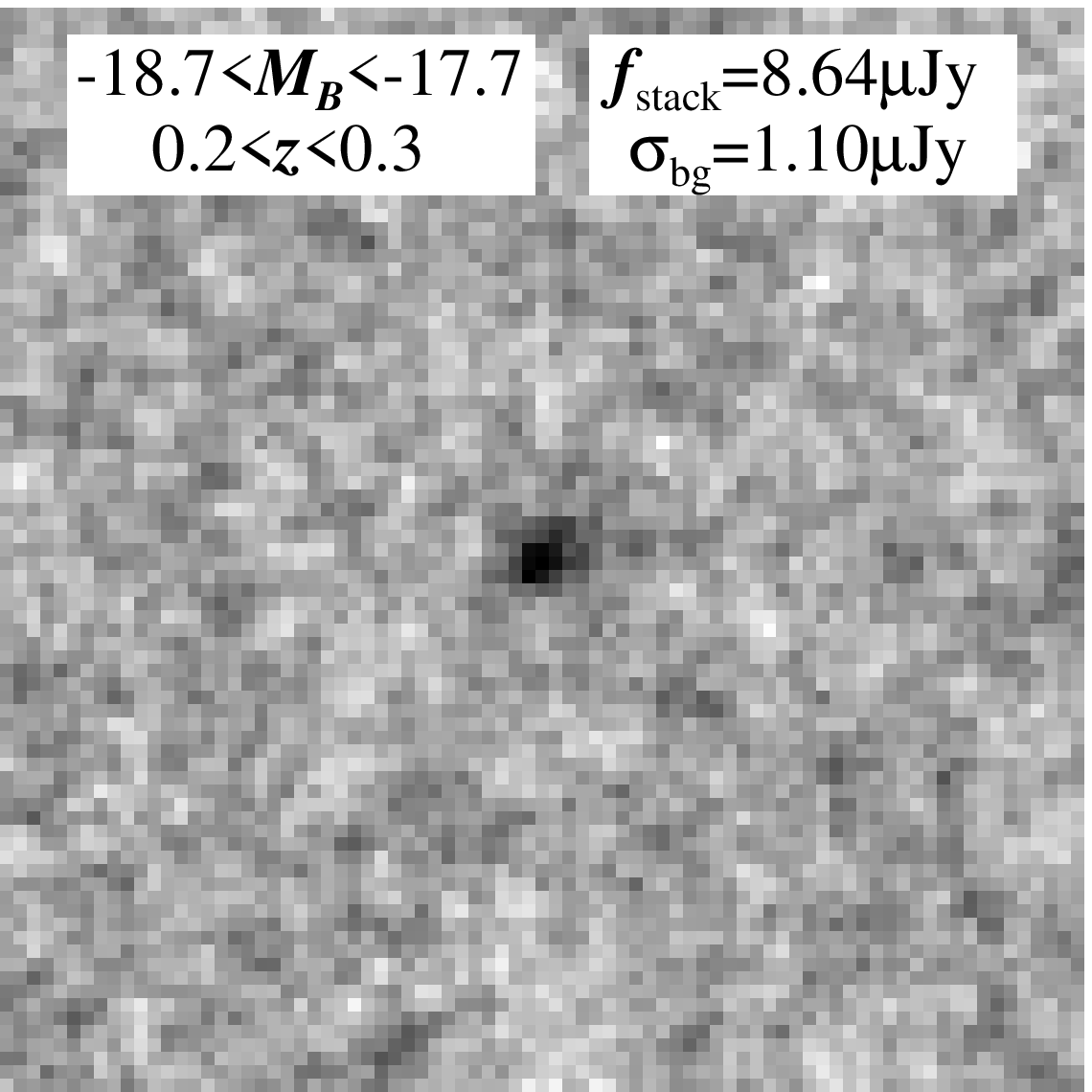}\hfill
\includegraphics[width=0.24\textwidth,clip]{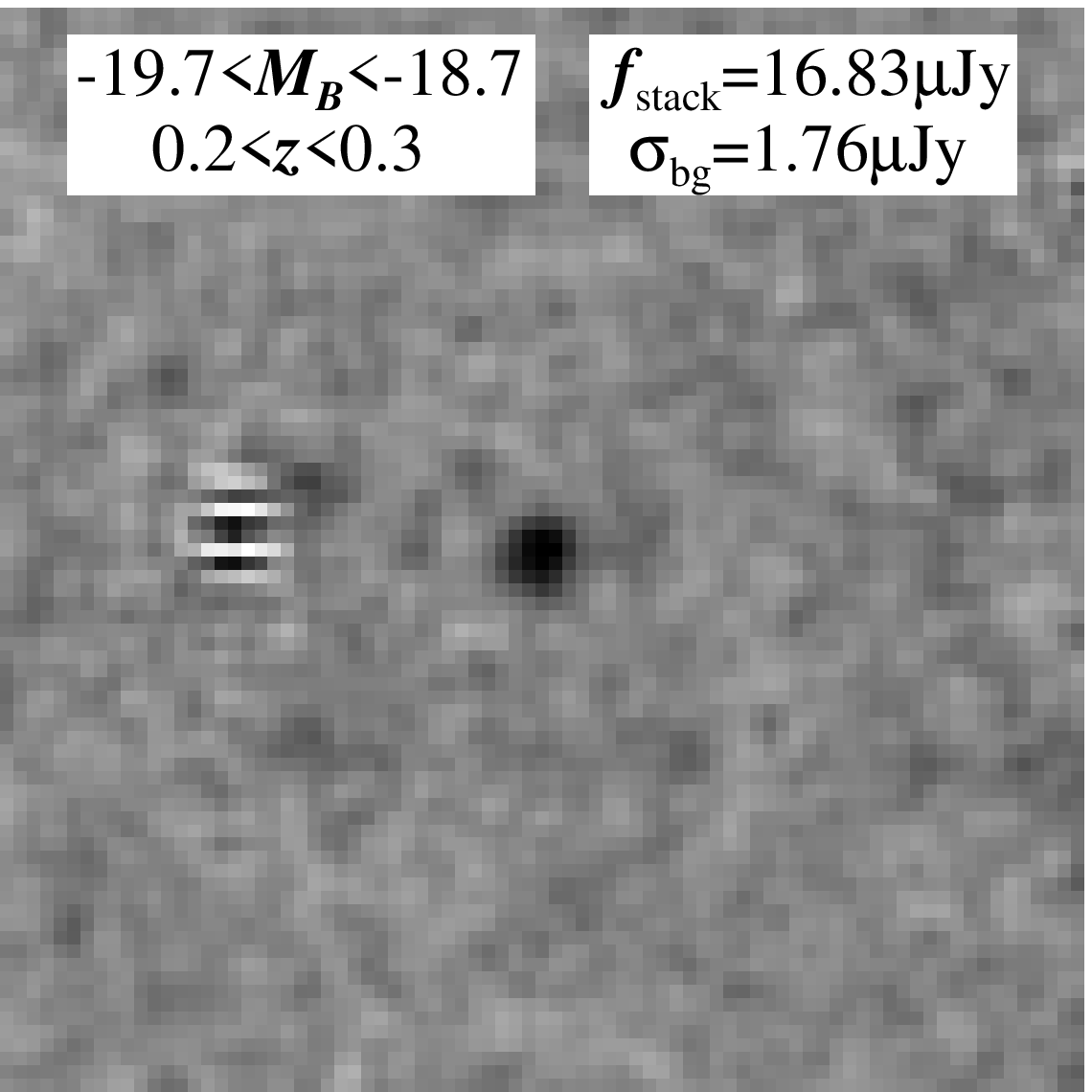}  \includegraphics[width=0.24\textwidth,clip]{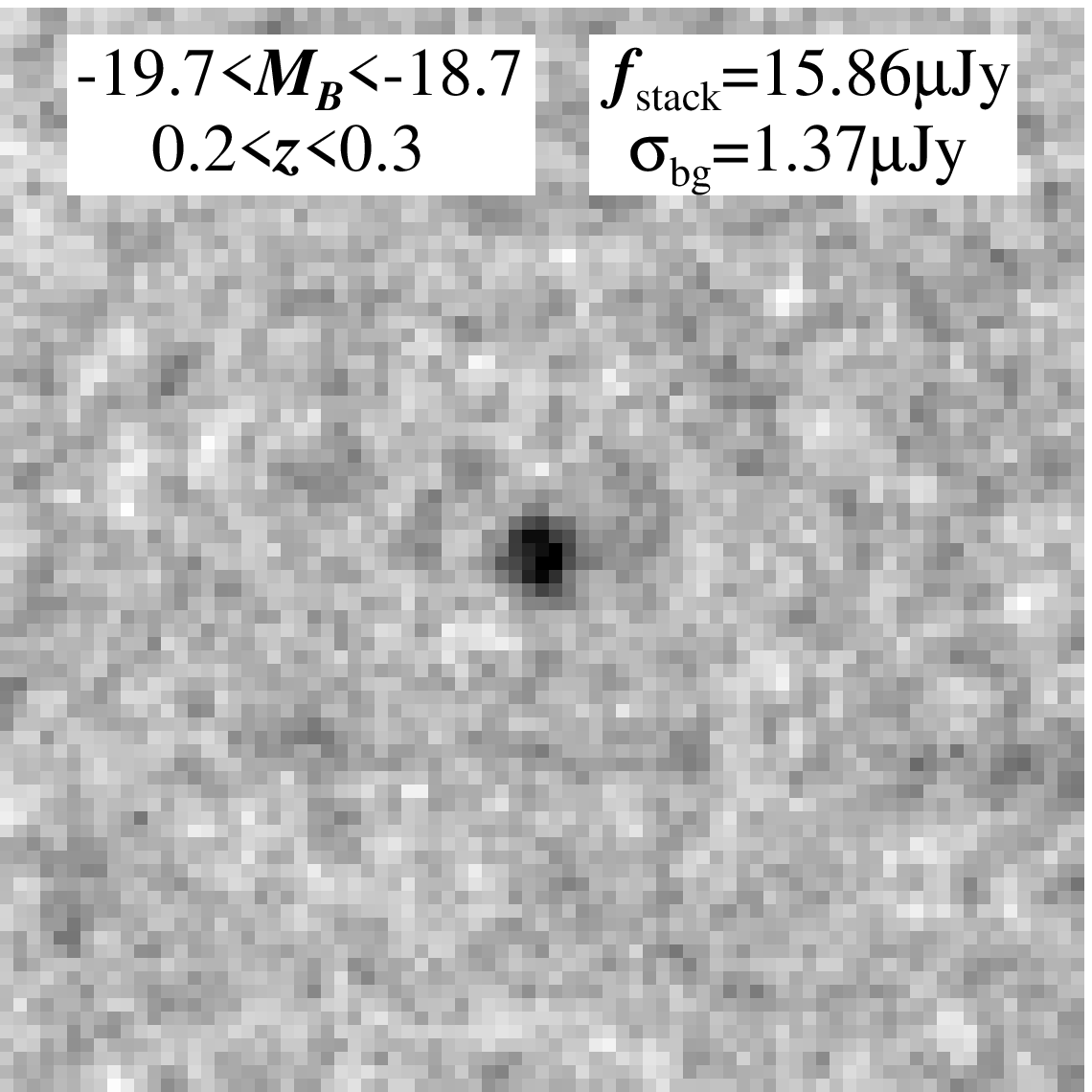}
\vskip 3mm
\includegraphics[width=0.24\textwidth,clip]{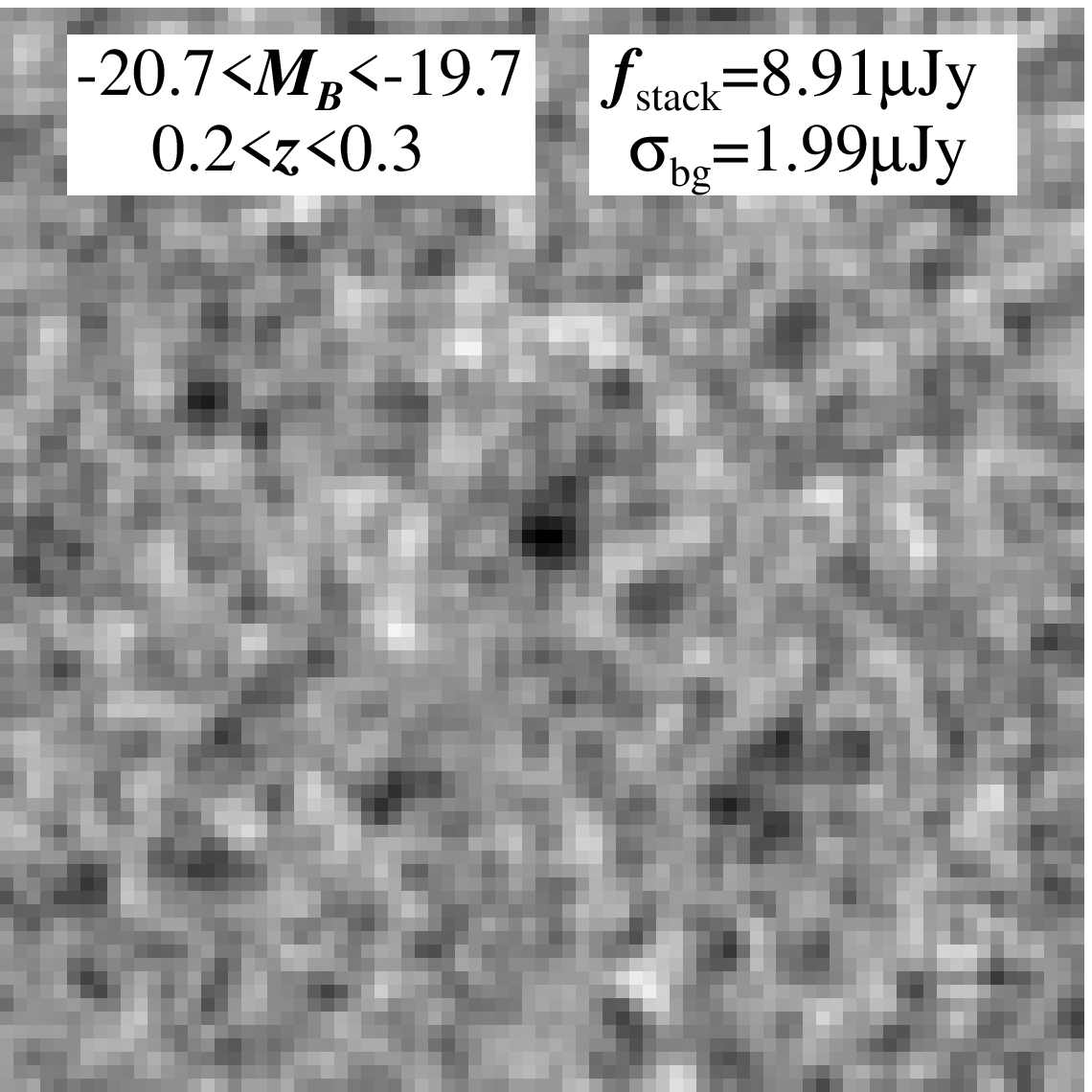}  \includegraphics[width=0.24\textwidth,clip]{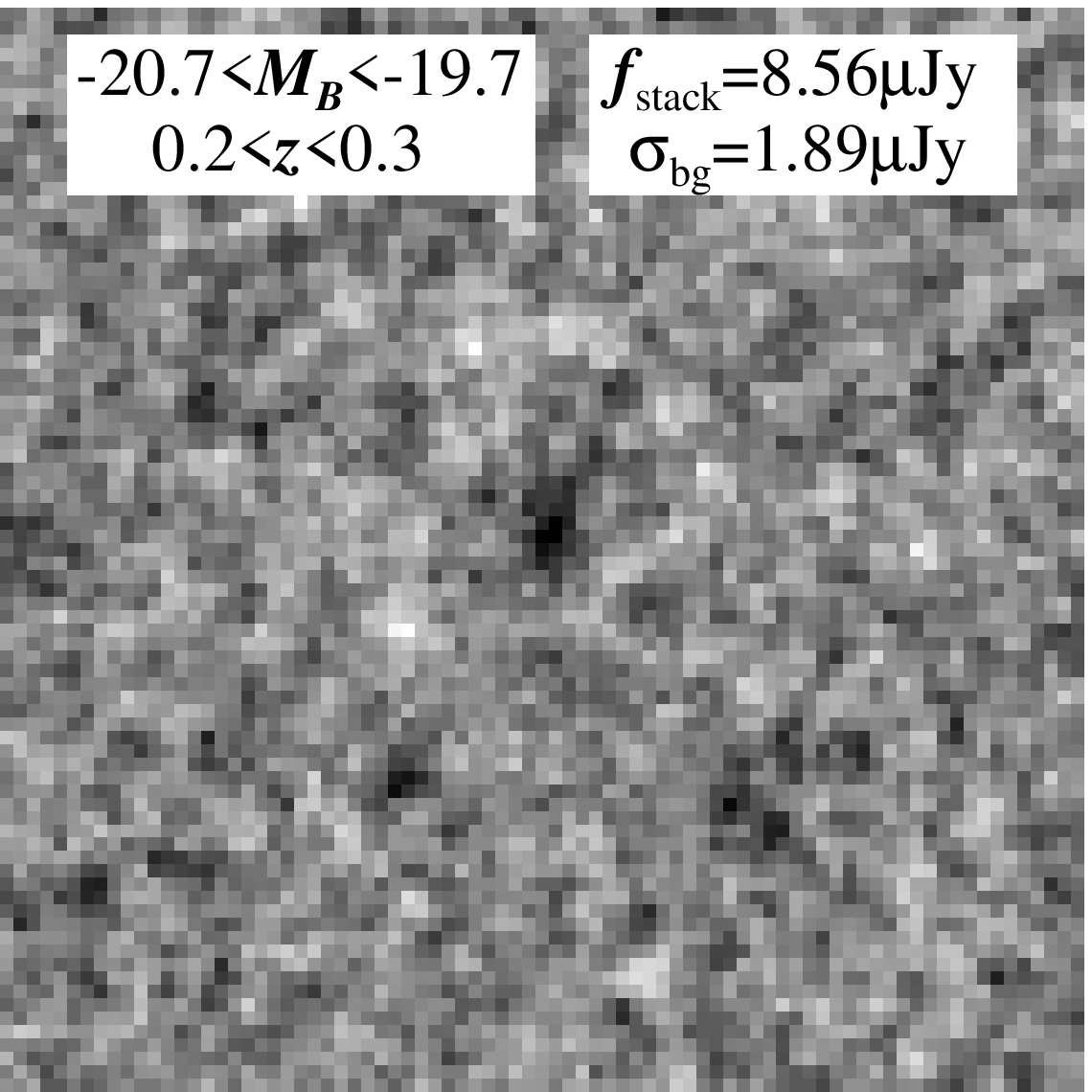}\hfill
\includegraphics[width=0.24\textwidth,clip]{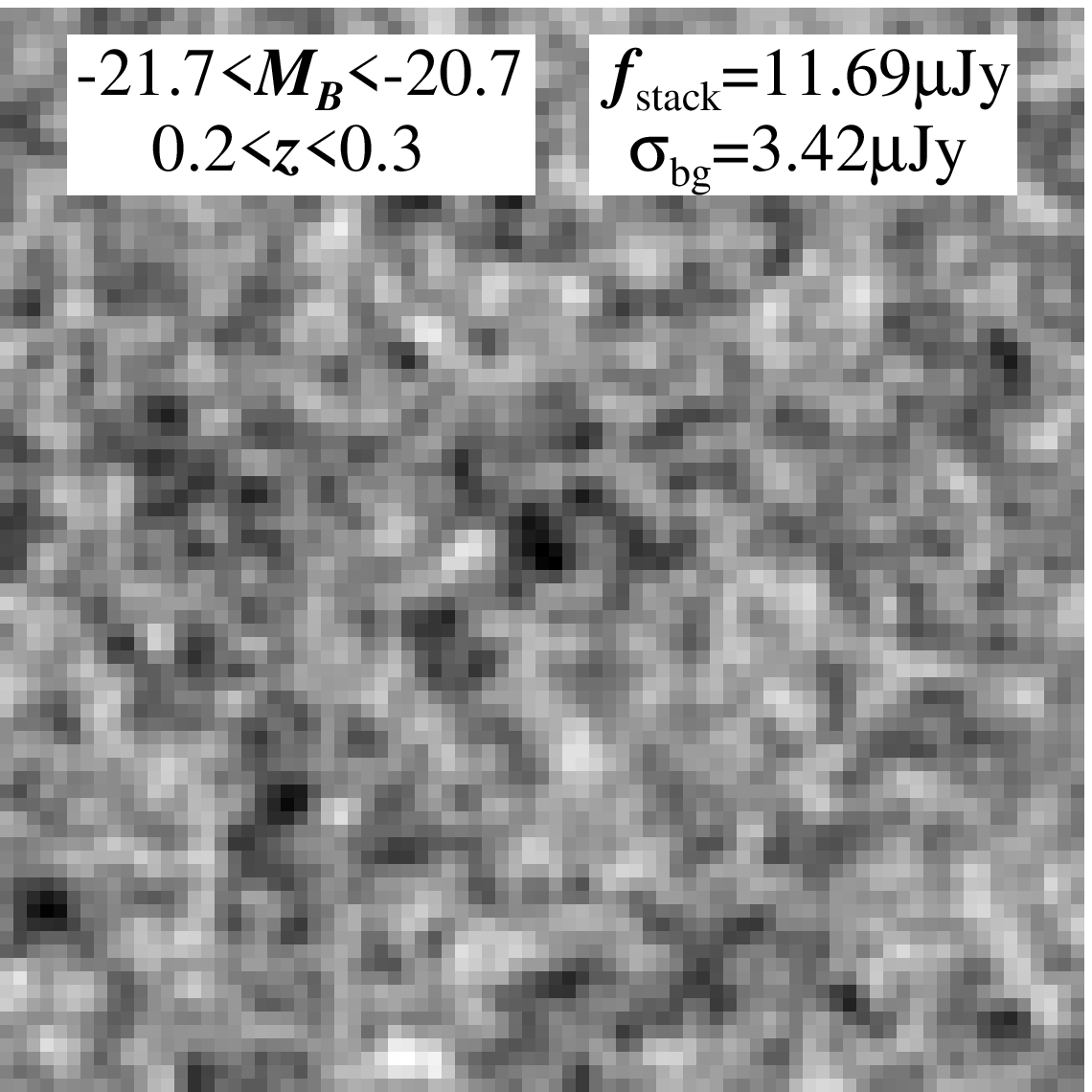}  \includegraphics[width=0.24\textwidth,clip]{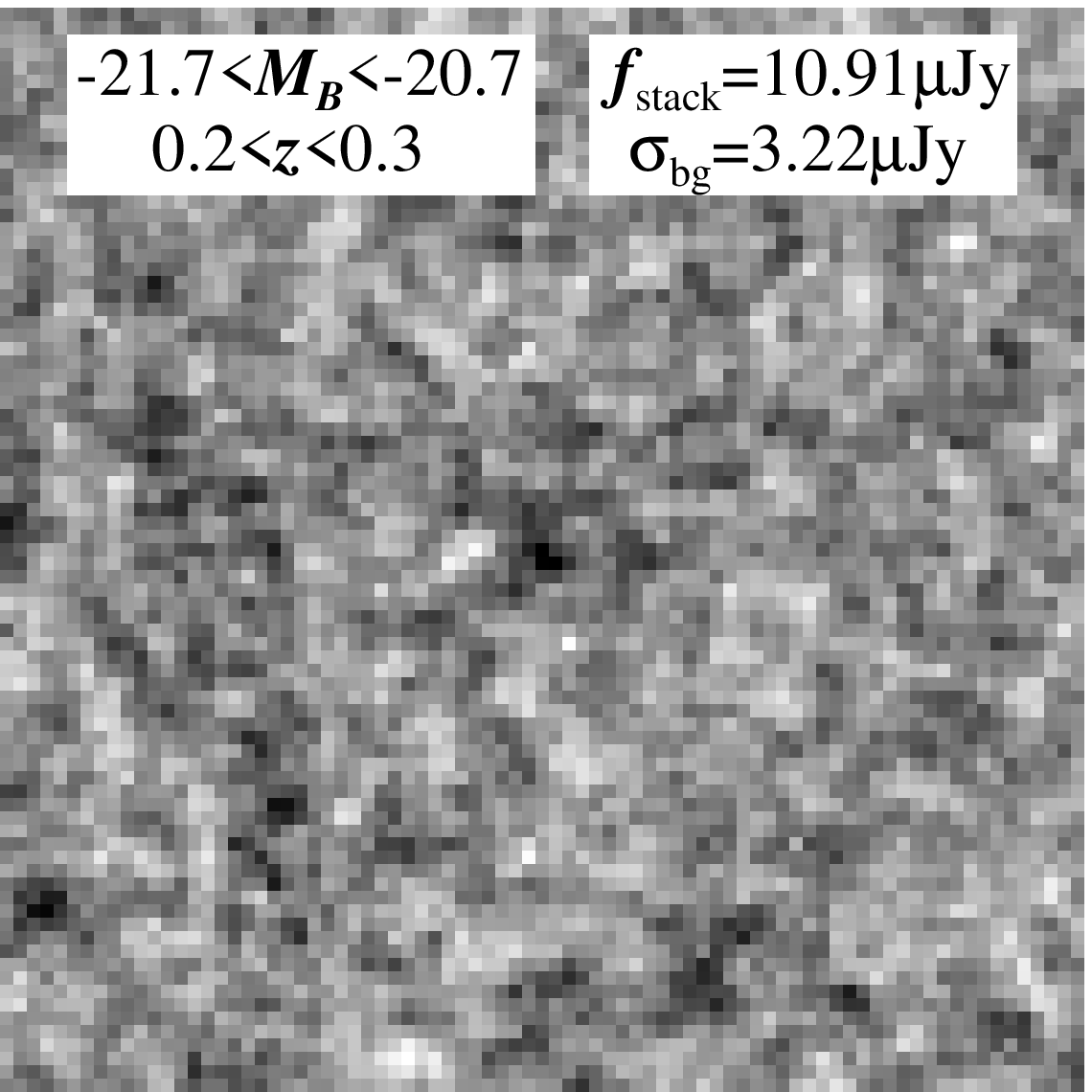}
\caption{Median-stacked (left in a pair) and mean-stacked (right in a pair)
  MIPS 24\,$\micron$ images for 37 stack bins in our sample. For each image,
  magnitude and redshift ranges are labeled at top-left; integrated central 
  flux within aperture of radius 5$\arcsec$, $f_{\rm stack}$, and background 
  RMS measured using the same aperture, $\sigma_{\rm bg}$, are labeled at 
  top-right.}
\end{figure*}

\addtocounter{figure}{-1}

\begin{figure*}[] 
\includegraphics[width=0.24\textwidth,clip]{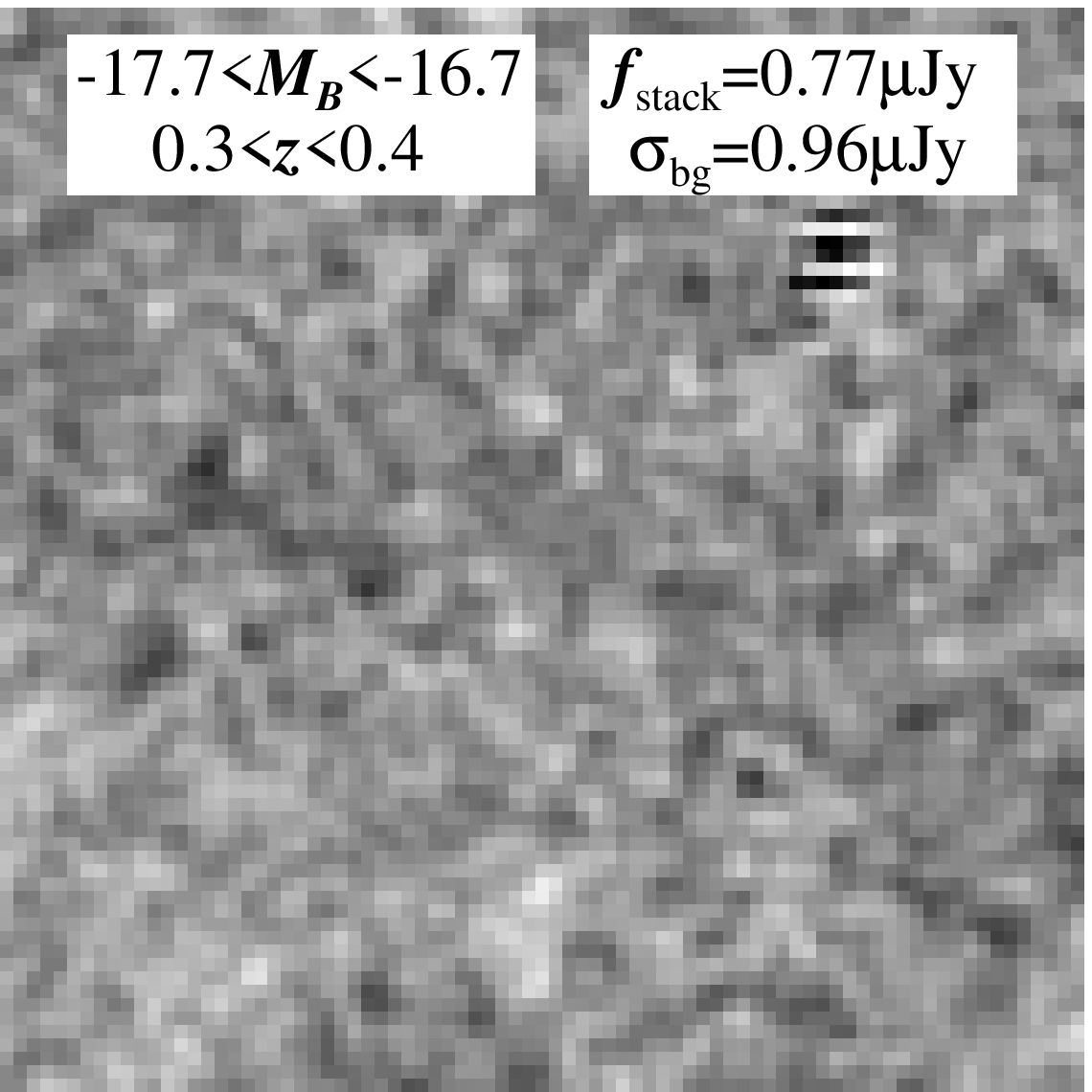}  \includegraphics[width=0.24\textwidth,clip]{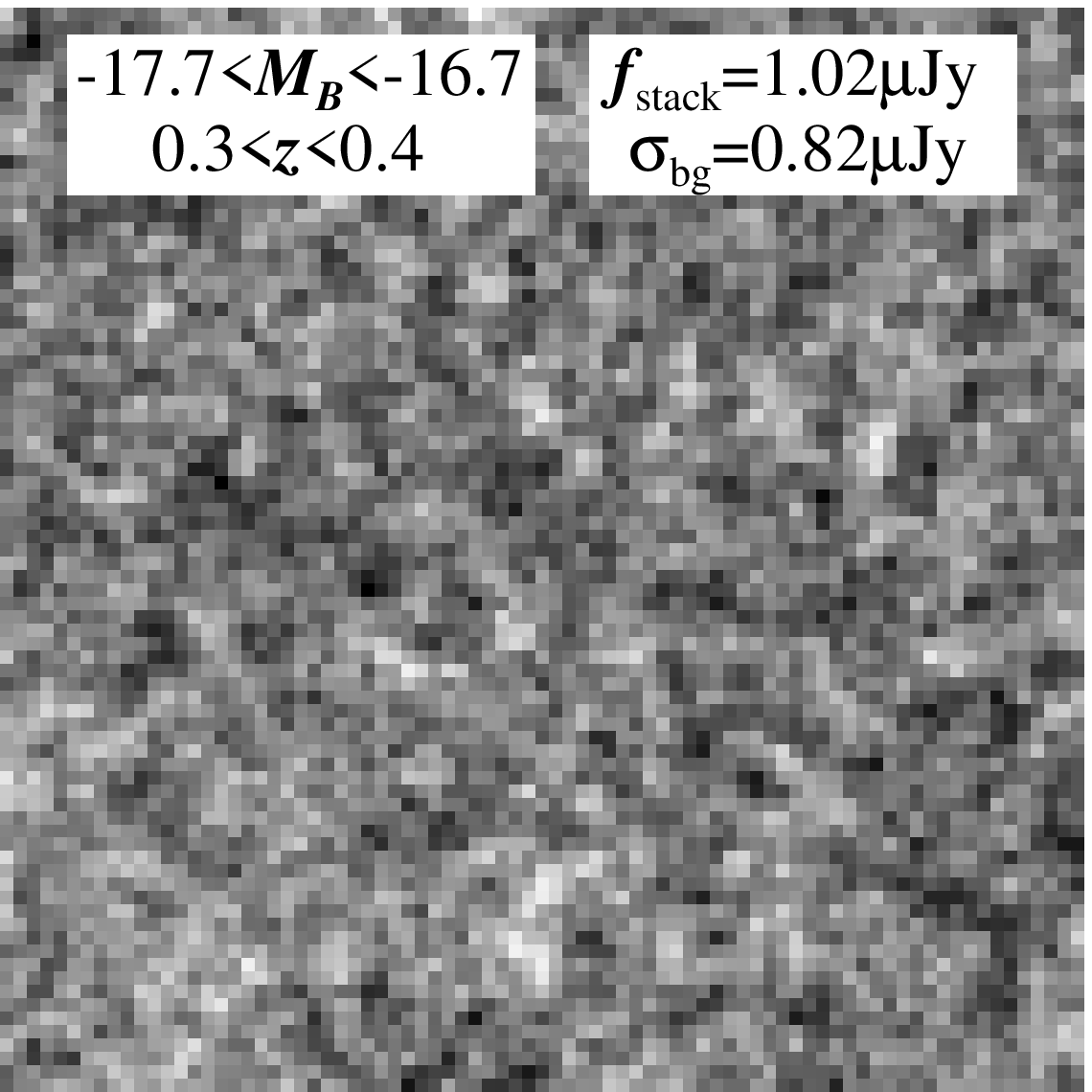}\hfill
\includegraphics[width=0.24\textwidth,clip]{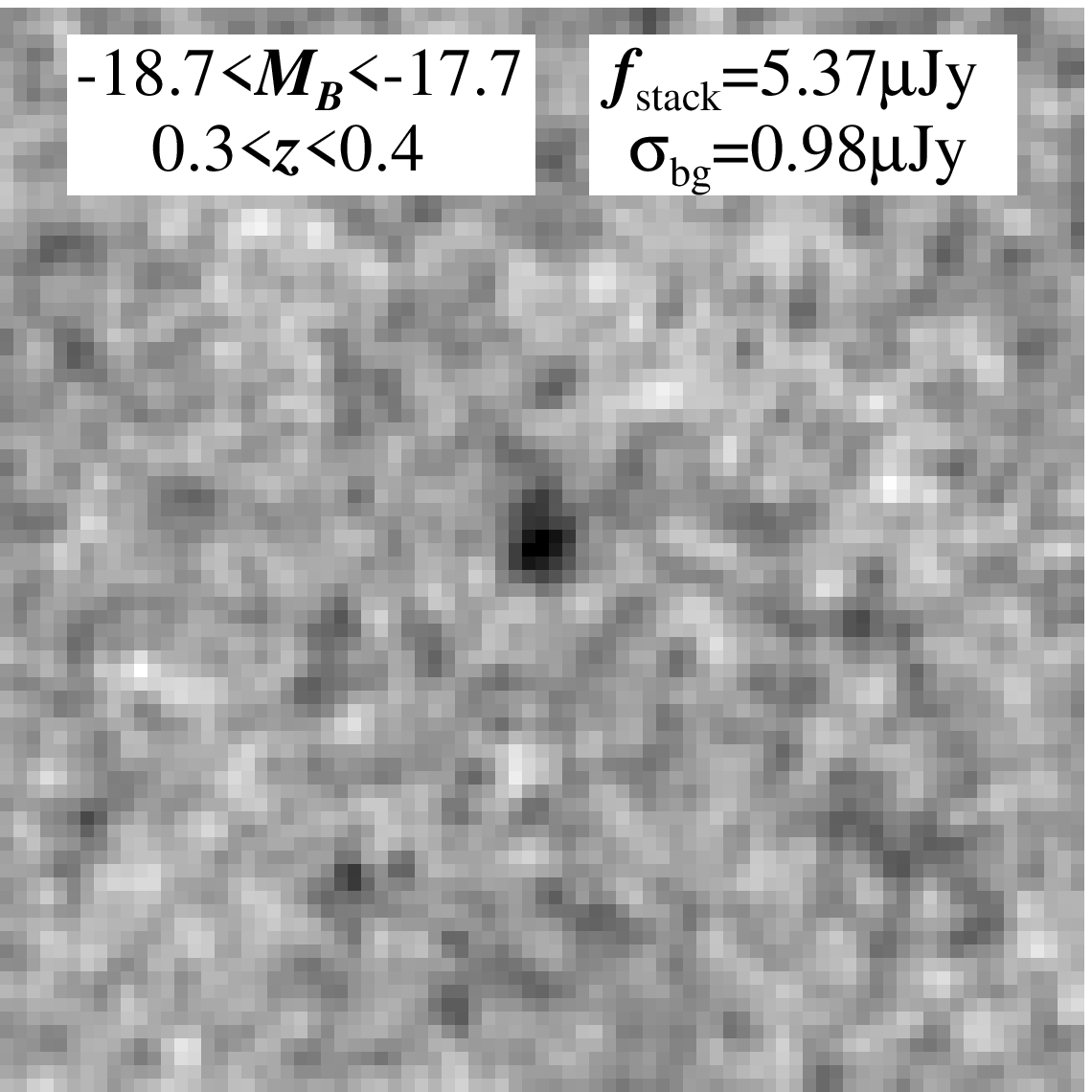}  \includegraphics[width=0.24\textwidth,clip]{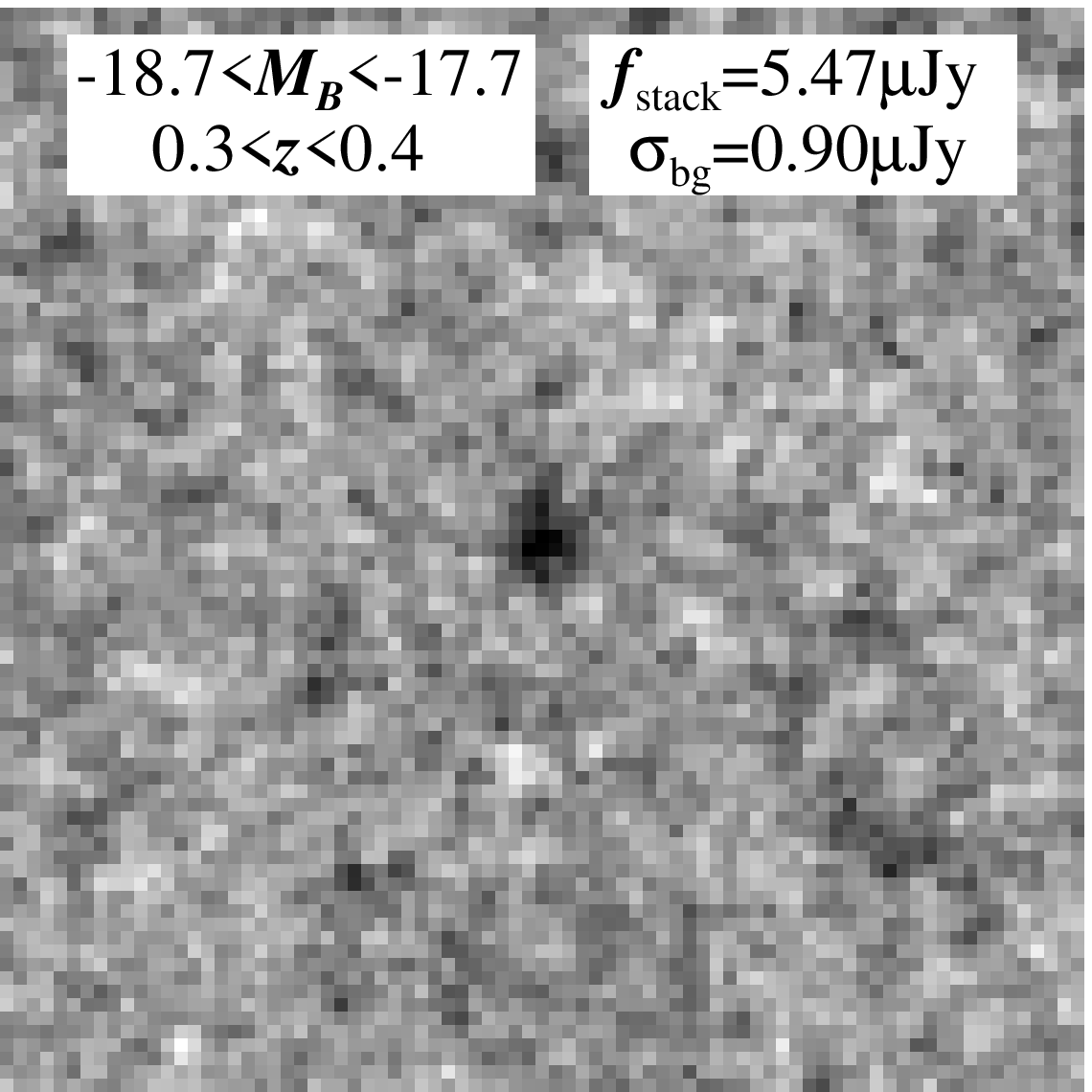}
\vskip 3mm
\includegraphics[width=0.24\textwidth,clip]{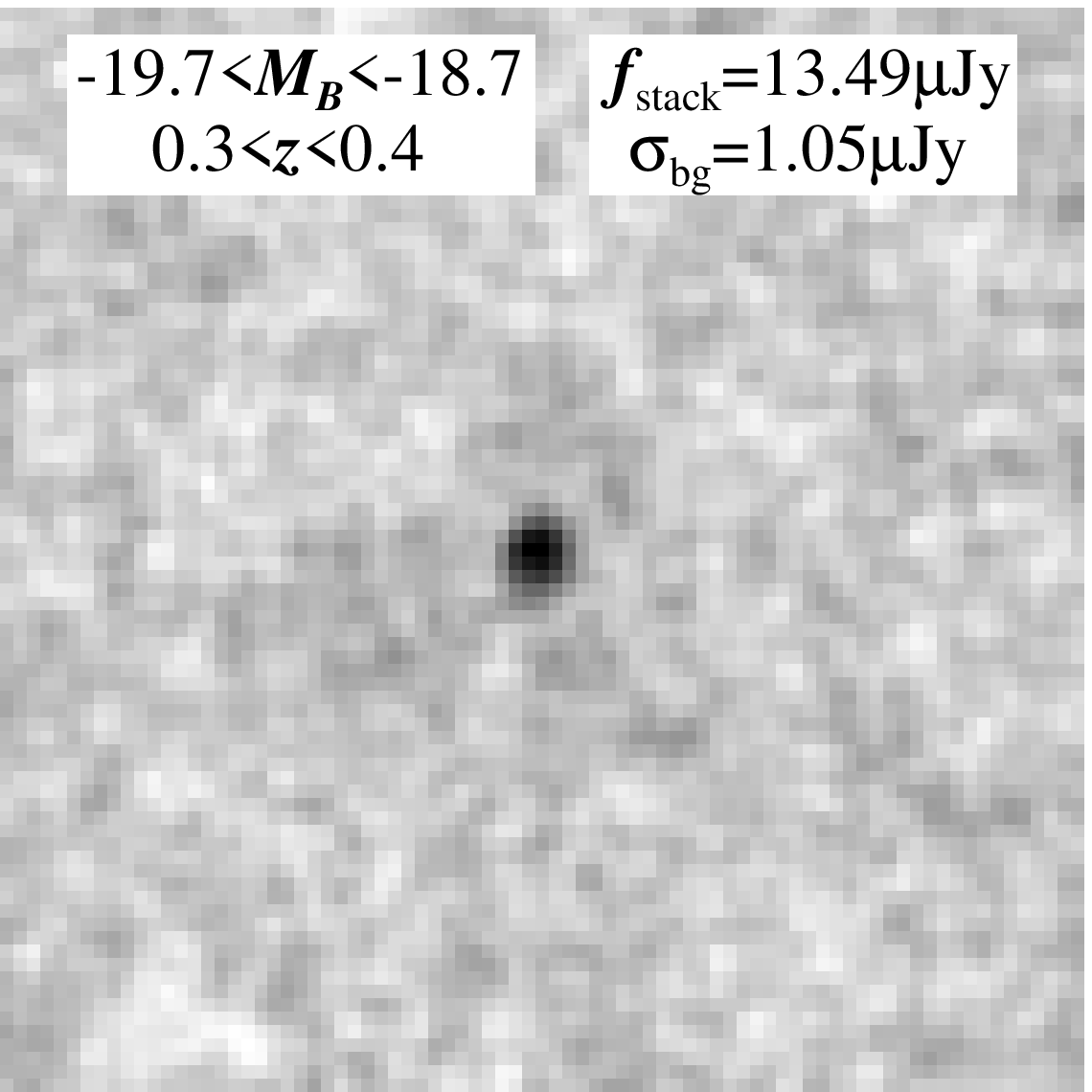}  \includegraphics[width=0.24\textwidth,clip]{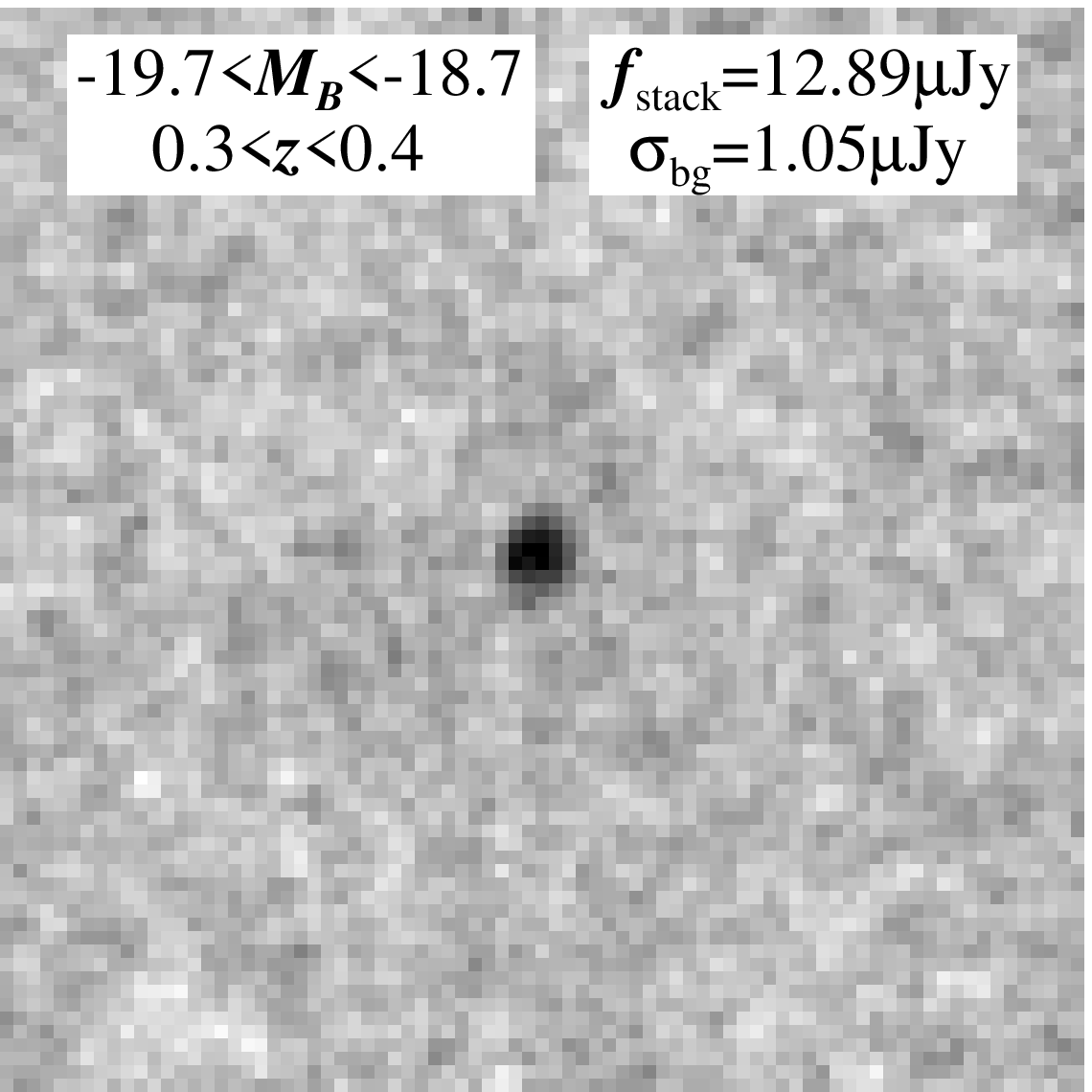}\hfill
\includegraphics[width=0.24\textwidth,clip]{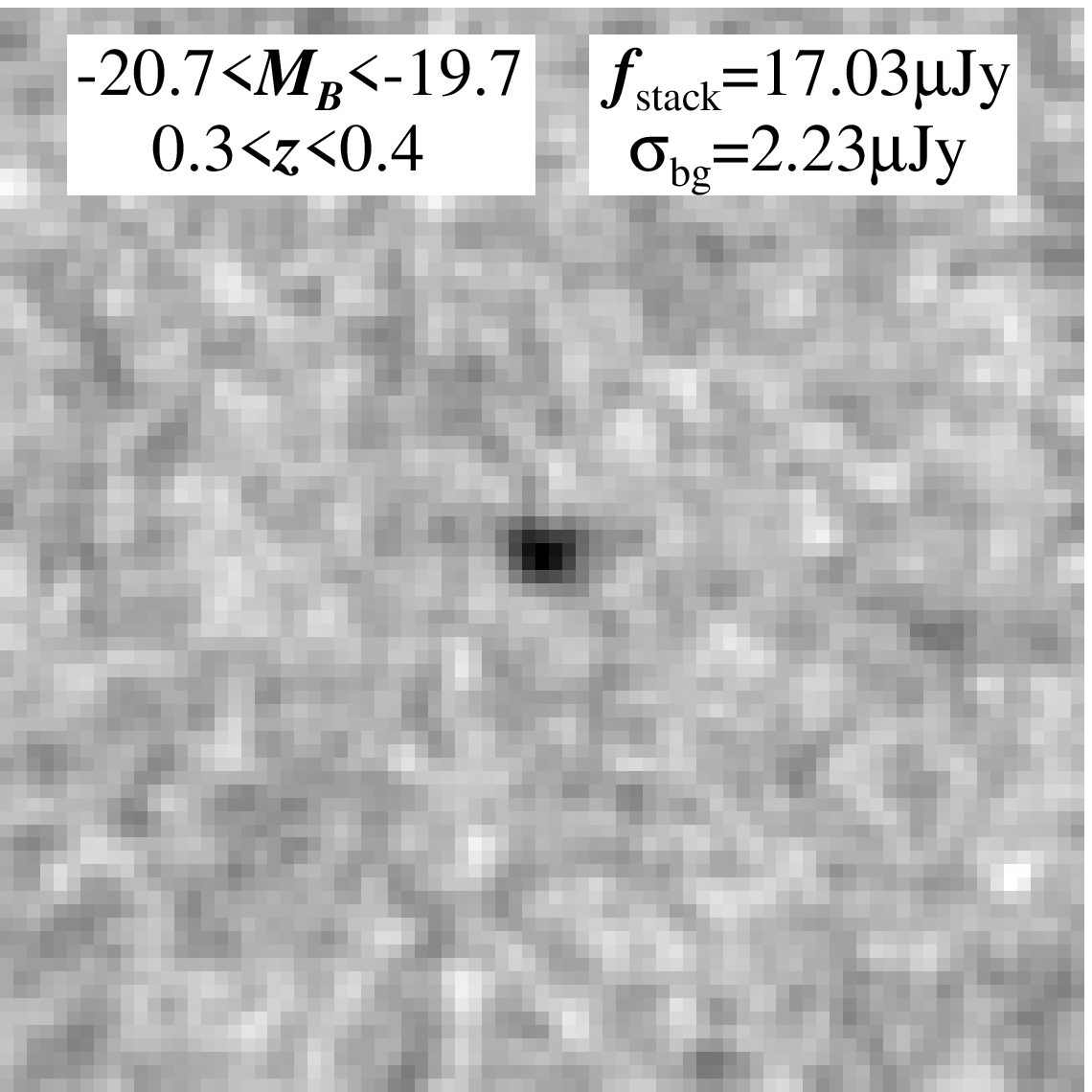}  \includegraphics[width=0.24\textwidth,clip]{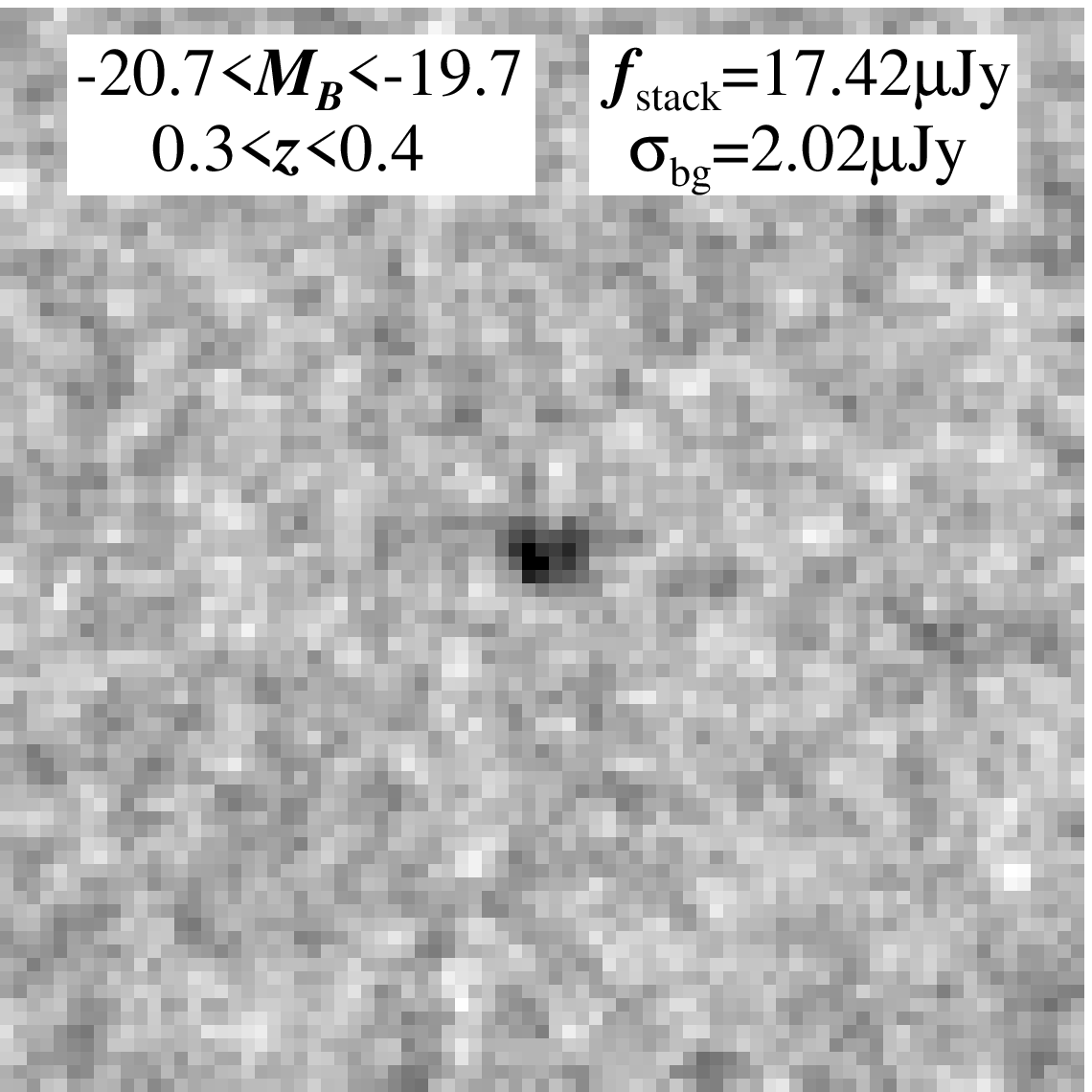}
\vskip 3mm
\includegraphics[width=0.24\textwidth,clip]{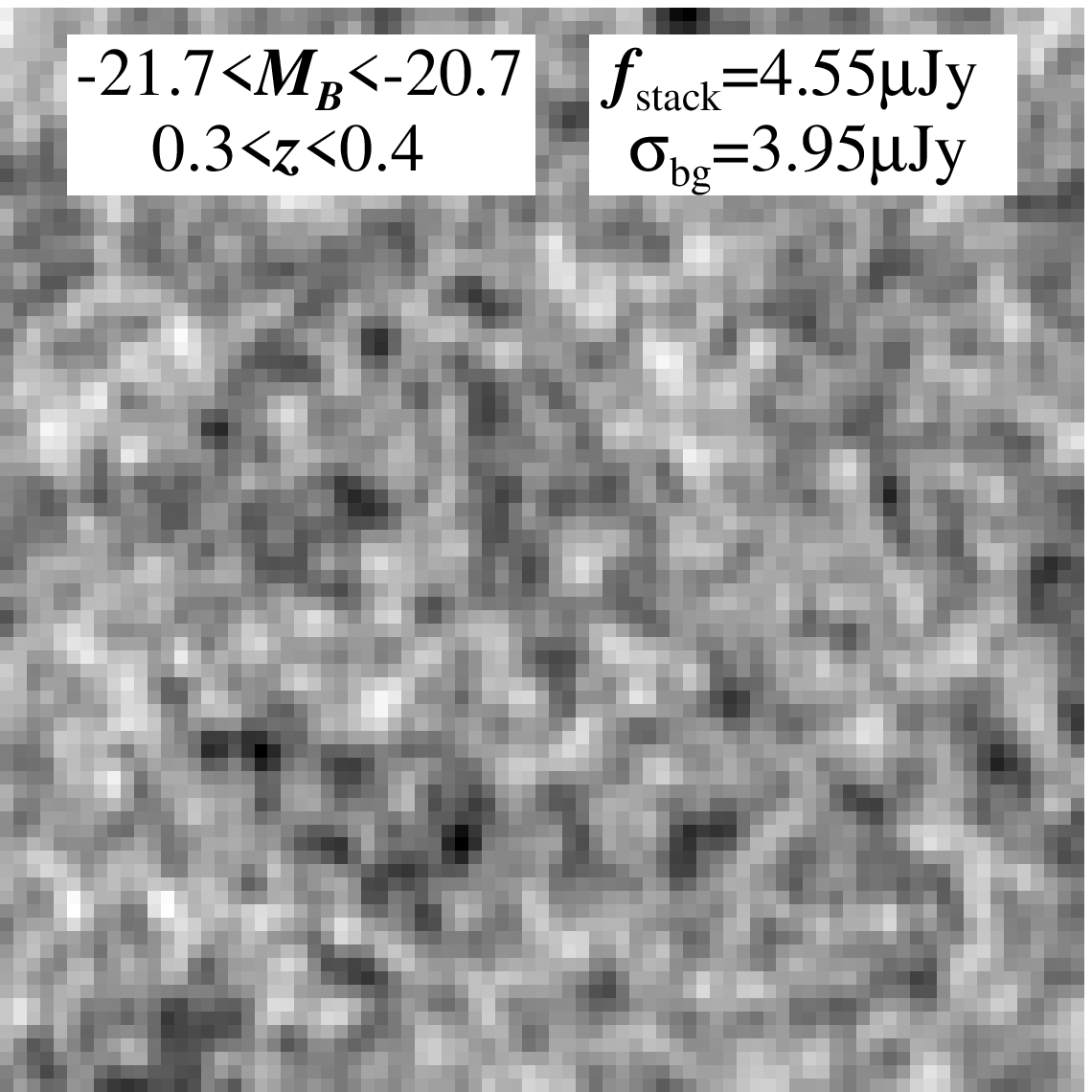}  \includegraphics[width=0.24\textwidth,clip]{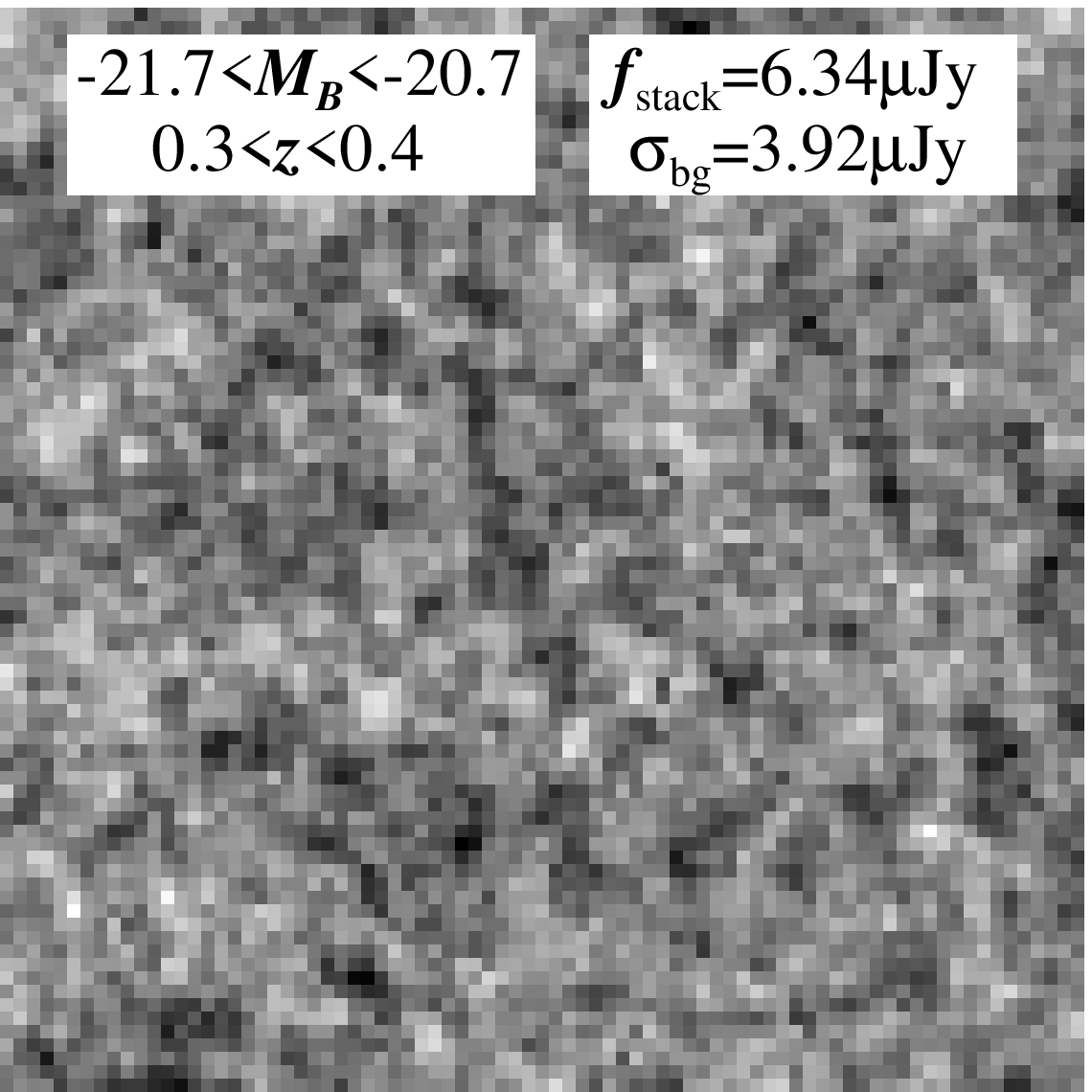}\hfill
\includegraphics[width=0.24\textwidth,clip]{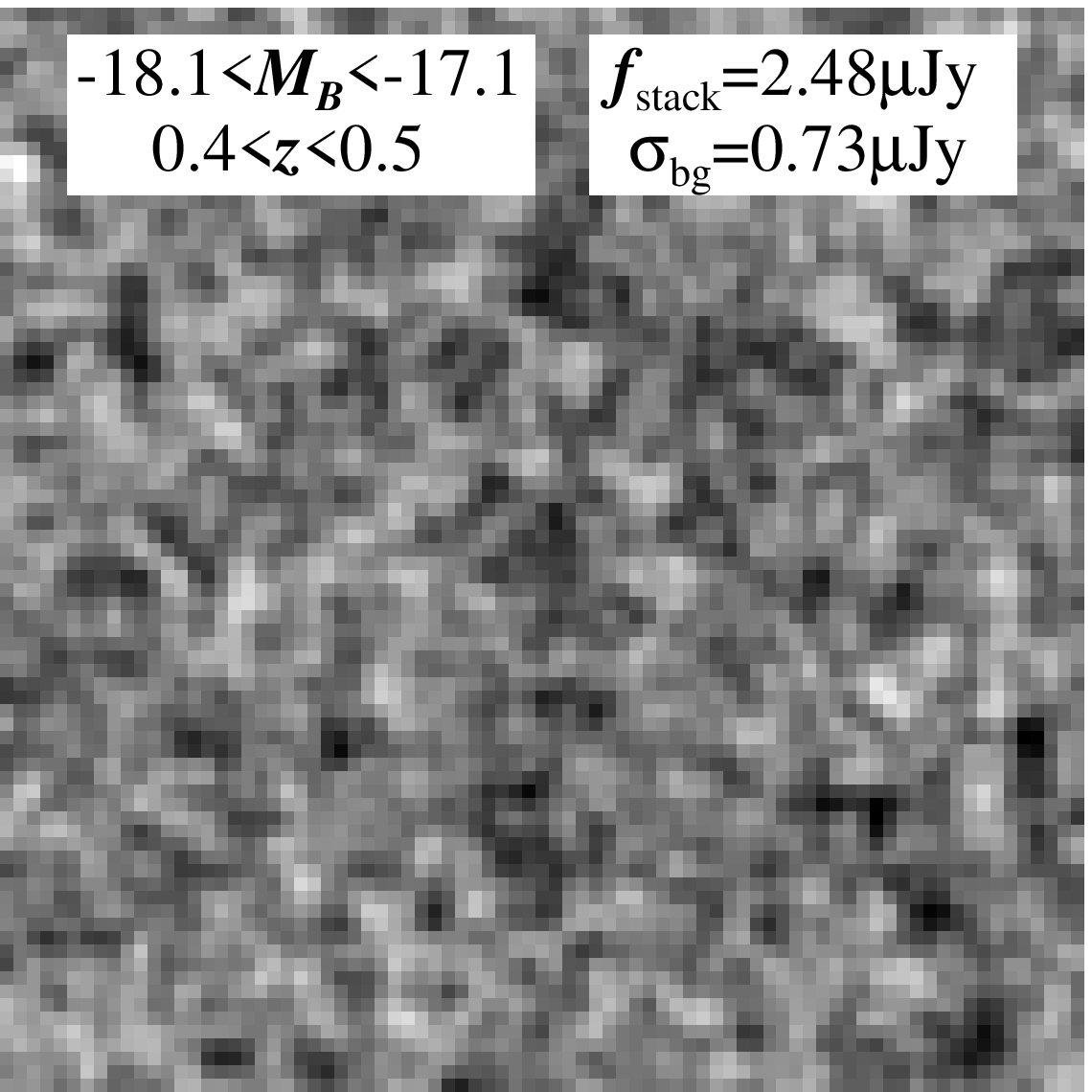}  \includegraphics[width=0.24\textwidth,clip]{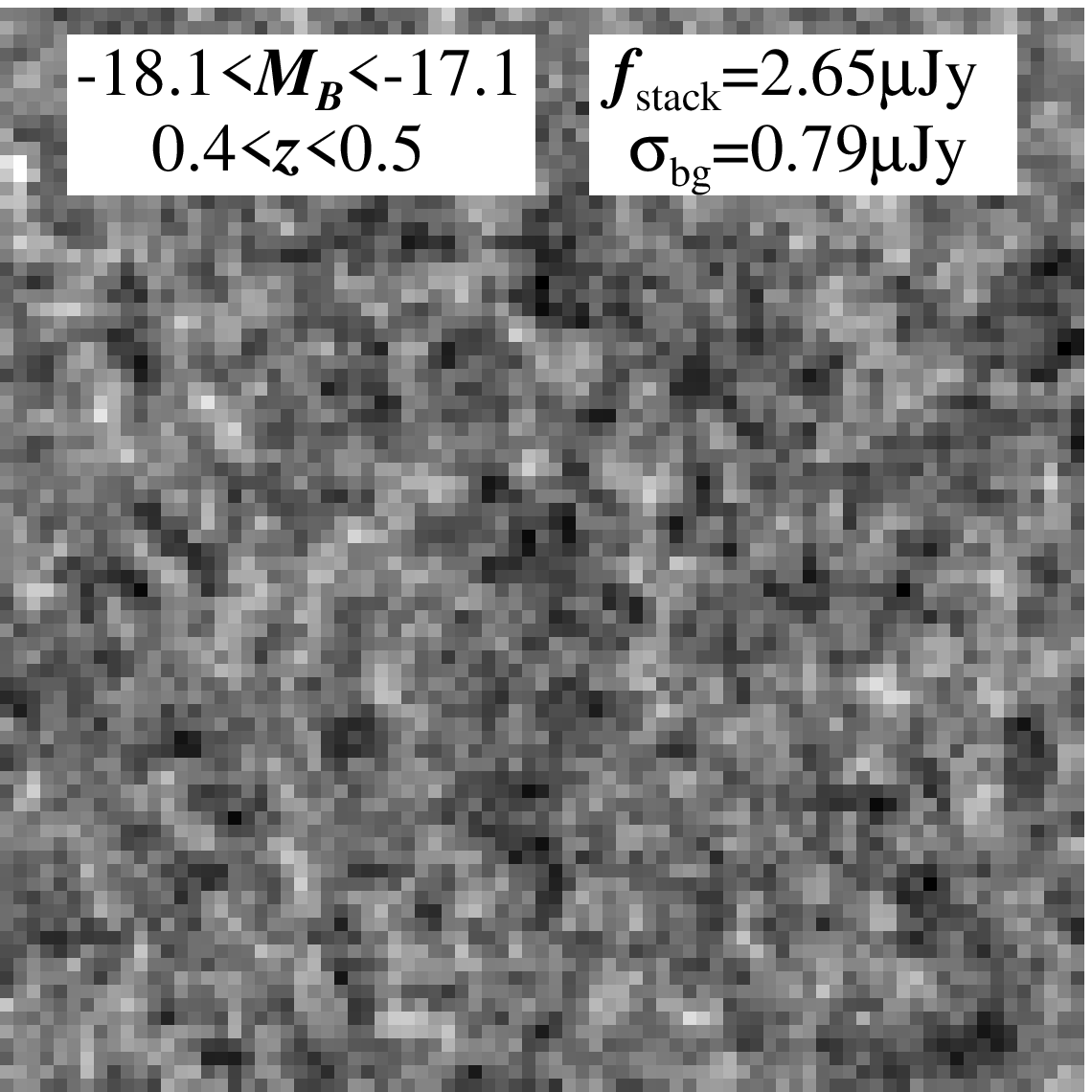}
\vskip 3mm
\includegraphics[width=0.24\textwidth,clip]{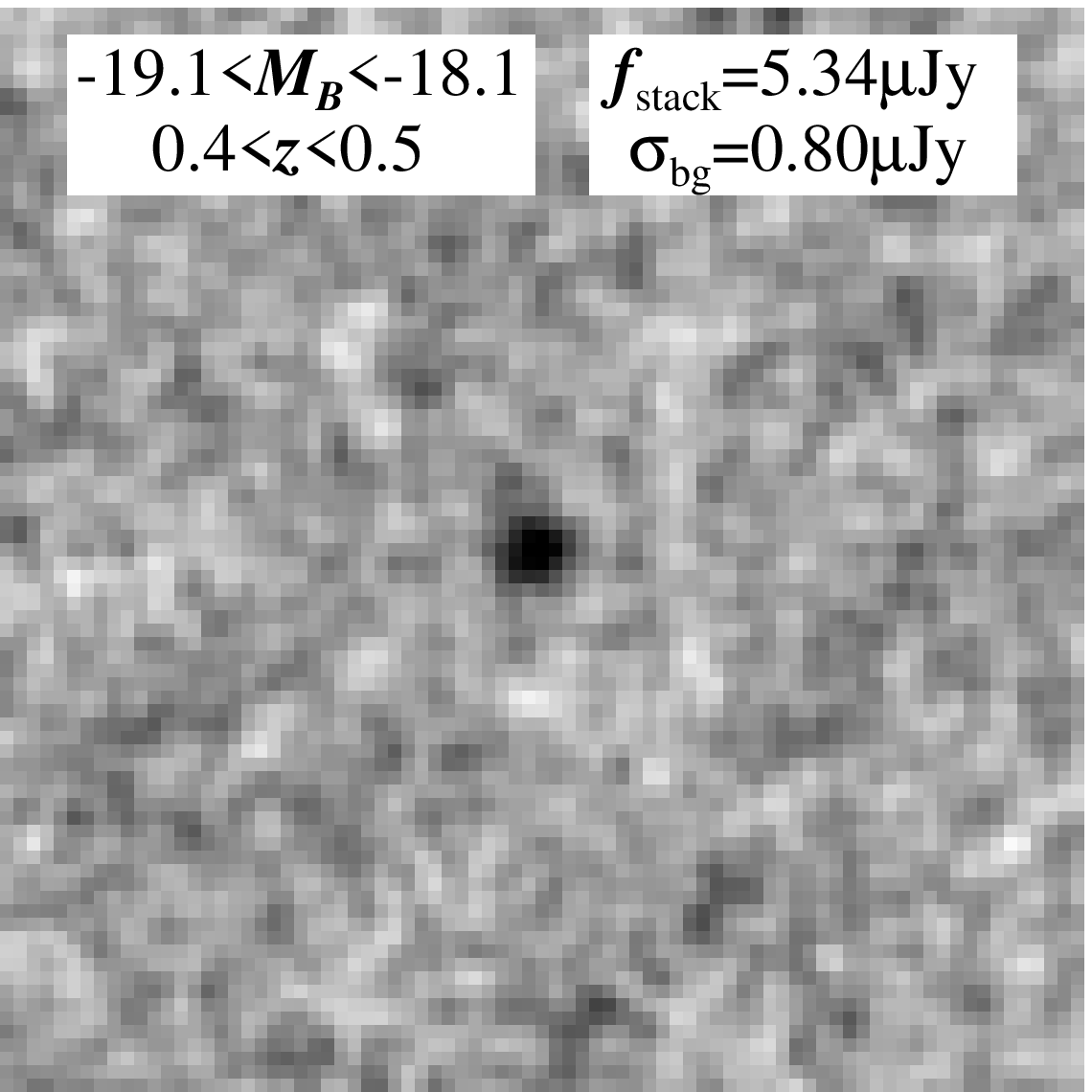}  \includegraphics[width=0.24\textwidth,clip]{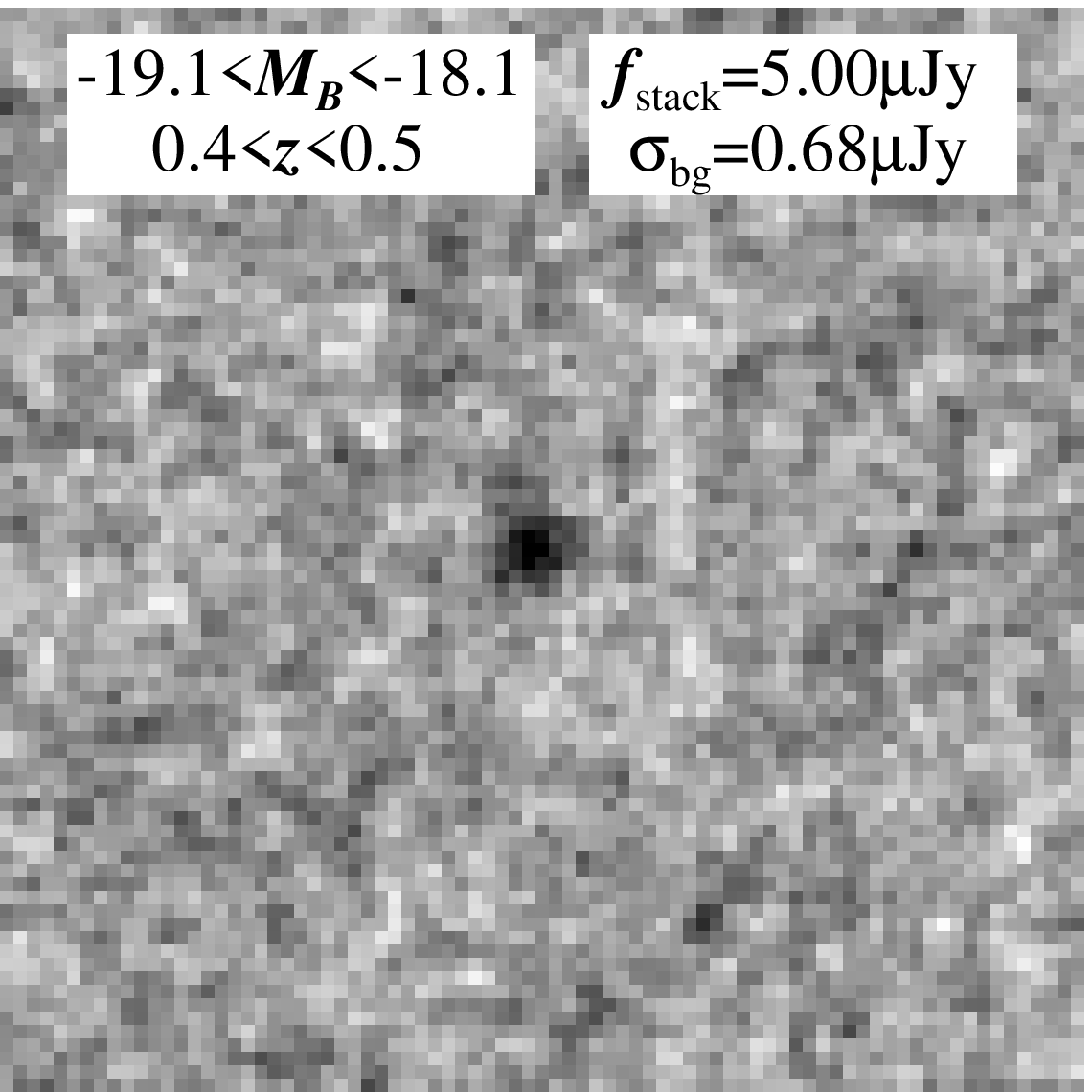}\hfill
\includegraphics[width=0.24\textwidth,clip]{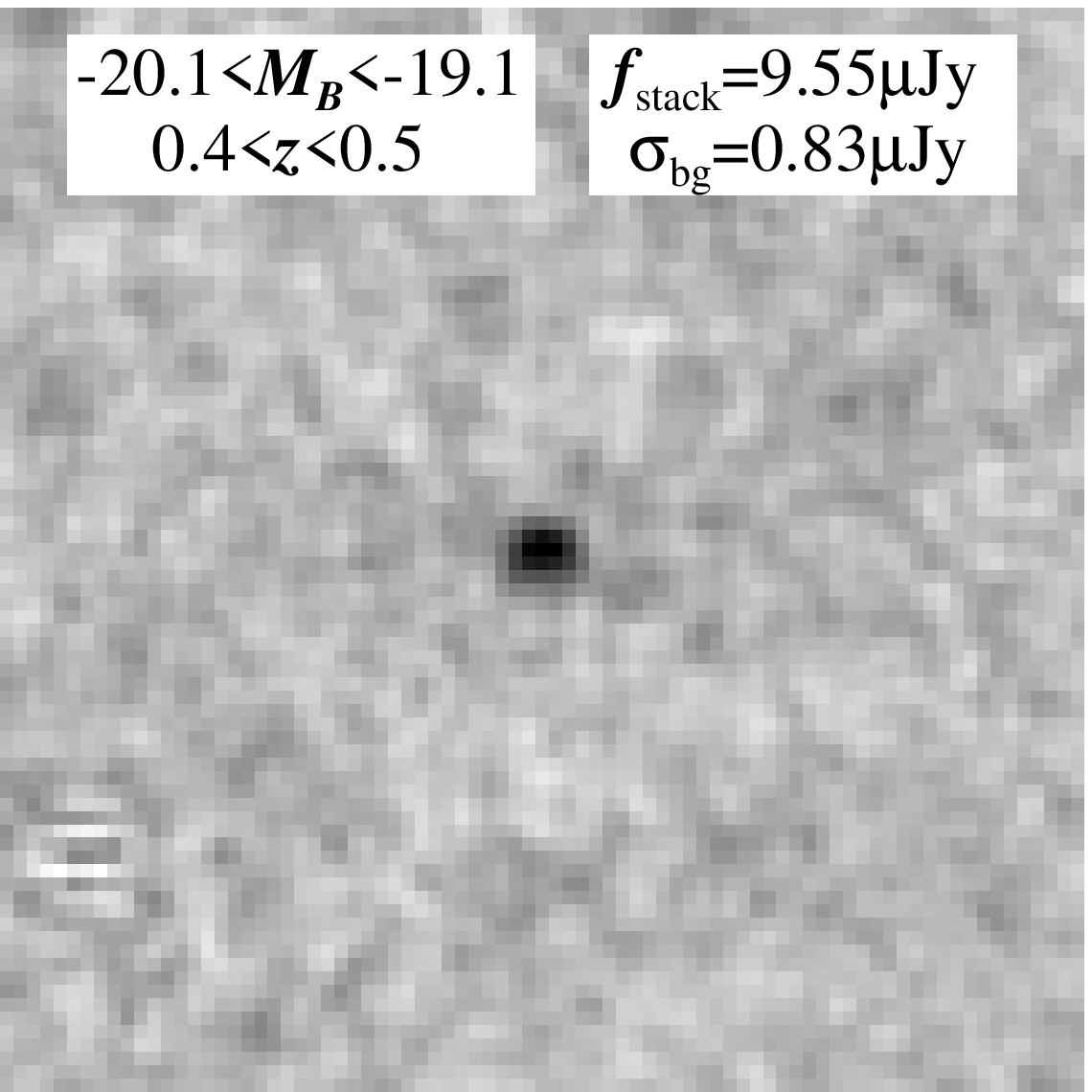}  \includegraphics[width=0.24\textwidth,clip]{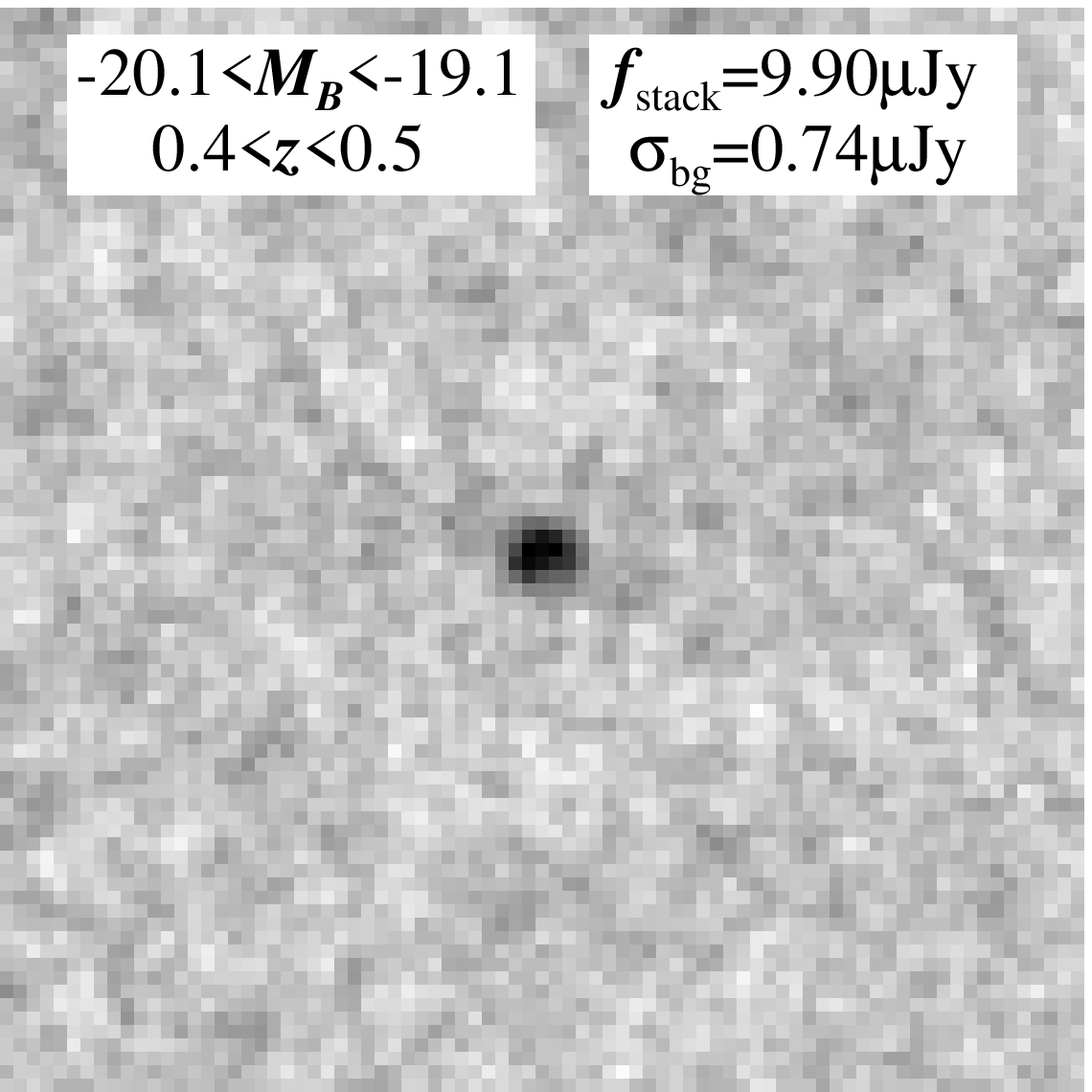}
\vskip 3mm
\includegraphics[width=0.24\textwidth,clip]{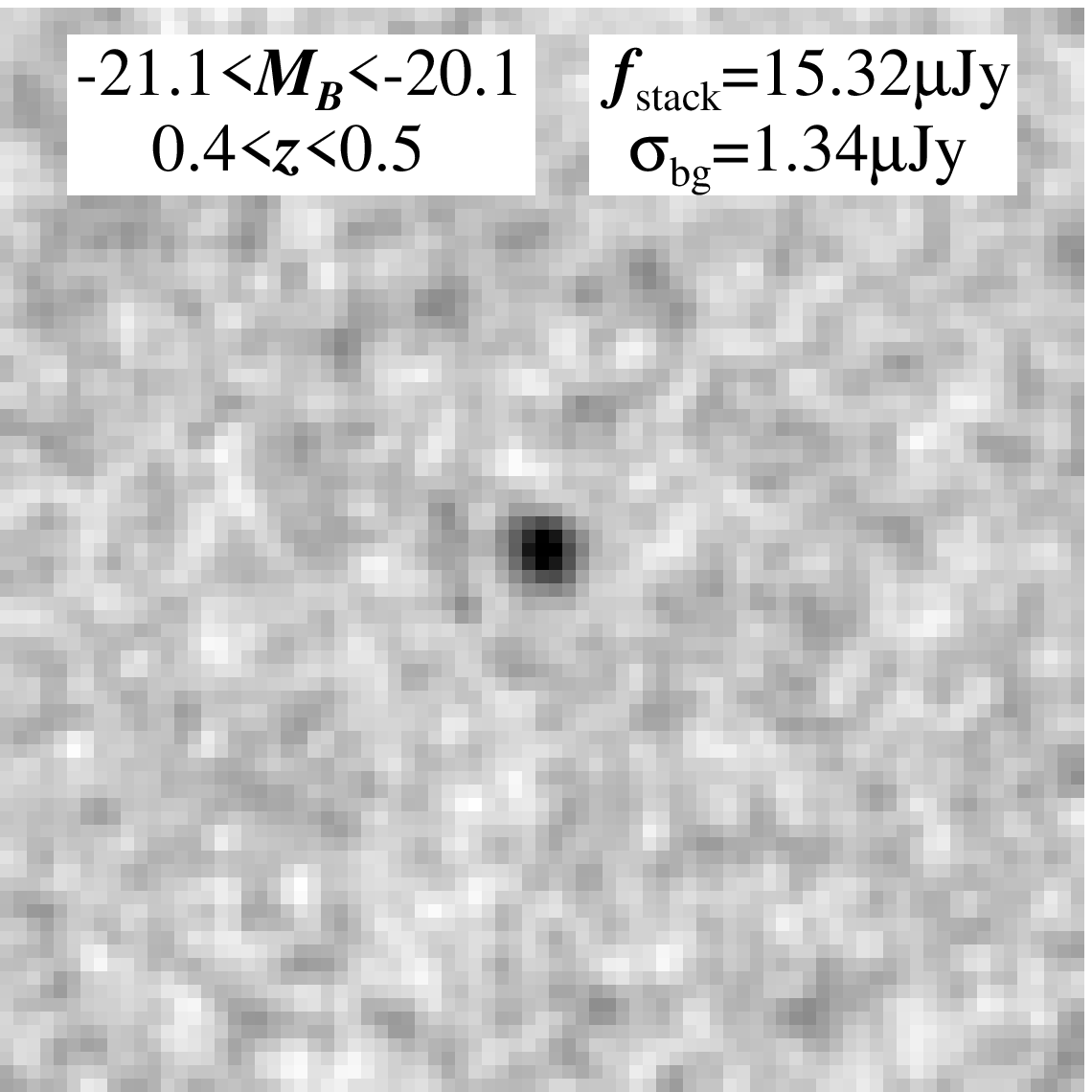}  \includegraphics[width=0.24\textwidth,clip]{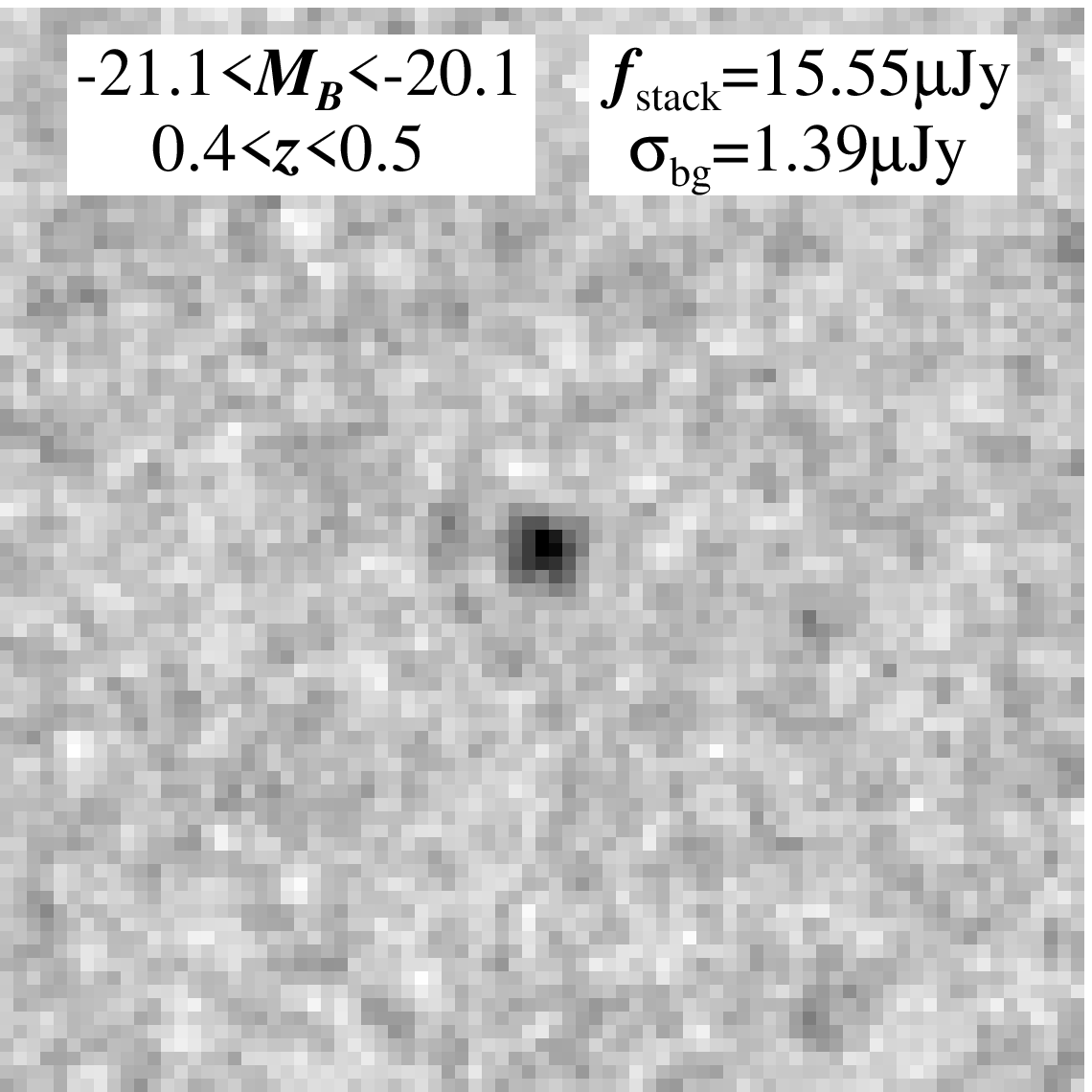}\hfill
\includegraphics[width=0.24\textwidth,clip]{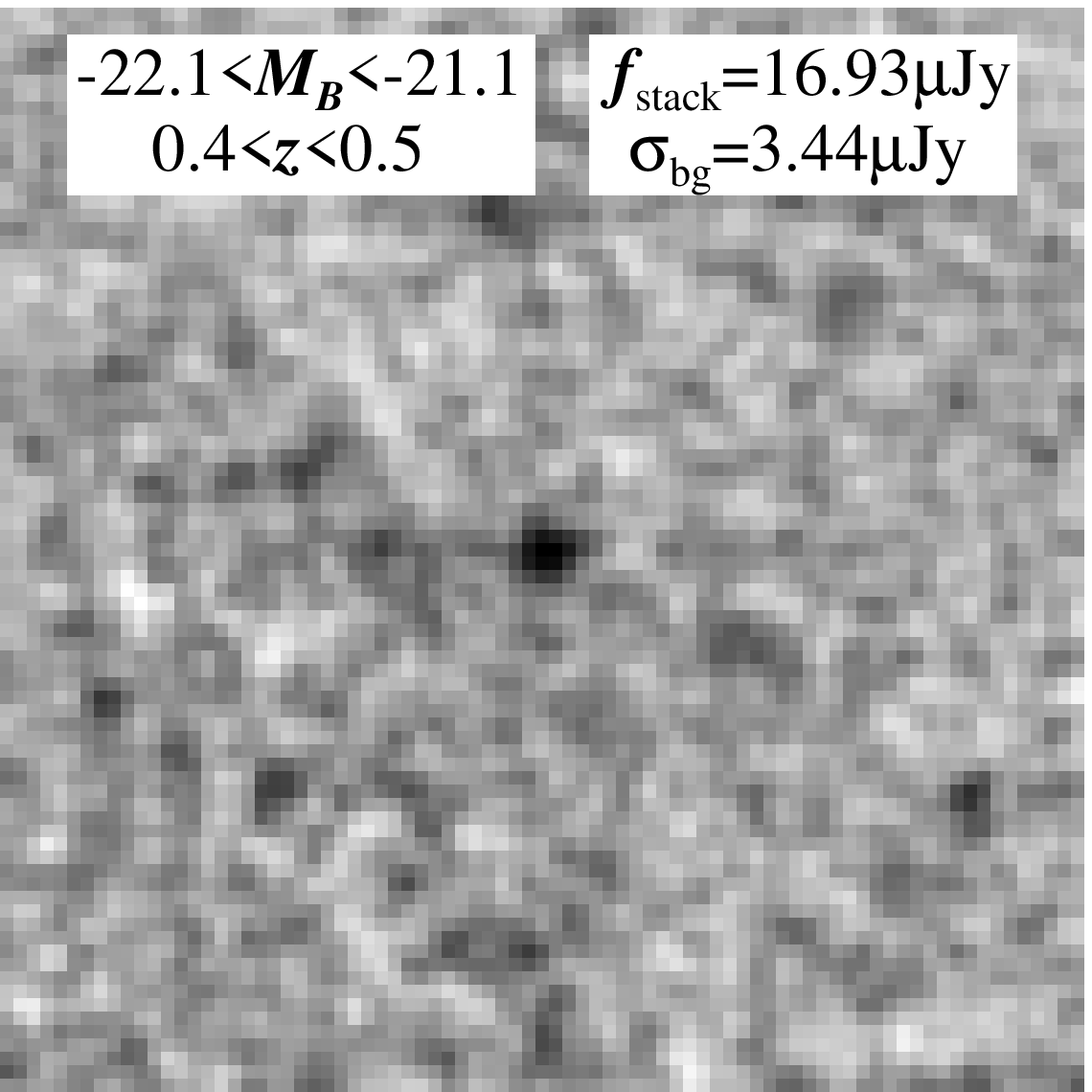}  \includegraphics[width=0.24\textwidth,clip]{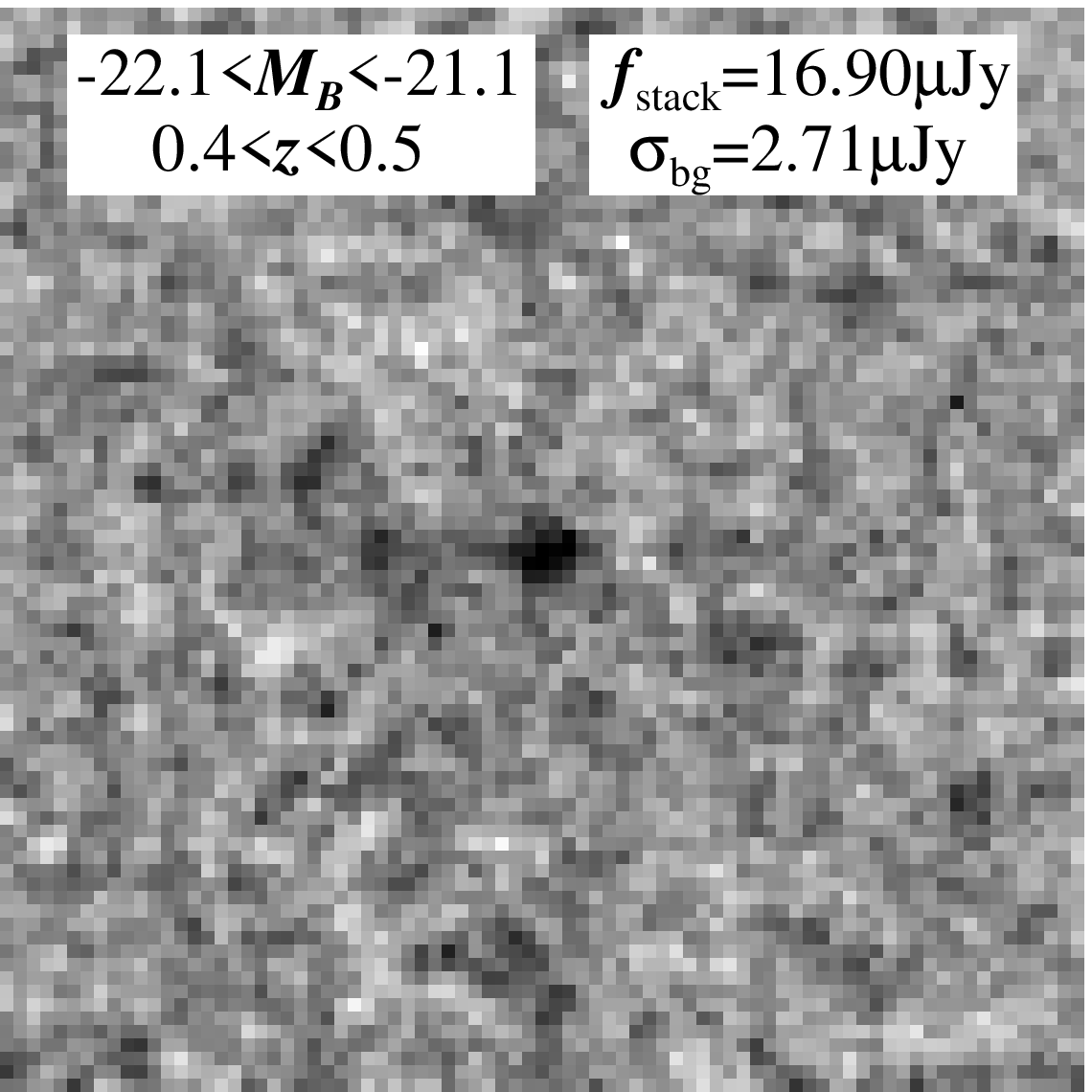}
\caption{Continued.}
\end{figure*}

\addtocounter{figure}{-1}

\begin{figure*}[] 
\includegraphics[width=0.24\textwidth,clip]{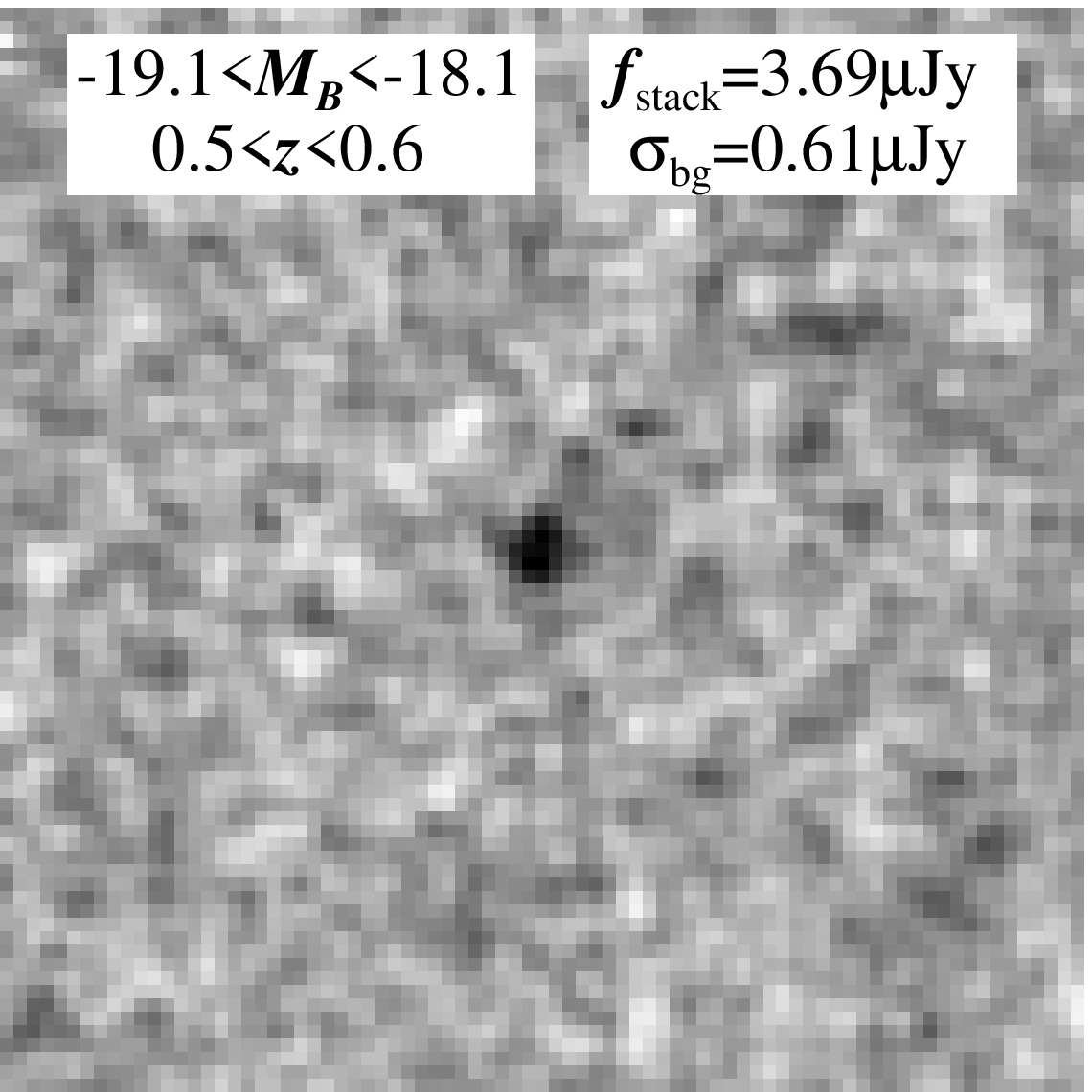}  \includegraphics[width=0.24\textwidth,clip]{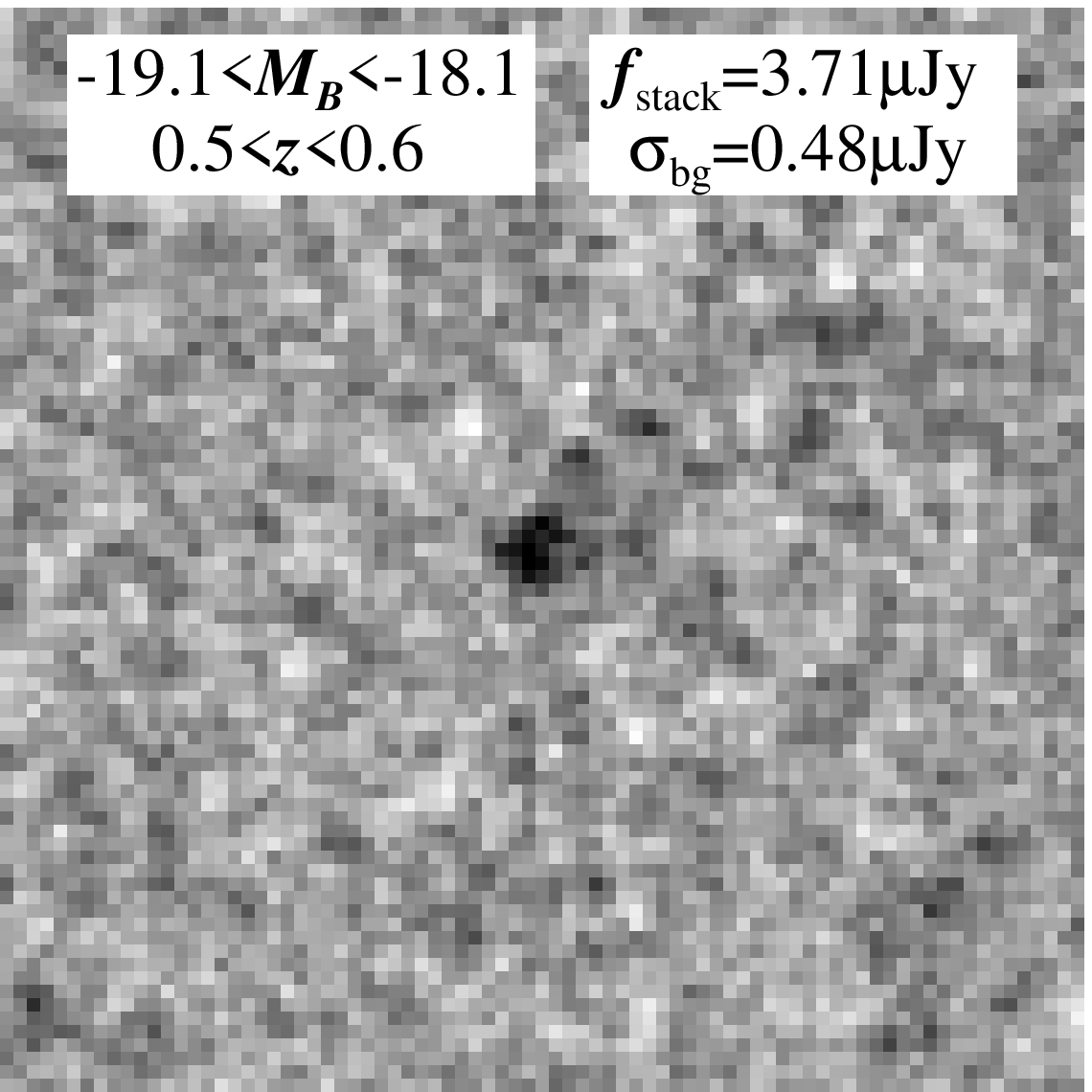}\hfill
\includegraphics[width=0.24\textwidth,clip]{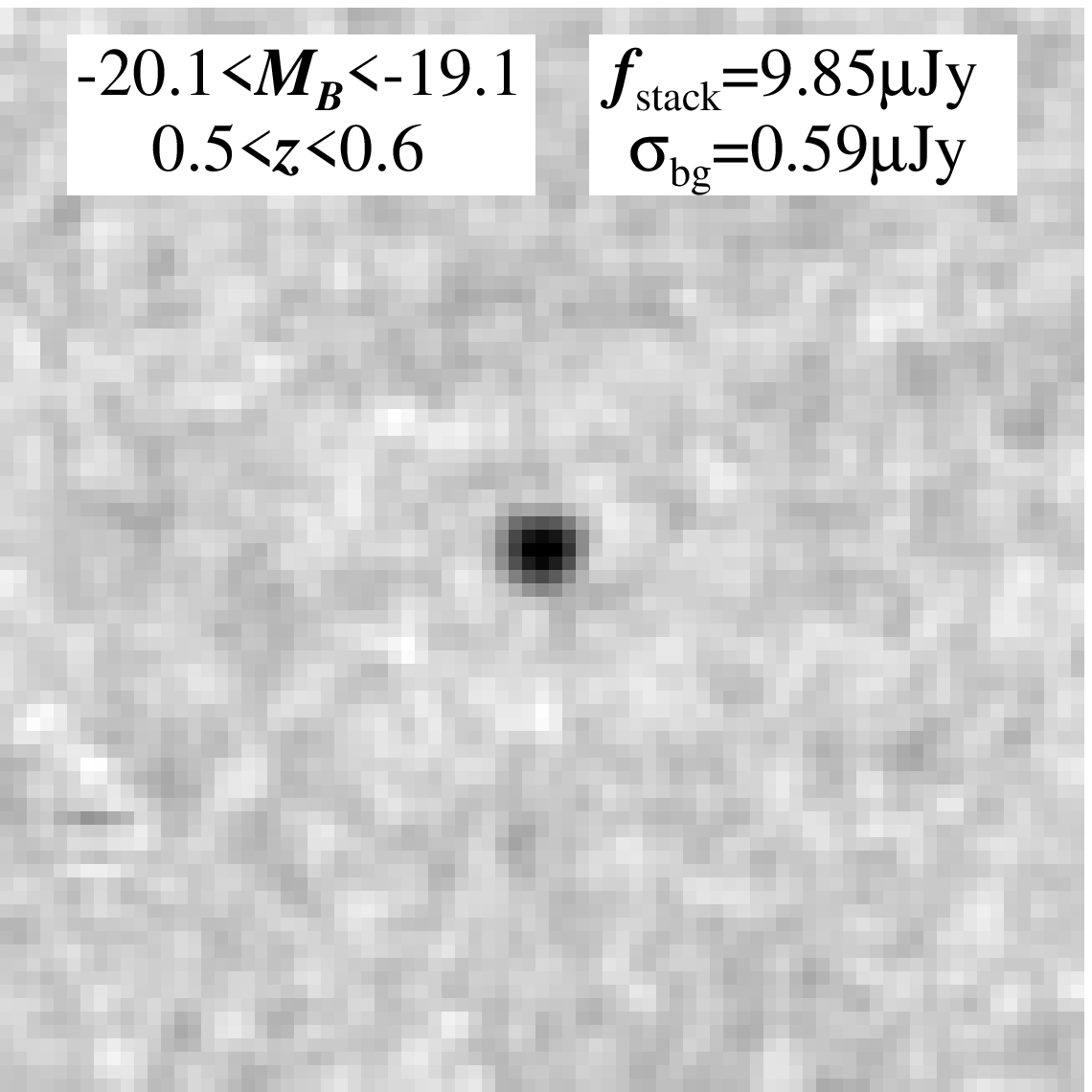}  \includegraphics[width=0.24\textwidth,clip]{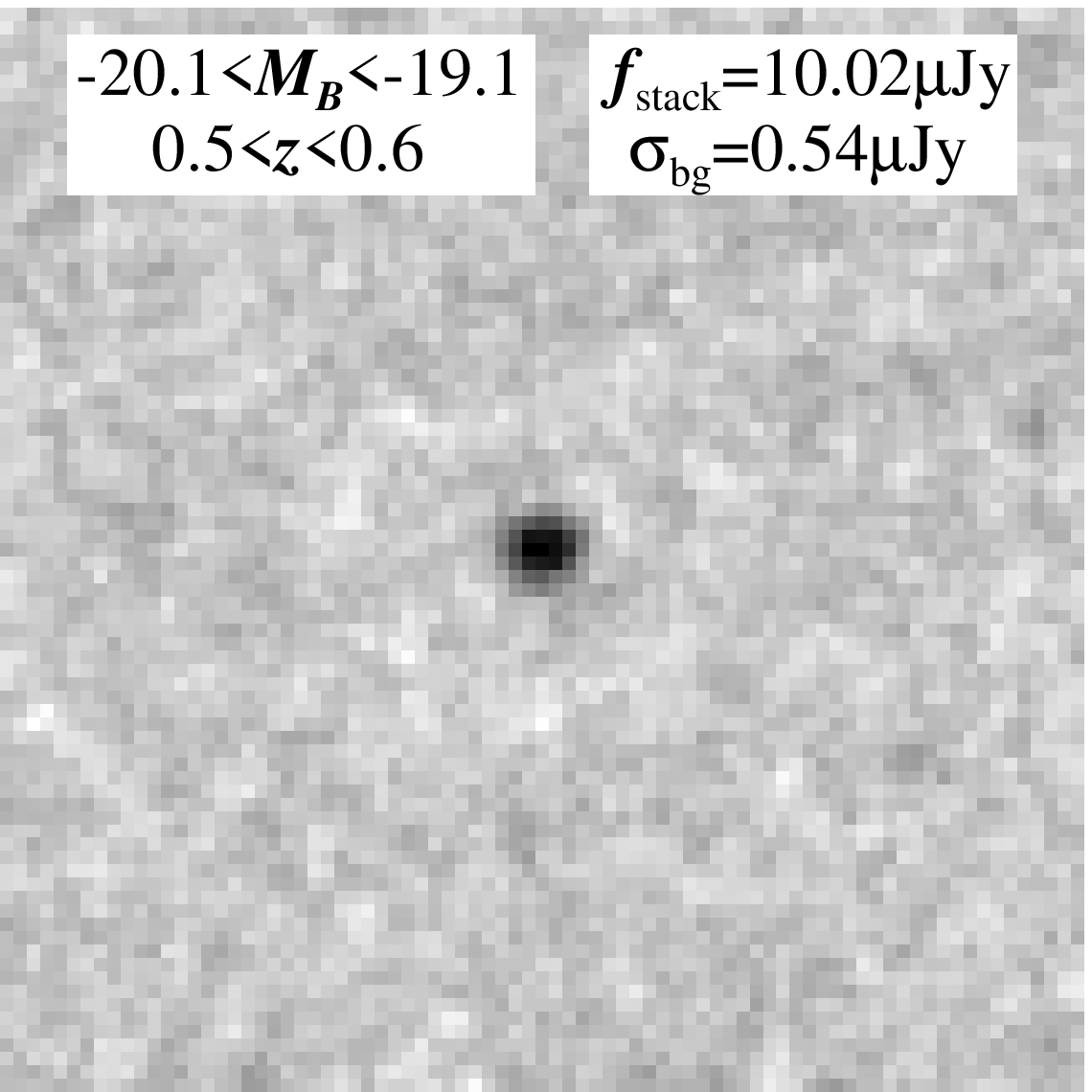}
\vskip 3mm
\includegraphics[width=0.24\textwidth,clip]{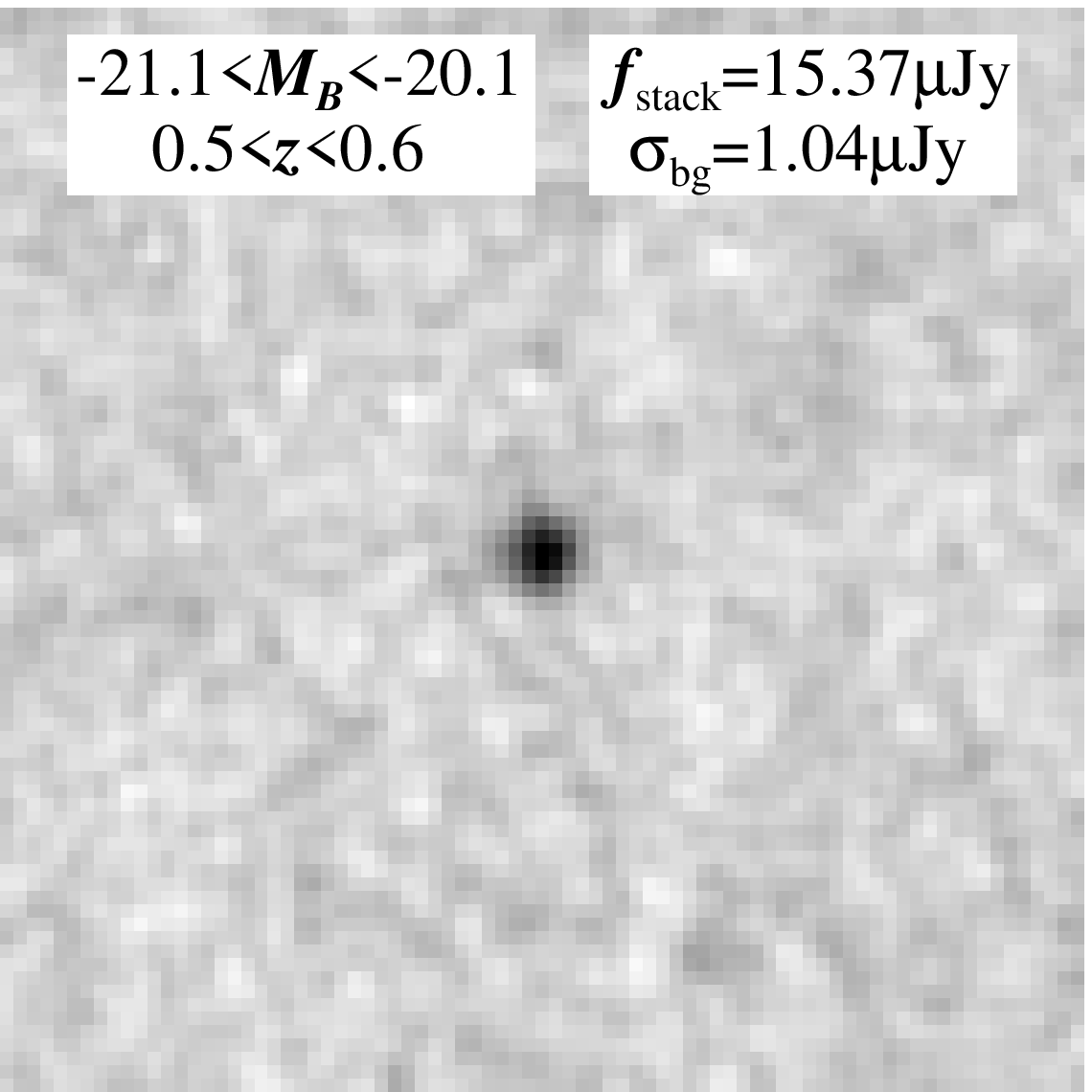}  \includegraphics[width=0.24\textwidth,clip]{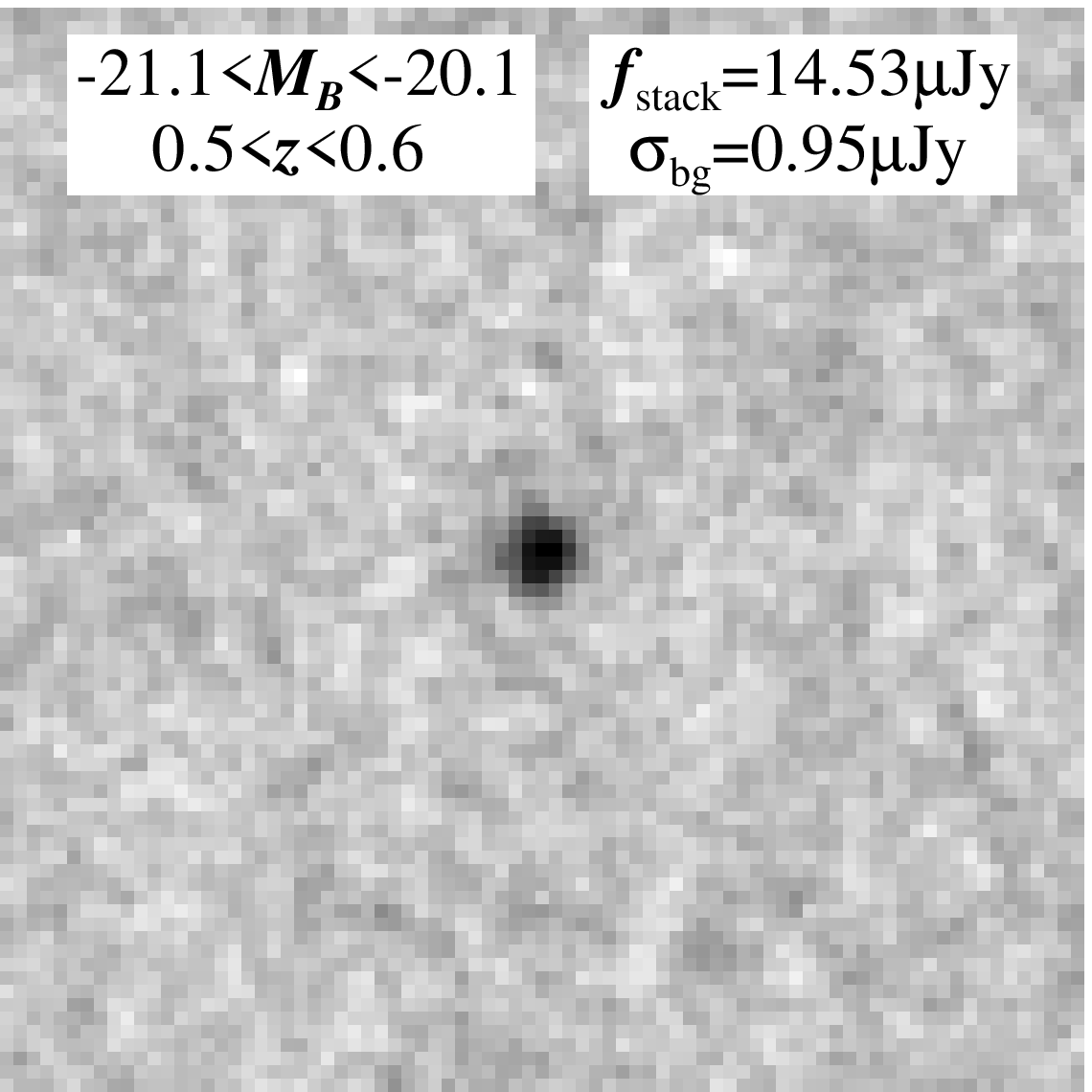}\hfill
\includegraphics[width=0.24\textwidth,clip]{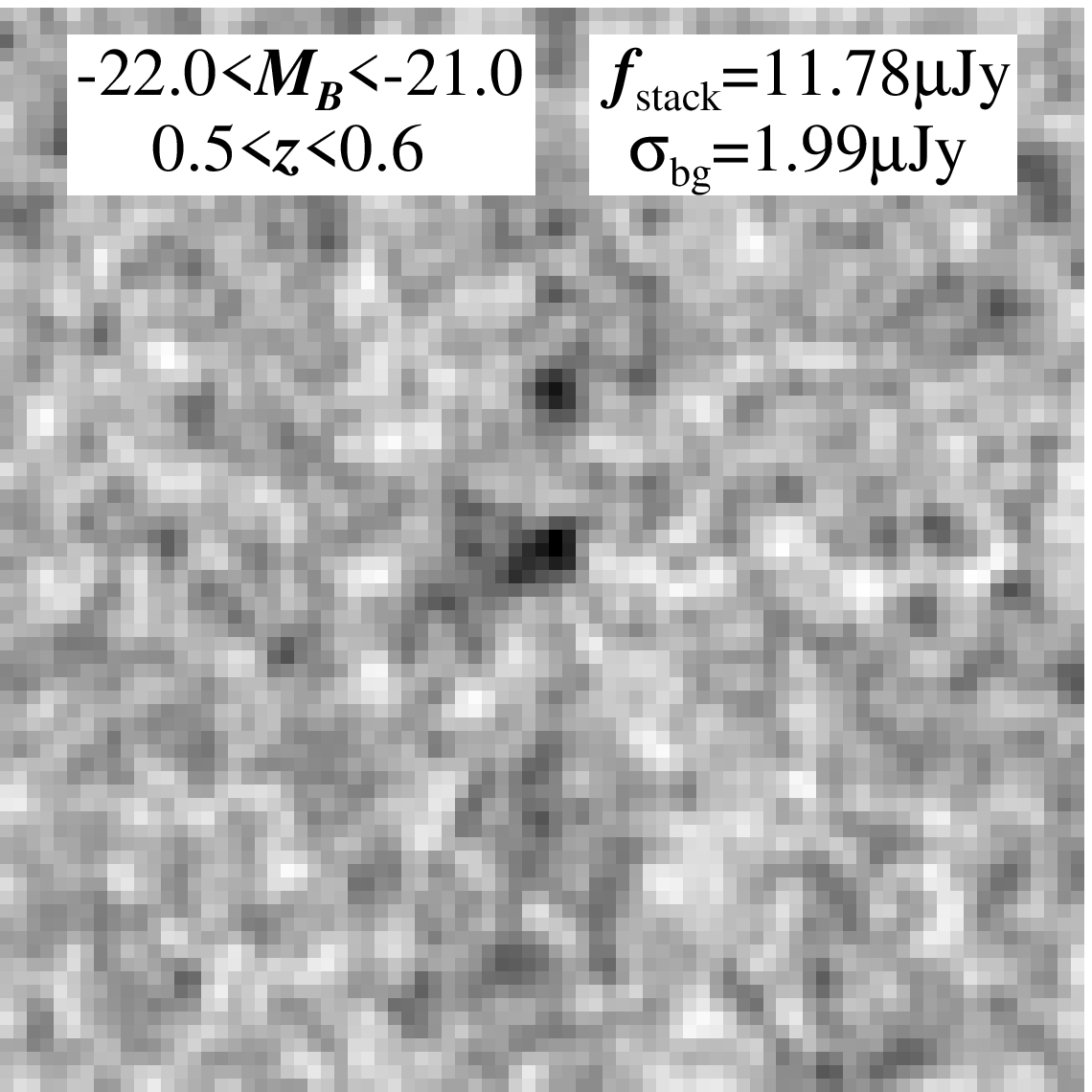}  \includegraphics[width=0.24\textwidth,clip]{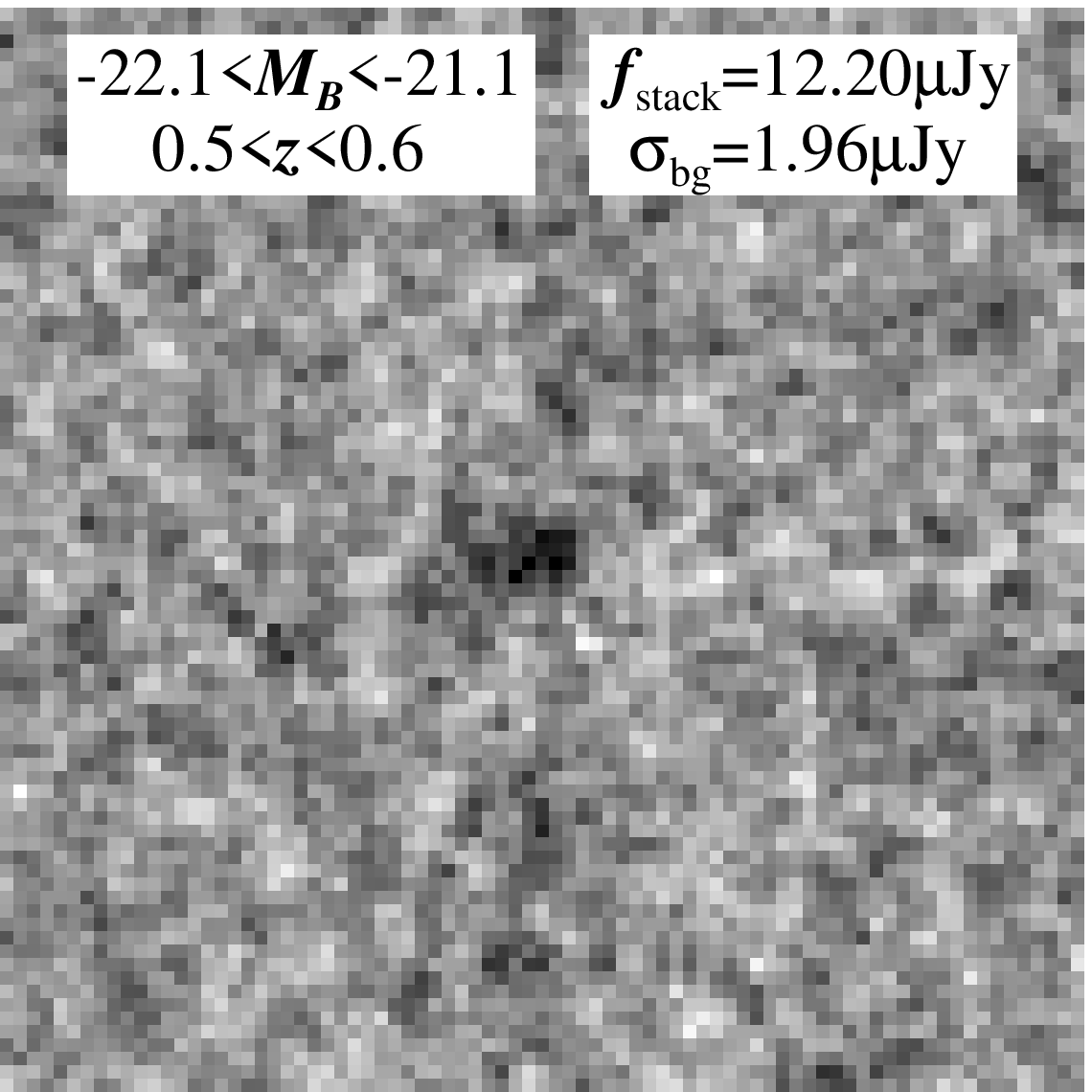}
\vskip 3mm
\includegraphics[width=0.24\textwidth,clip]{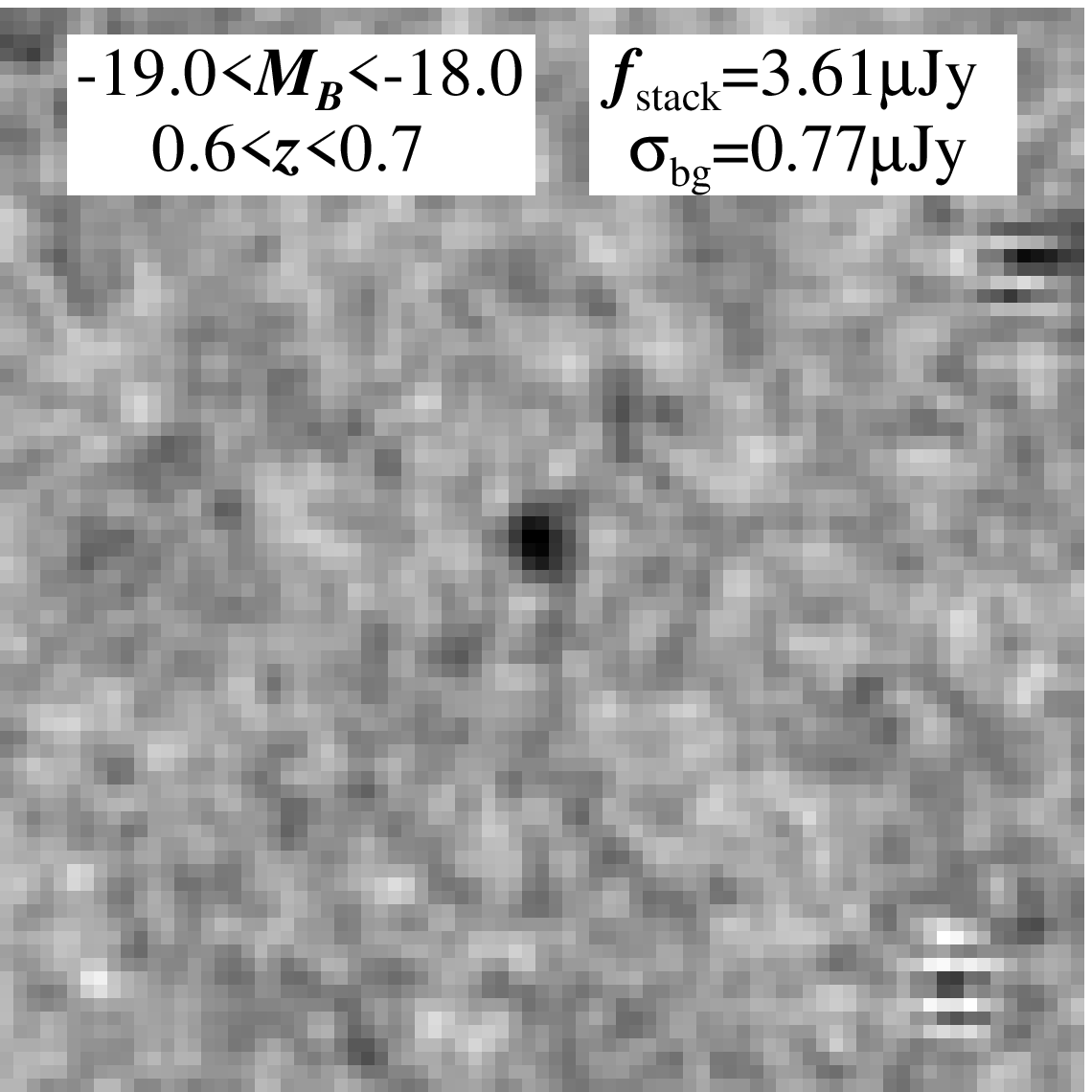}  \includegraphics[width=0.24\textwidth,clip]{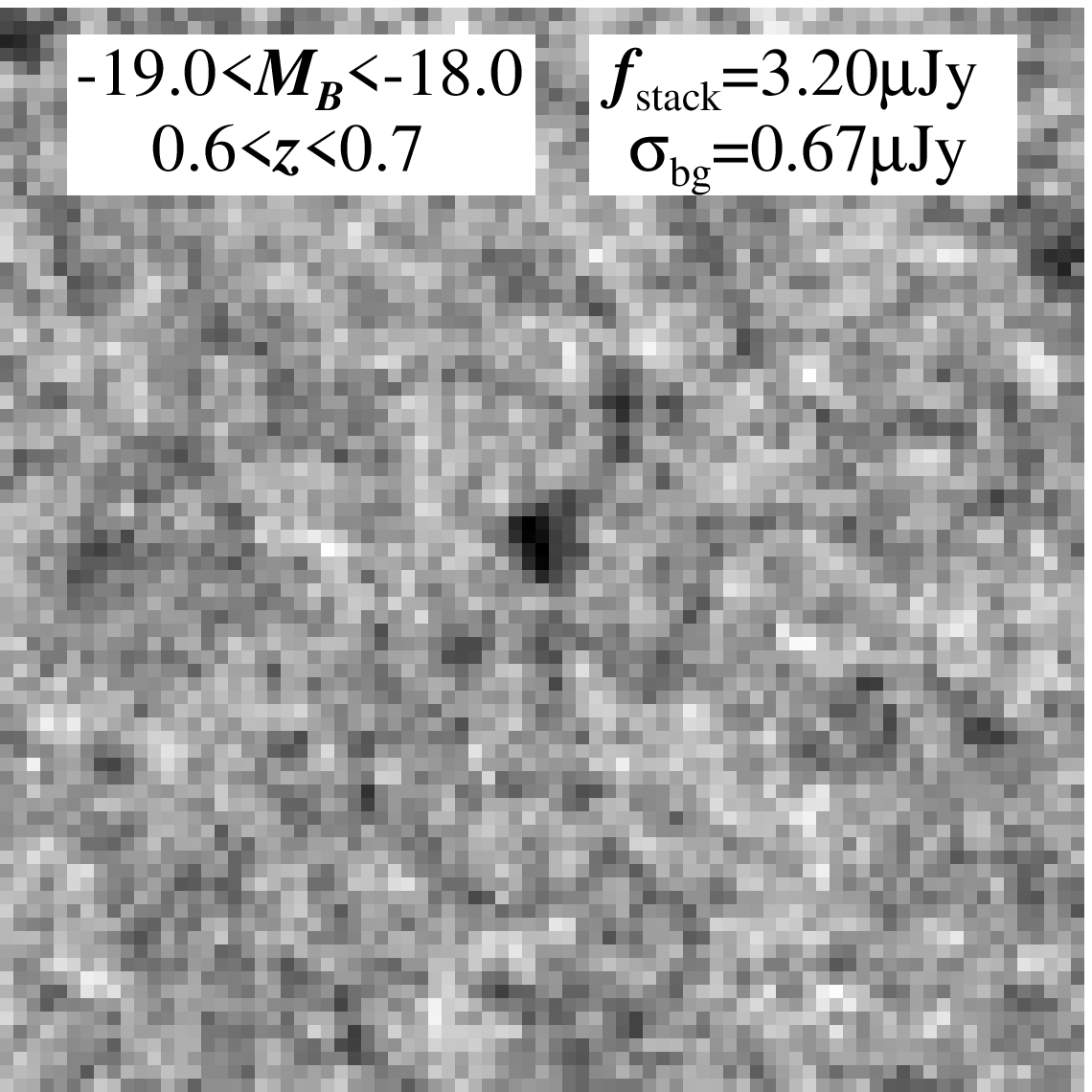}\hfill
\includegraphics[width=0.24\textwidth,clip]{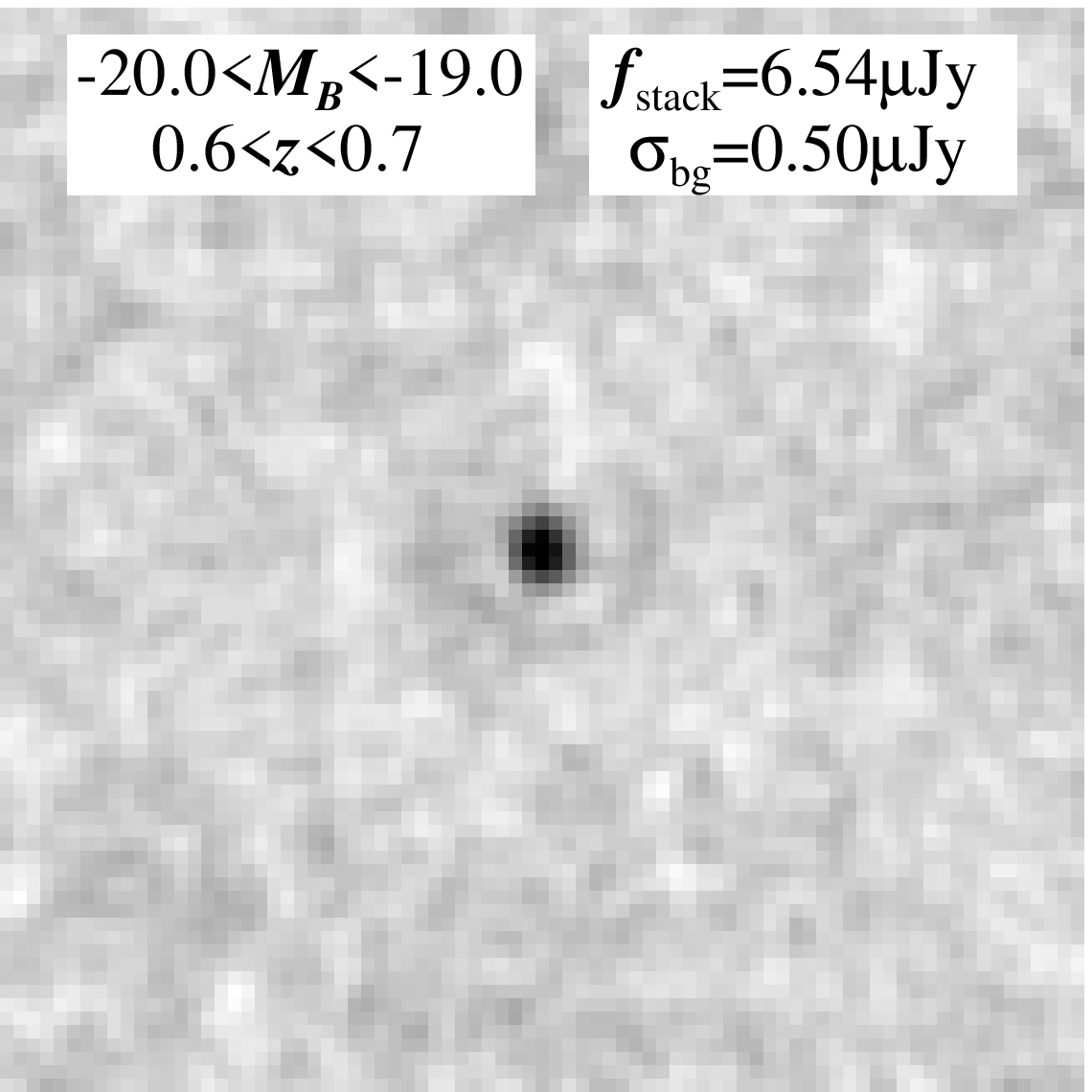}  \includegraphics[width=0.24\textwidth,clip]{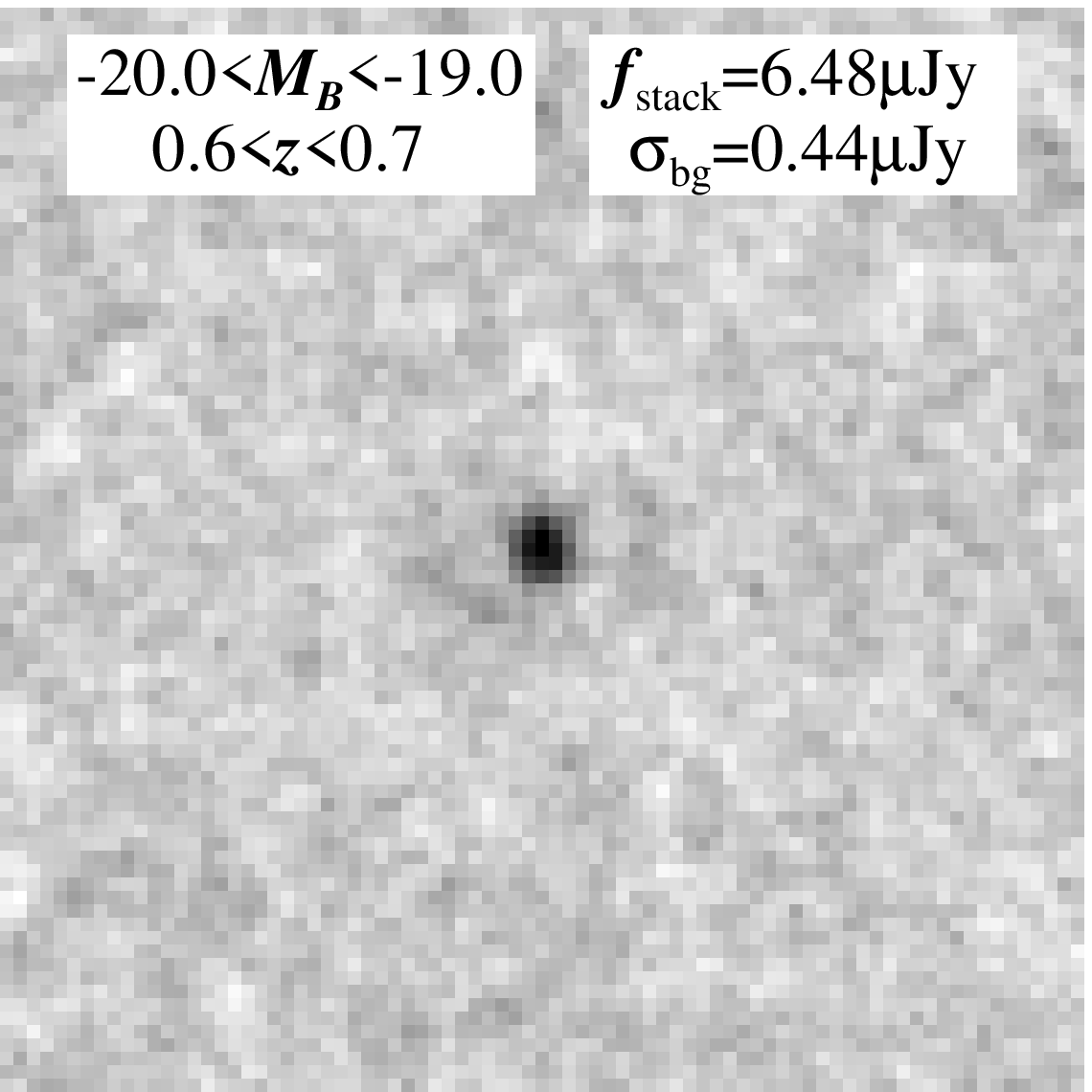}
\vskip 3mm
\includegraphics[width=0.24\textwidth,clip]{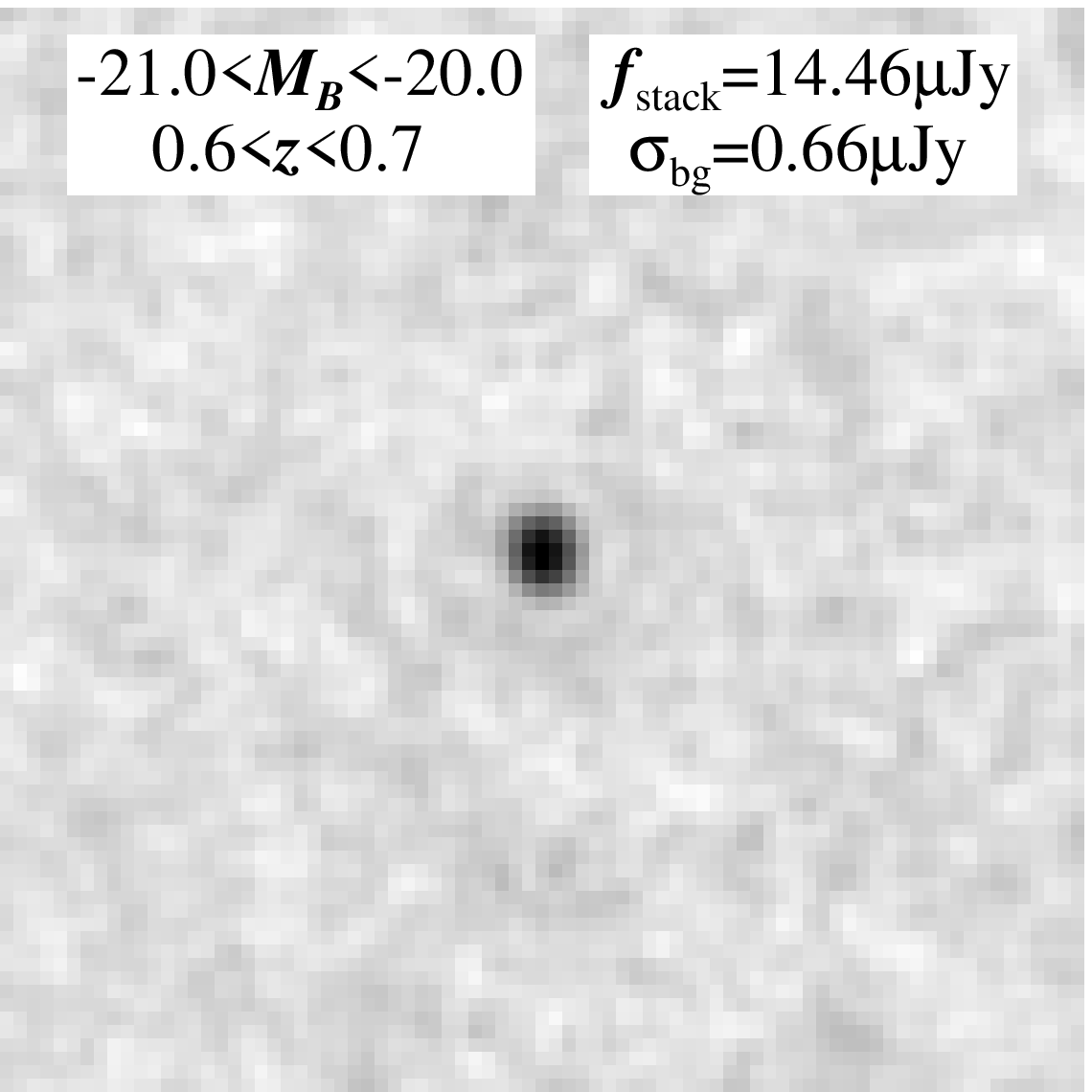} \includegraphics[width=0.24\textwidth,clip]{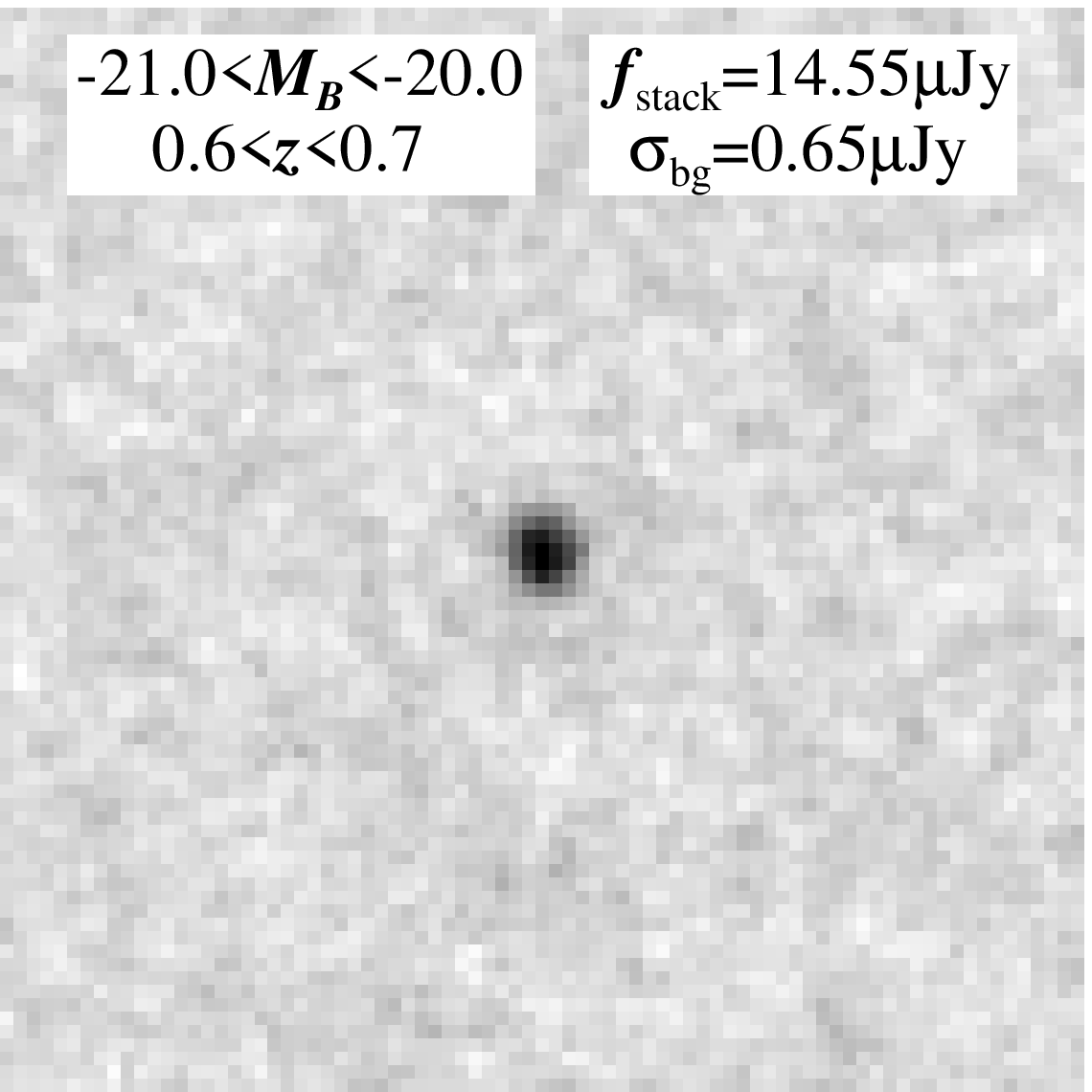}\hfill
\includegraphics[width=0.24\textwidth,clip]{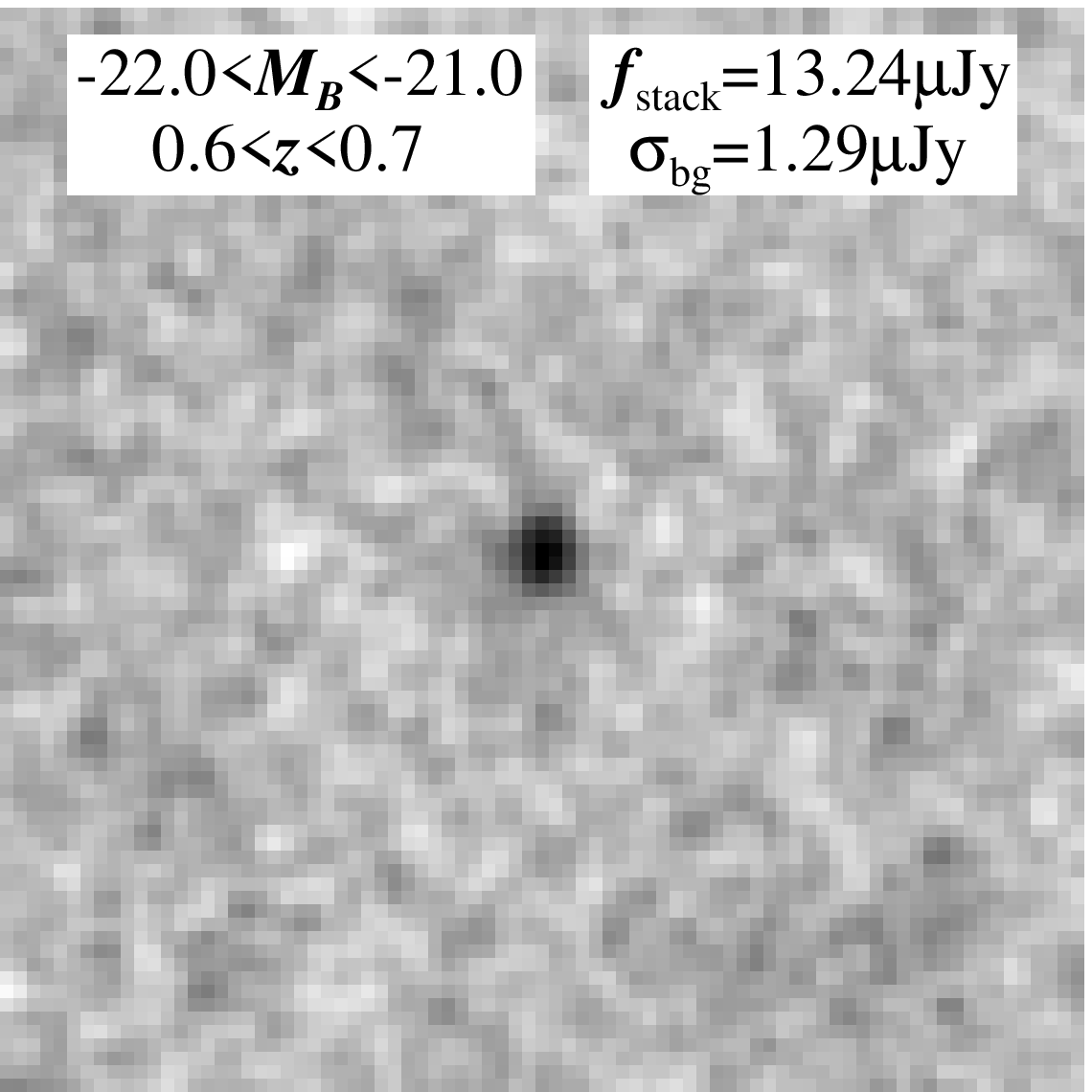} \includegraphics[width=0.24\textwidth,clip]{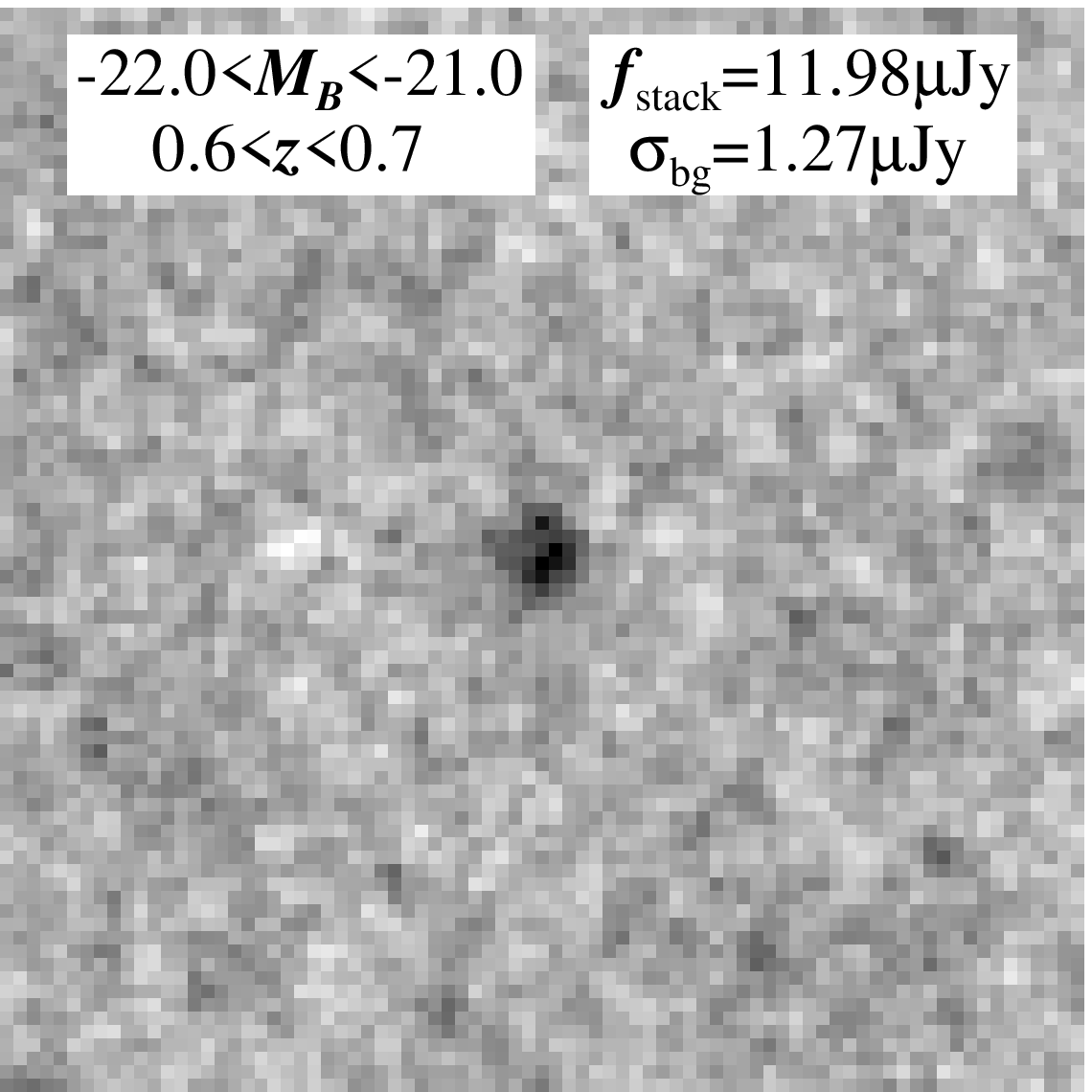}
\vskip 3mm
\includegraphics[width=0.24\textwidth,clip]{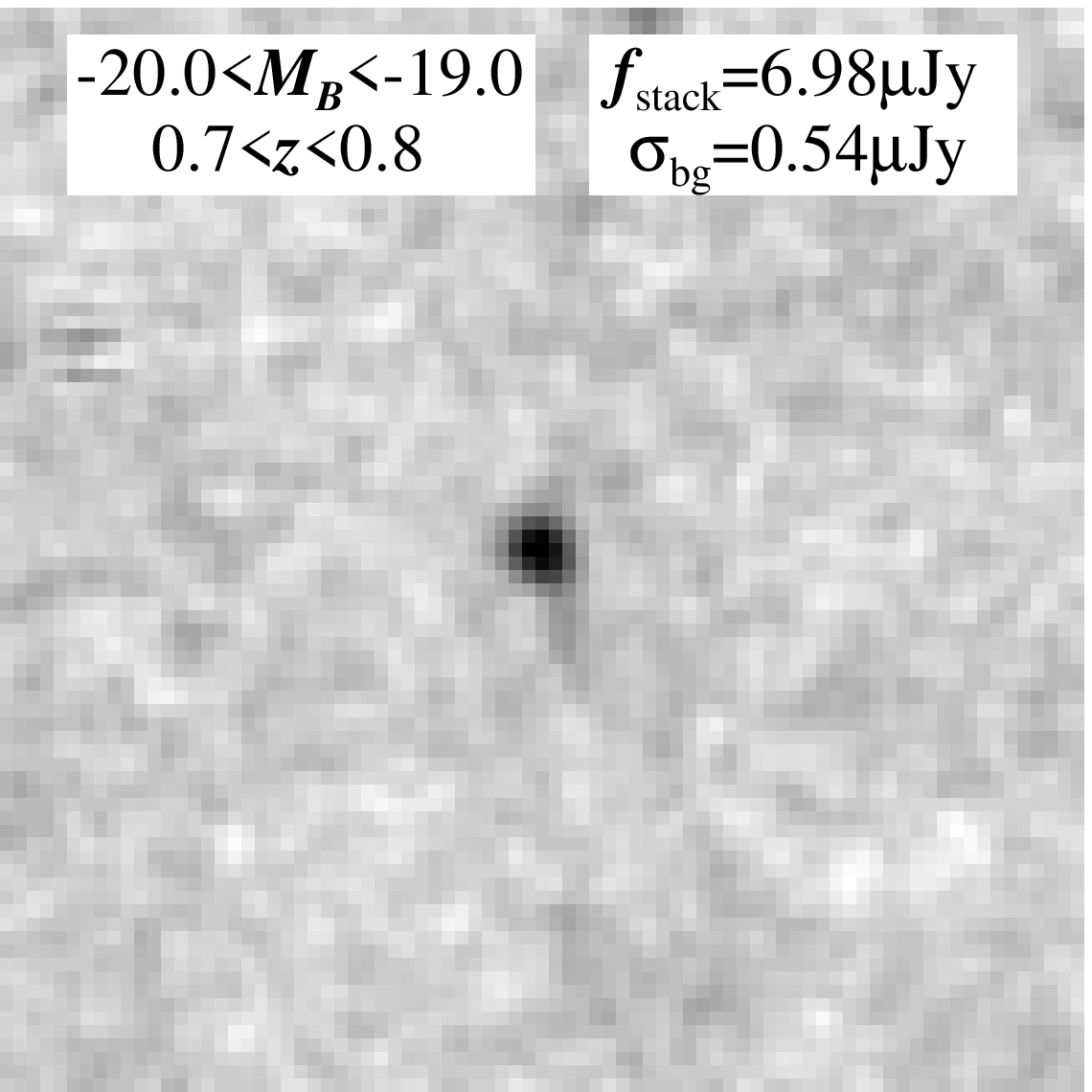} \includegraphics[width=0.24\textwidth,clip]{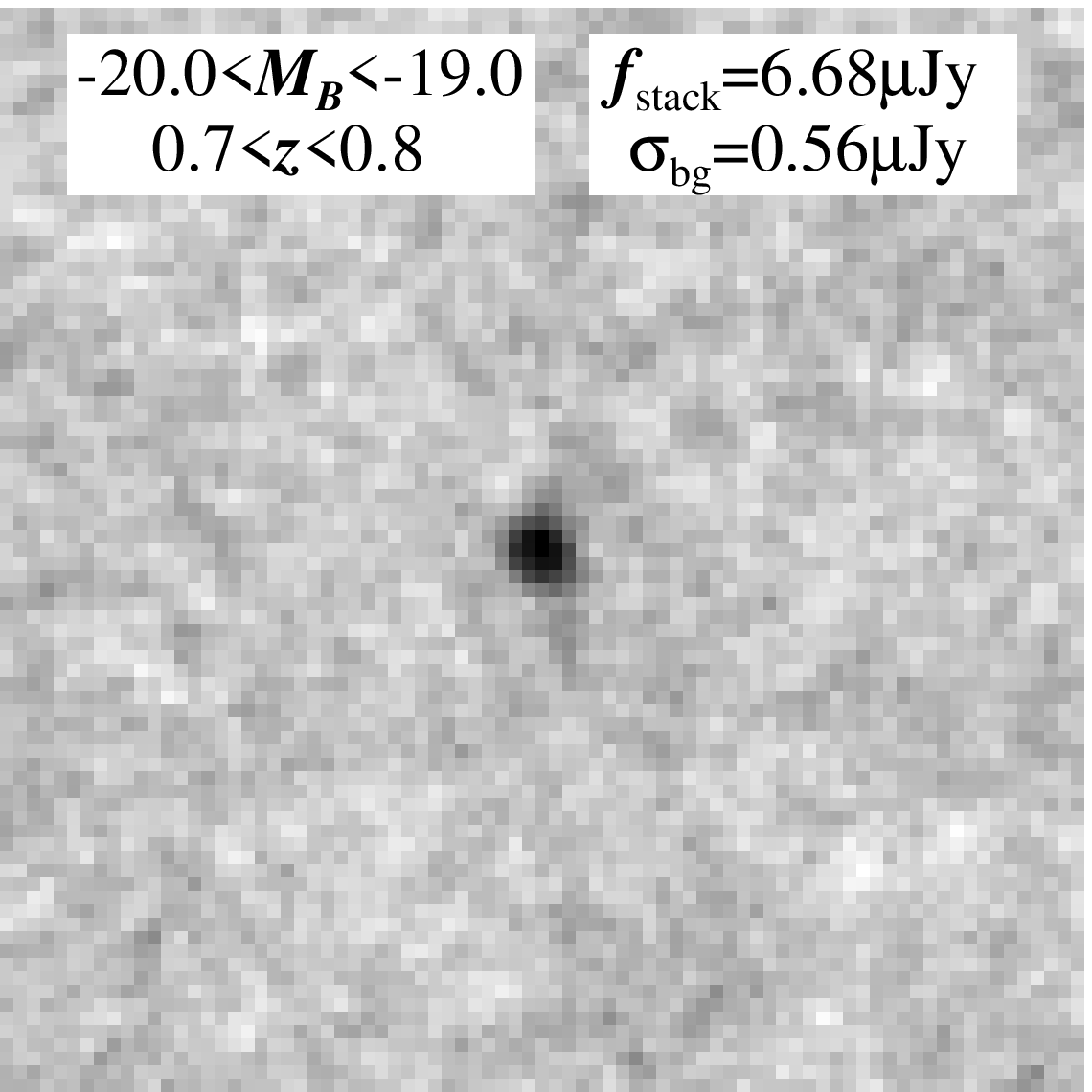}\hfill
\includegraphics[width=0.24\textwidth,clip]{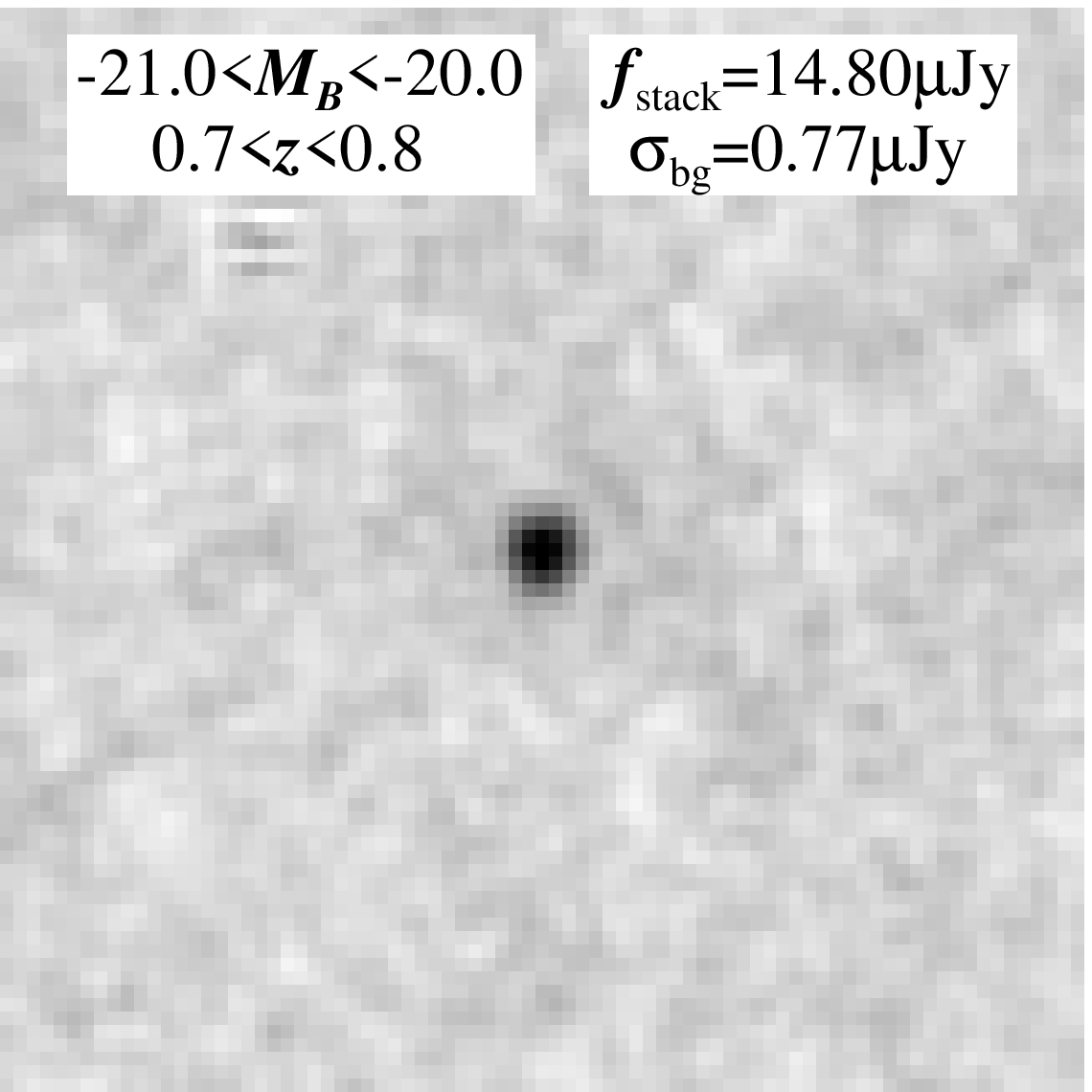} \includegraphics[width=0.24\textwidth,clip]{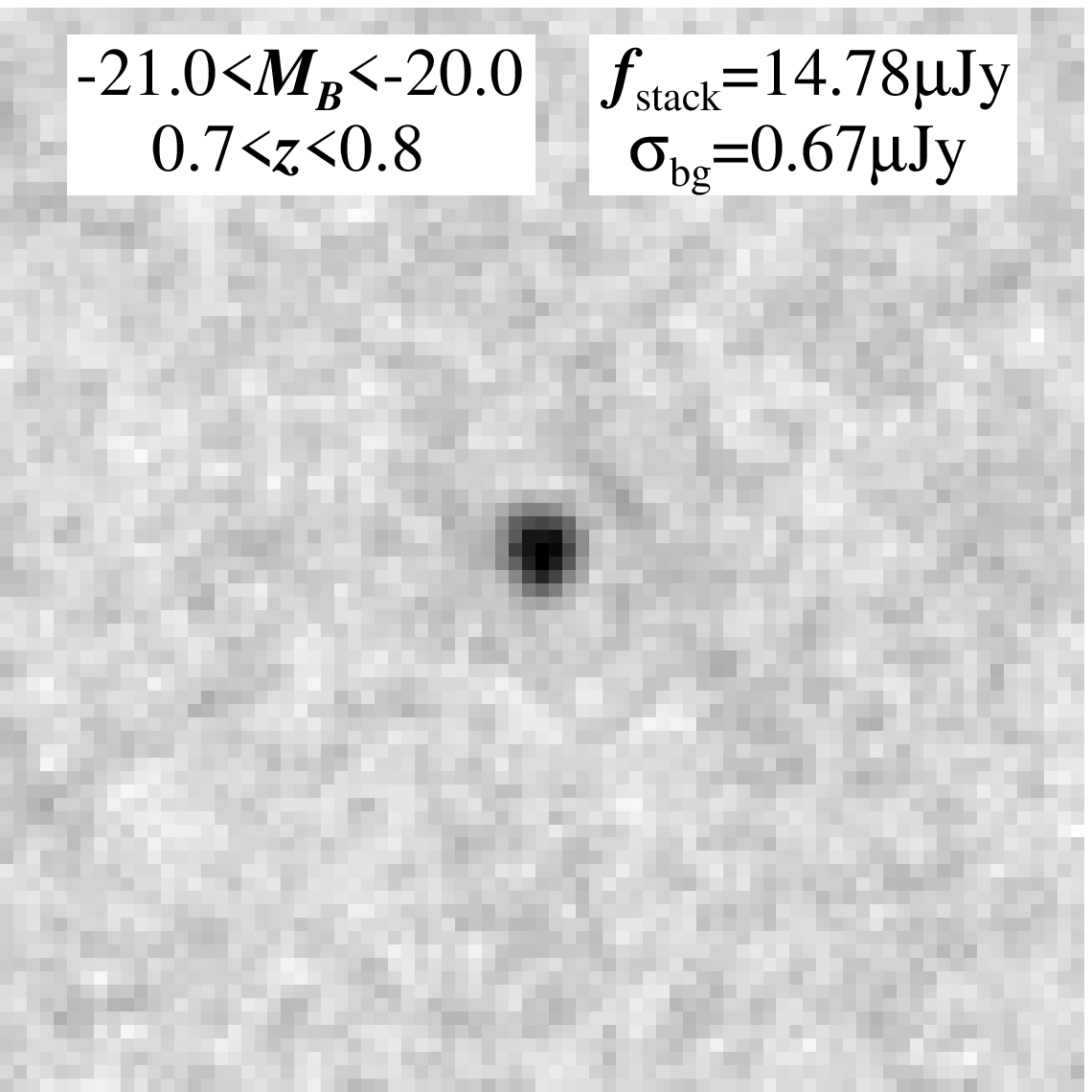}
\caption{Continued.}
\end{figure*}

\addtocounter{figure}{-1}

\begin{figure*}[] 
\includegraphics[width=0.24\textwidth,clip]{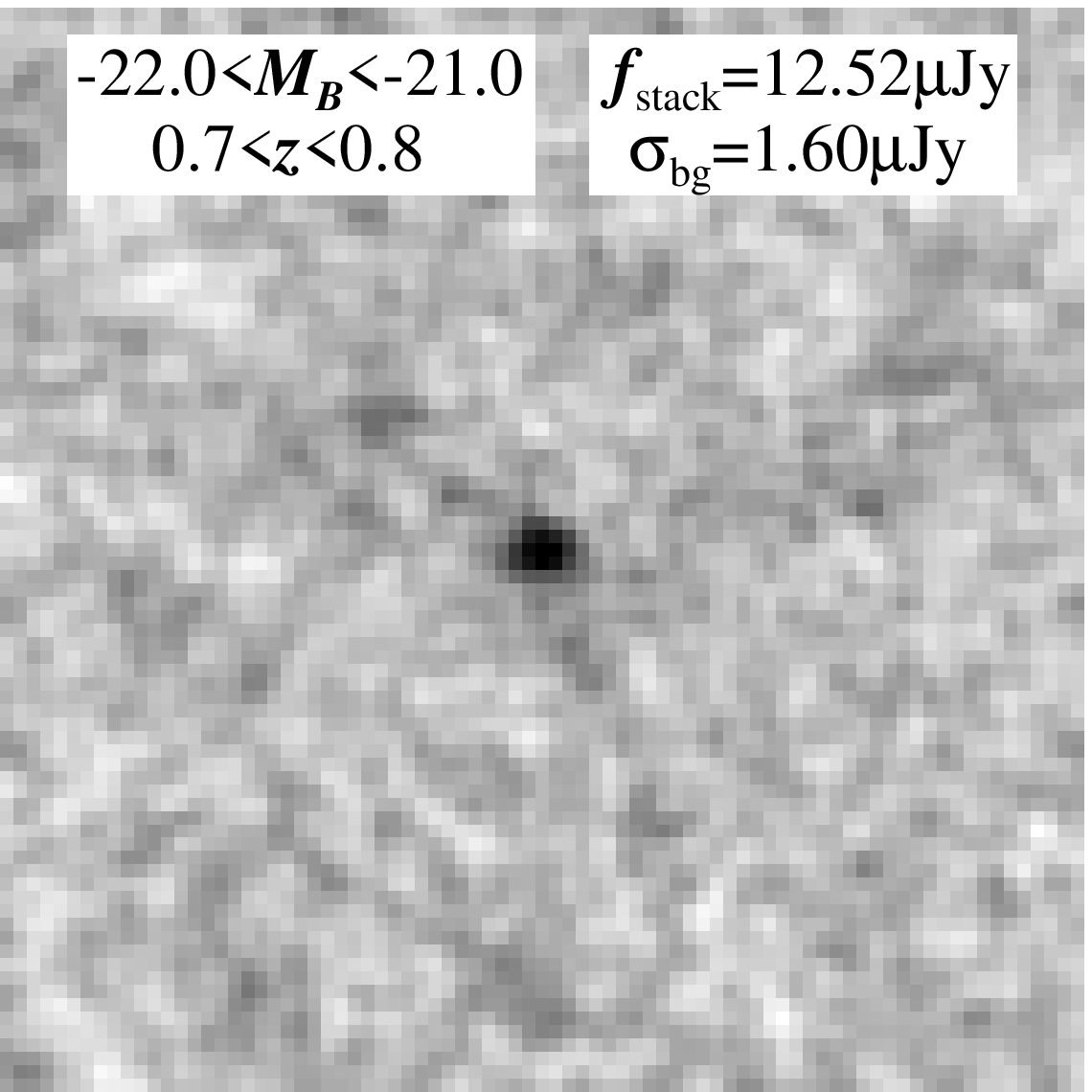} \includegraphics[width=0.24\textwidth,clip]{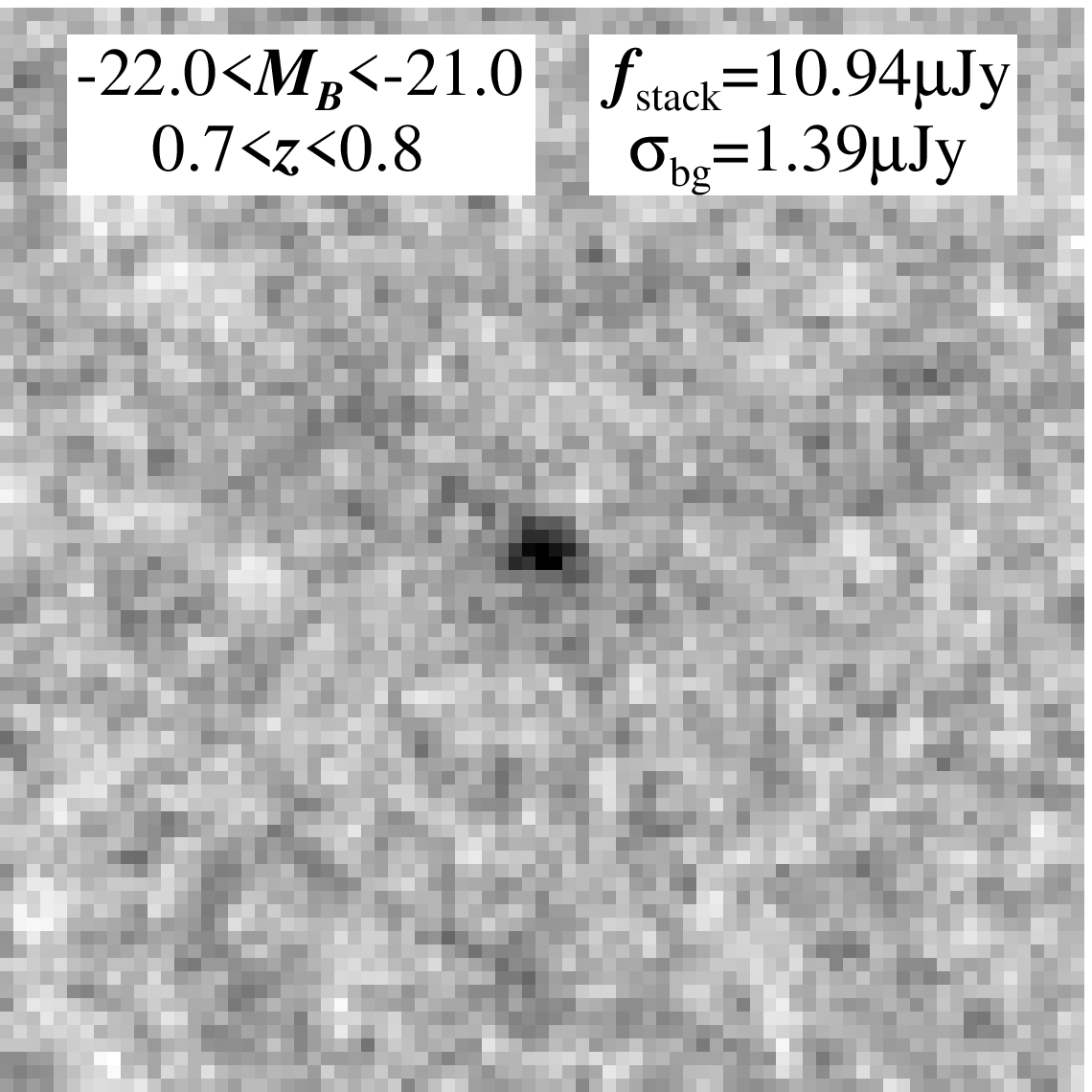}\hfill
\includegraphics[width=0.24\textwidth,clip]{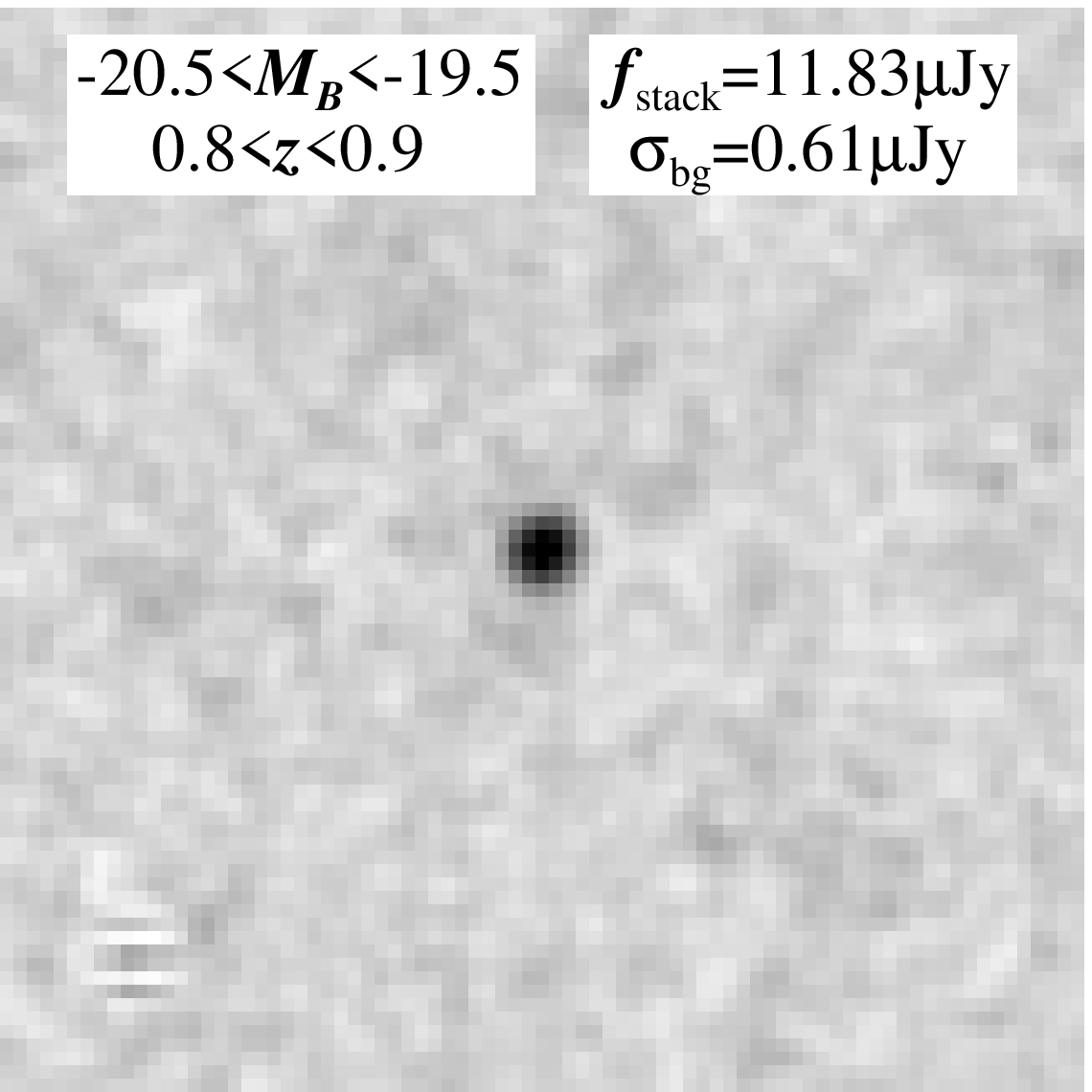} \includegraphics[width=0.24\textwidth,clip]{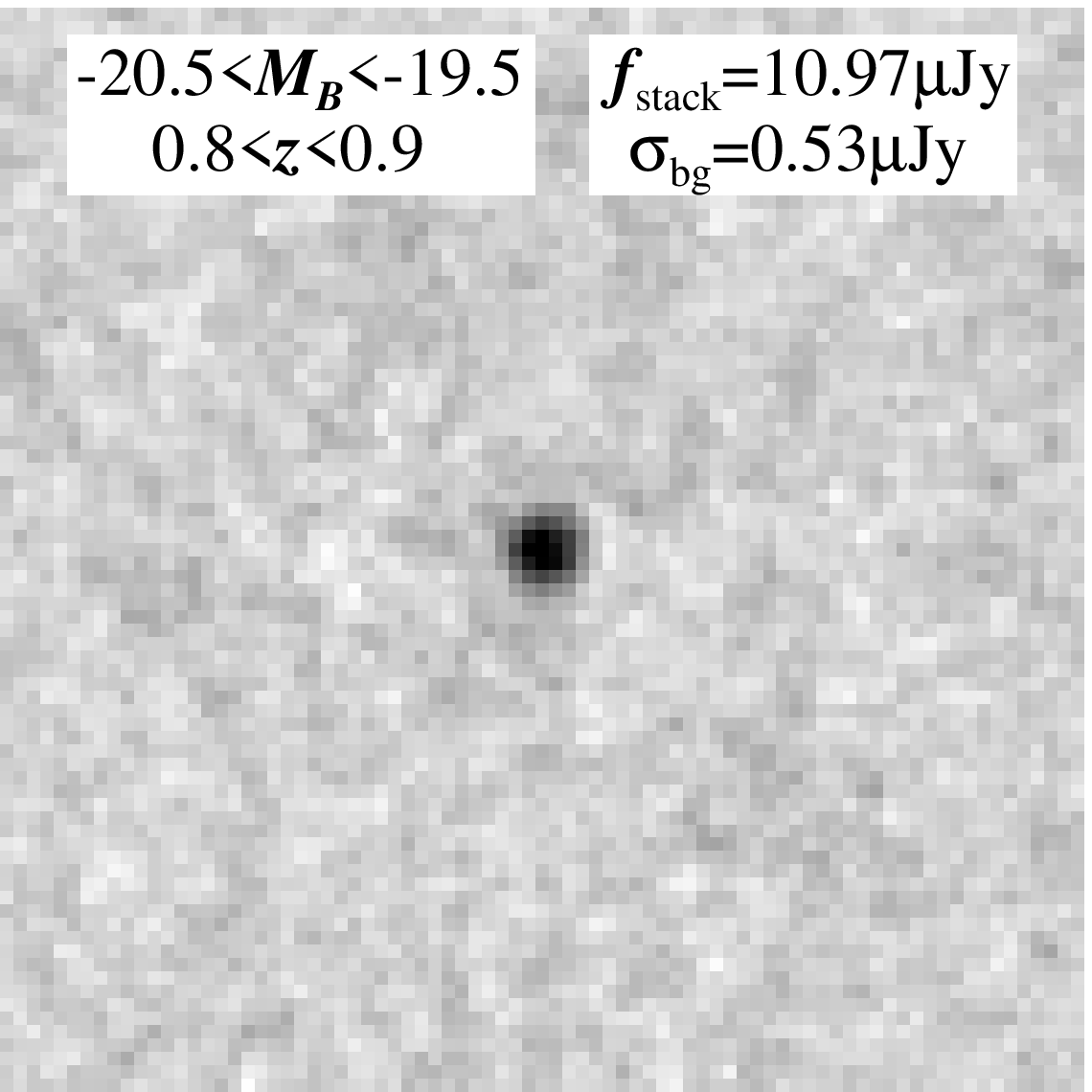}
\vskip 3mm
\includegraphics[width=0.24\textwidth,clip]{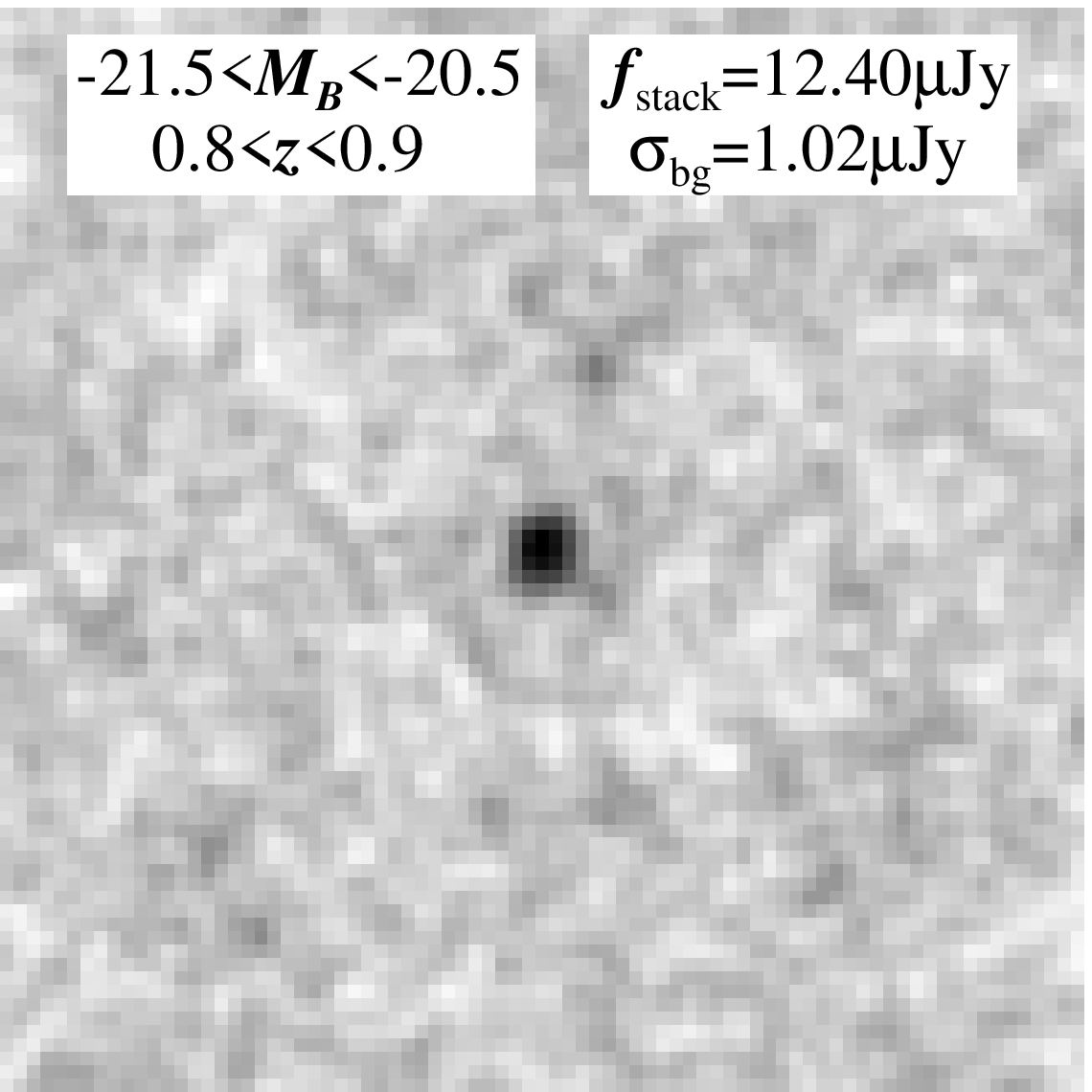} \includegraphics[width=0.24\textwidth,clip]{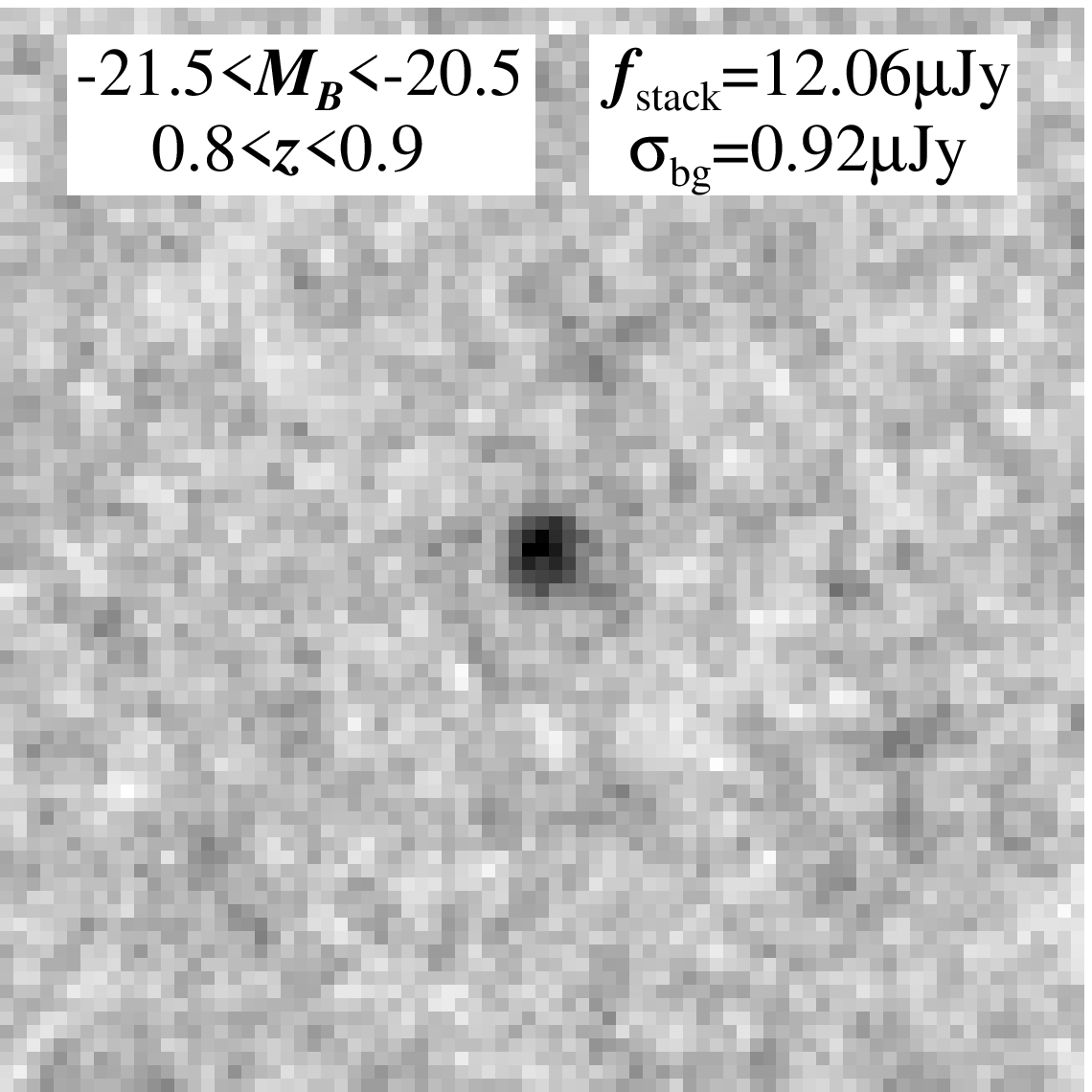}\hfill
\includegraphics[width=0.24\textwidth,clip]{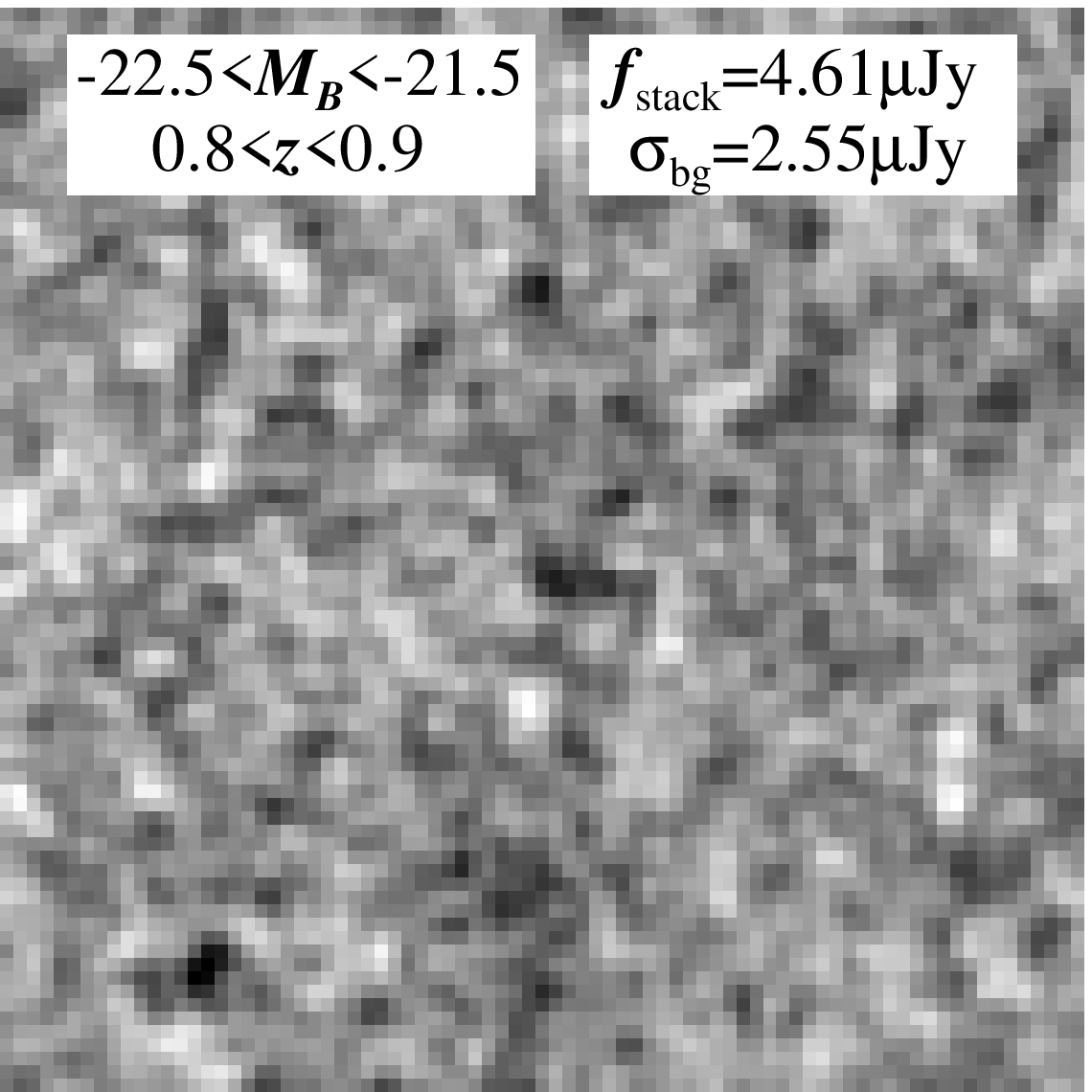} \includegraphics[width=0.24\textwidth,clip]{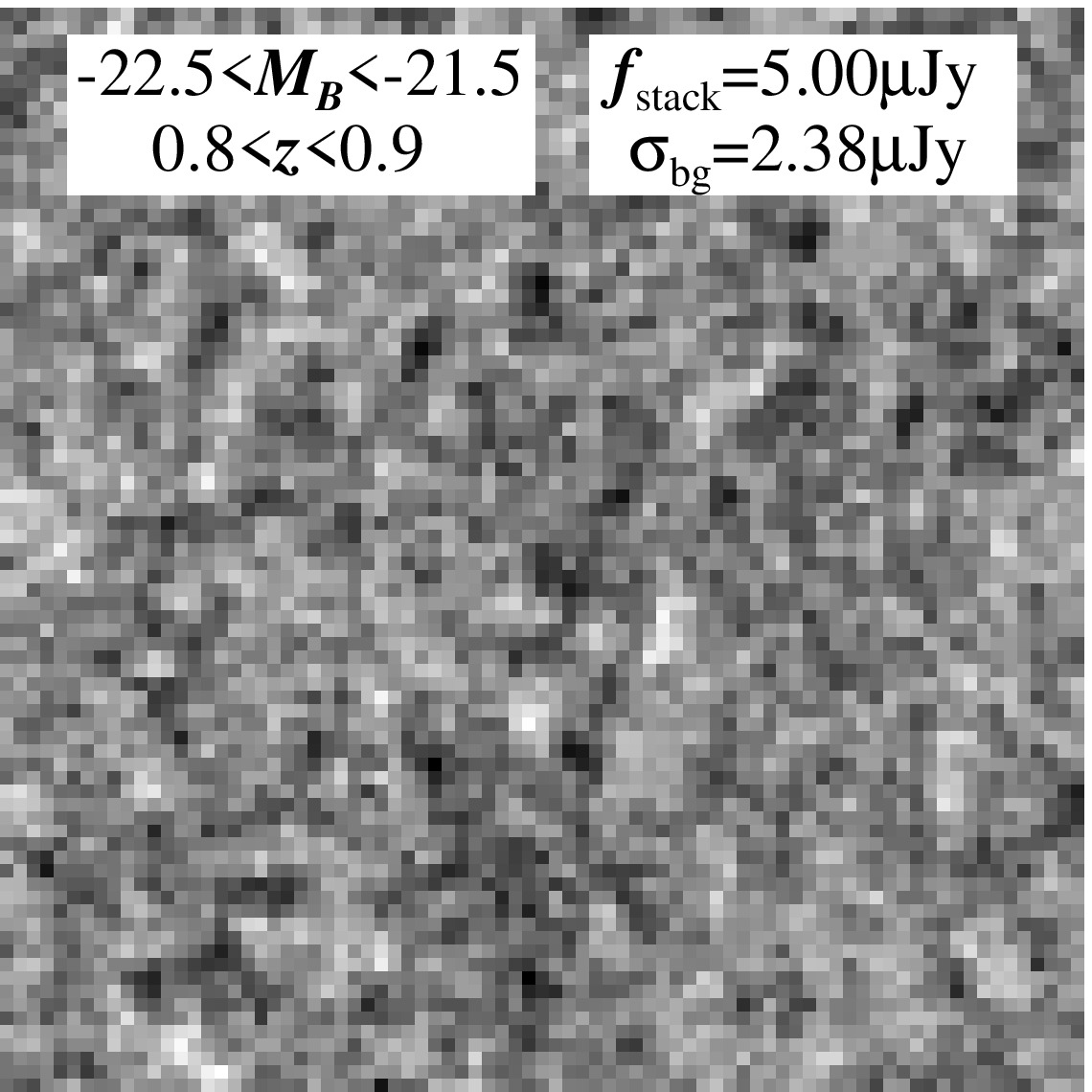}
\vskip 3mm
\includegraphics[width=0.24\textwidth,clip]{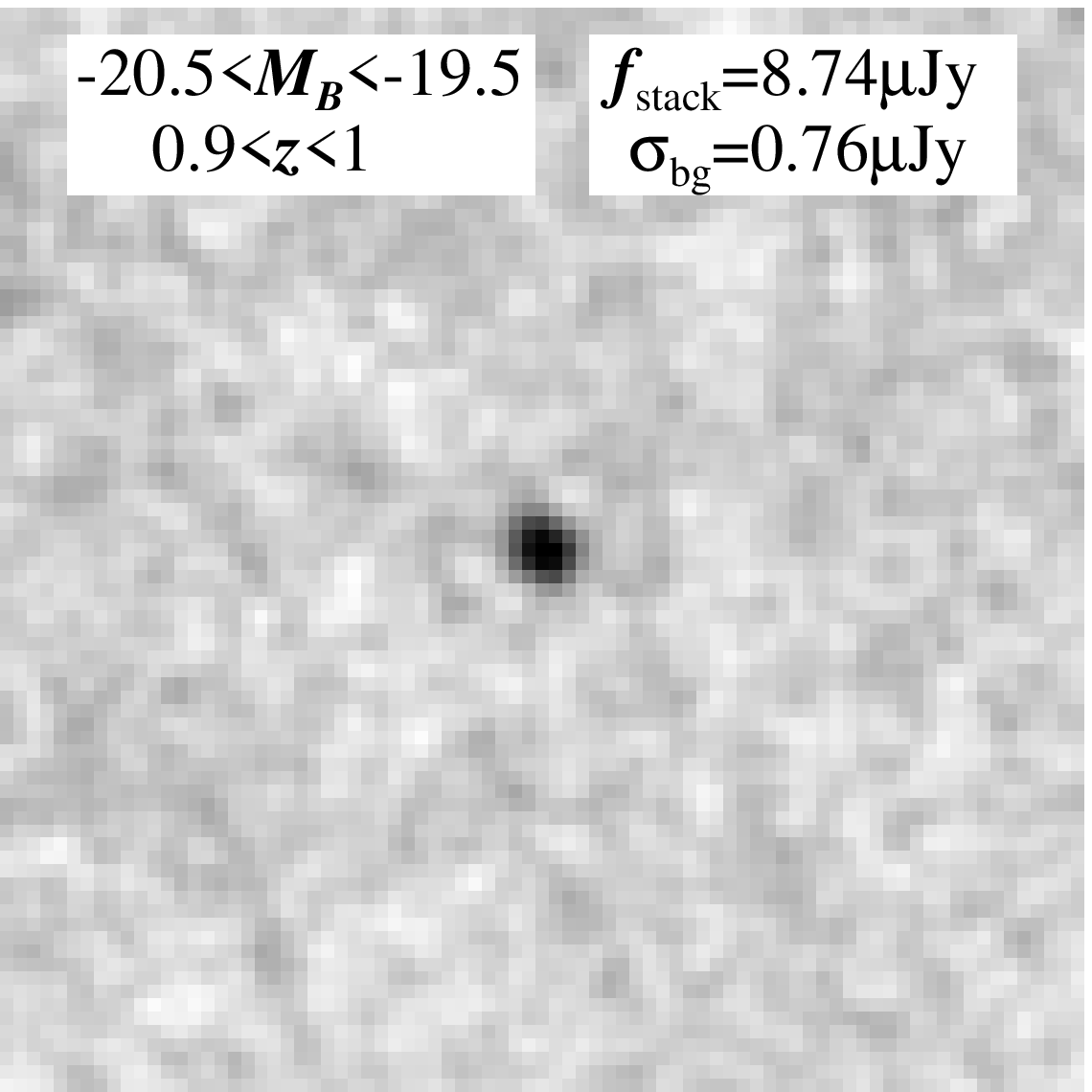} \includegraphics[width=0.24\textwidth,clip]{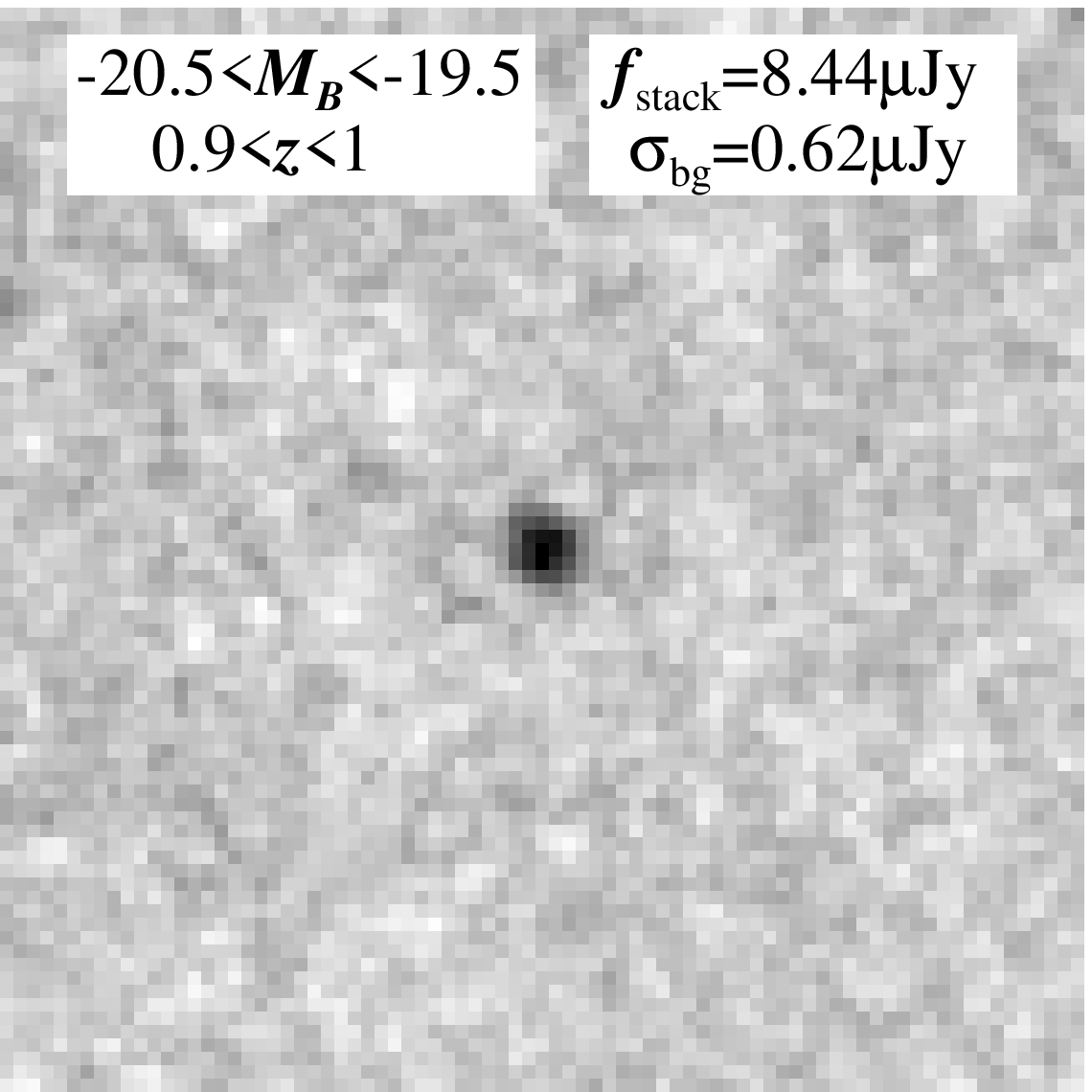}\hfill
\includegraphics[width=0.24\textwidth,clip]{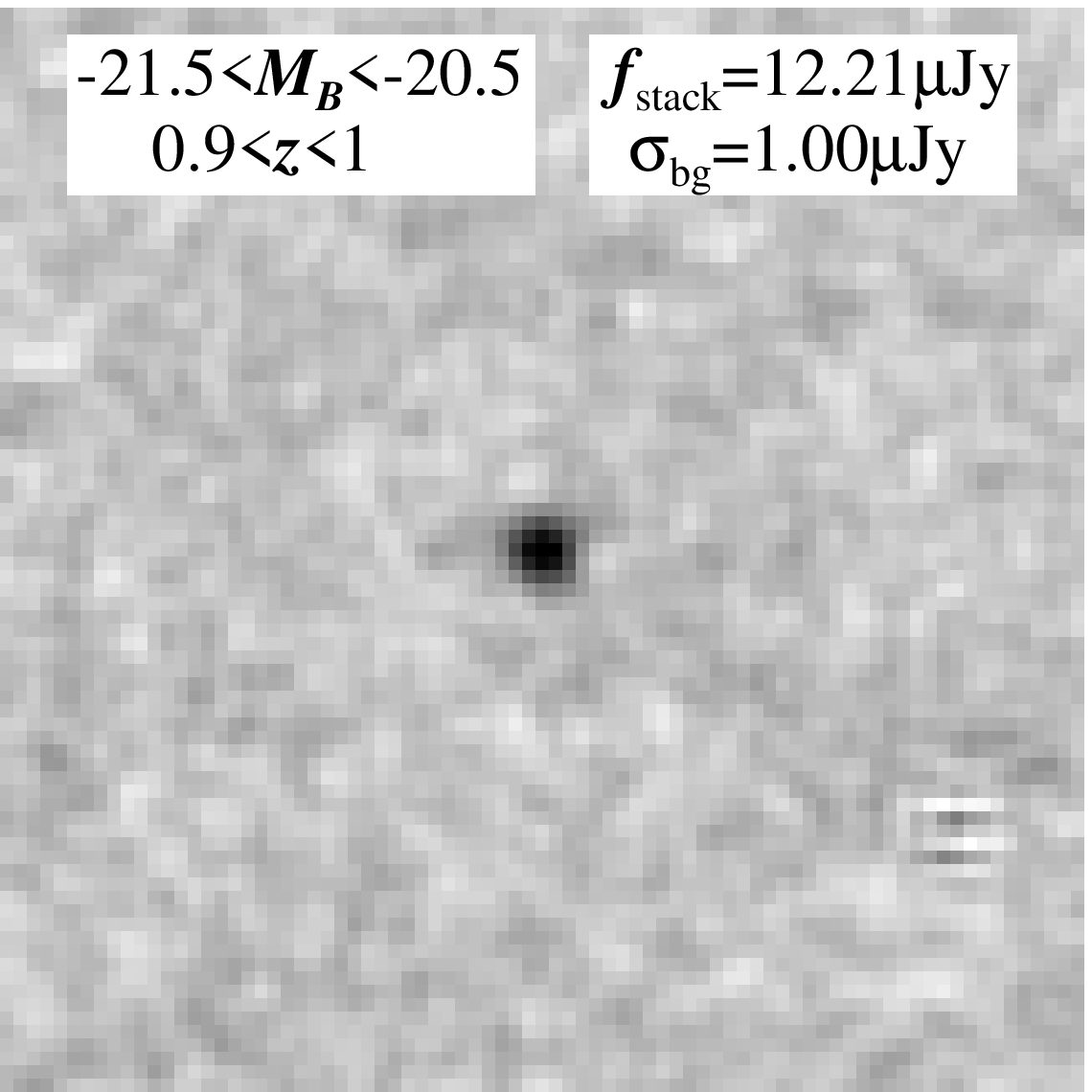} \includegraphics[width=0.24\textwidth,clip]{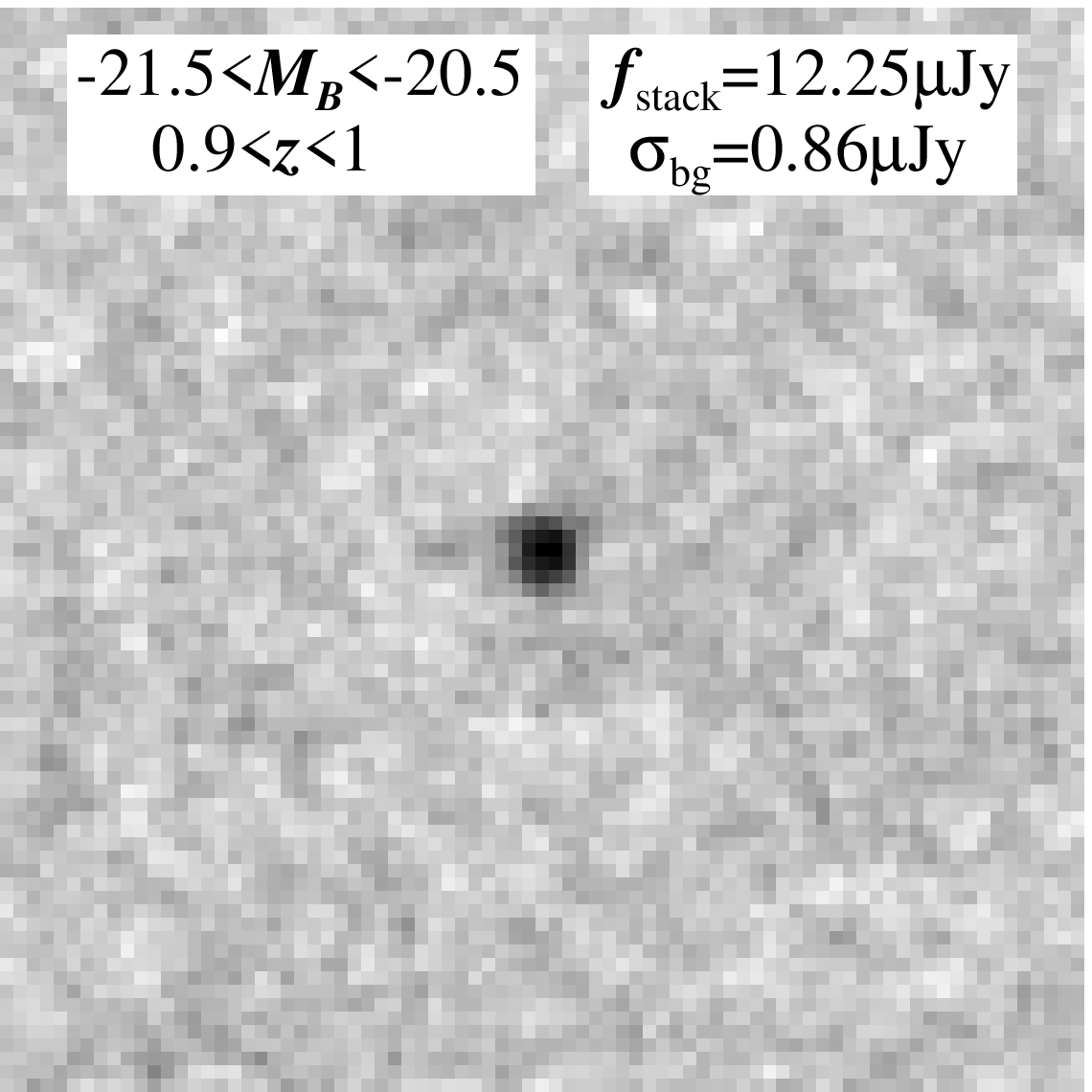}
\vskip 3mm
\includegraphics[width=0.24\textwidth,clip]{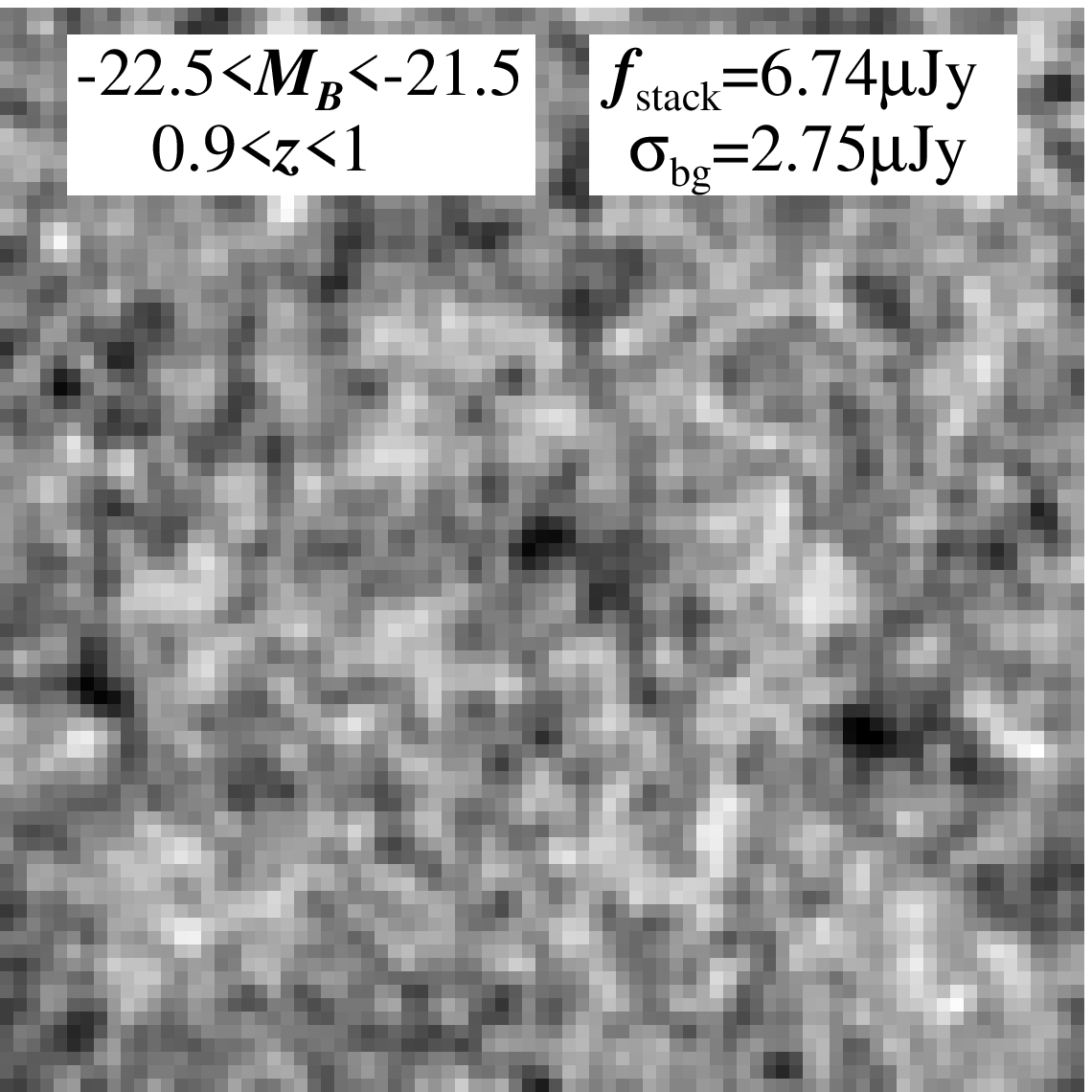} \includegraphics[width=0.24\textwidth,clip]{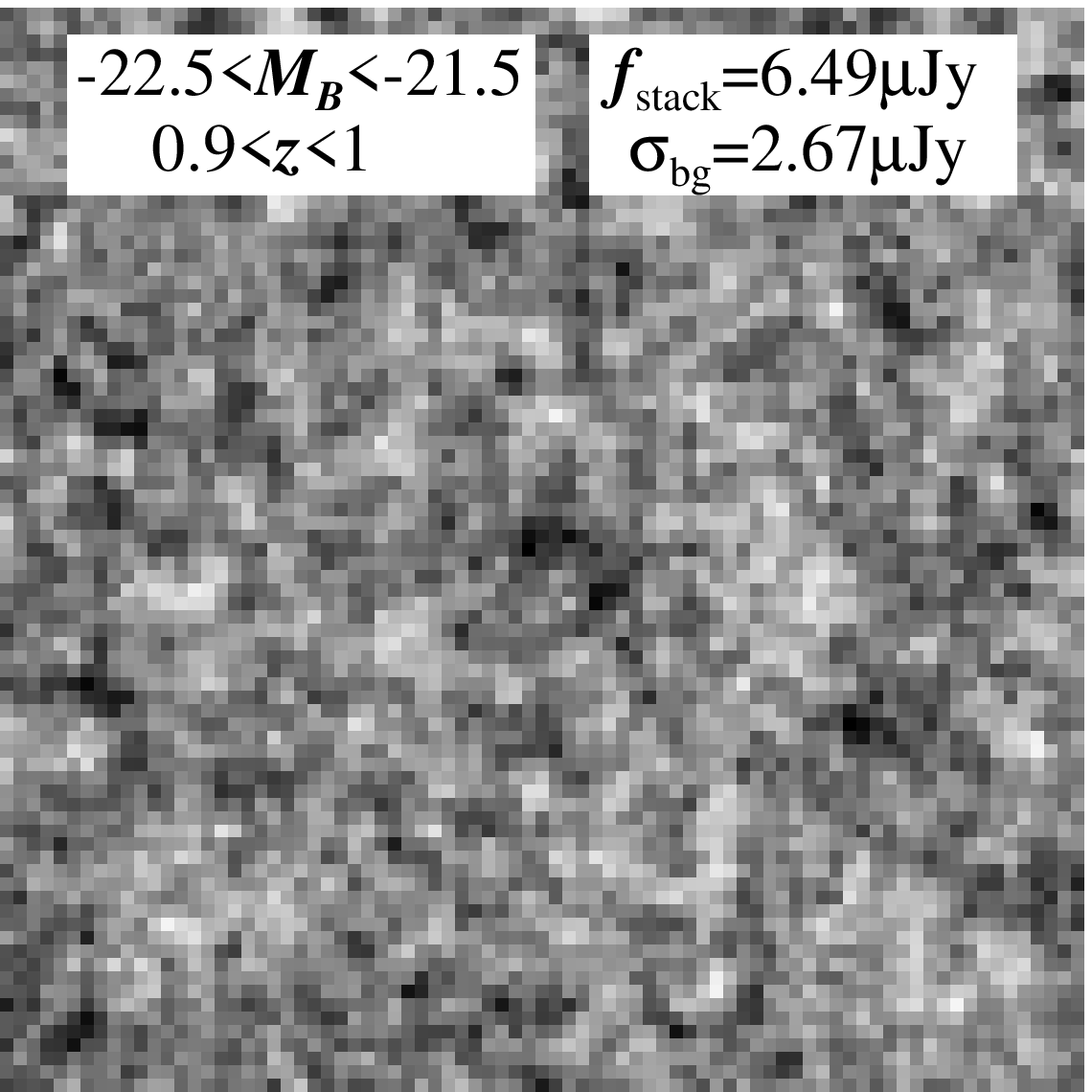}\hfill
\caption{Continued.}
\end{figure*}

%% file: ms_apj.bbl
\begin{thebibliography}{}

\bibitem[Adelberger \& Steidel (2000)]{Adelberger00} Adelberger, K. L., \& Steidel, C. C.
        2000, ApJ, 544, 218
\bibitem[Bauer et al. (2005)]{Bauer05} Bauer, A. E., Drory, N., Hill, G. J., \& Feulner, G.
        2005, ApJ, 621, L89 
\bibitem[Bell (2003)]{Bell03} Bell, E. F.
        2003, ApJ, 586, 794
\bibitem[Bell et al. (2004)]{Bell04} Bell, E. F., et al.
        2004, ApJ, 608, 752
\bibitem[Bell et al. (2005)]{Bell05} Bell, E. F., et al.
        2005, ApJ, 625, 23
\bibitem[Bertin \& Arnouts (1996)]{Bertin96} Bertin, E., \& Arnouts, S.
        1996, A\&AS, 117, 393
\bibitem[Brinchmann \& Ellis (2000)]{Brinchmann00} Brinchmann, J., \& Ellis, R. S.
        2000, ApJ, 536, L77
\bibitem[Buat et al. (2005)]{Buat05} Buat, V., et al.
        2005, ApJ, 619, L51
\bibitem[Bundy et al. (2005)]{Bundy05} Bundy, K., Ellis, R. S., \& Conselice, C. J.
        2005, ApJ, 625, 621
\bibitem[Calzetti \& Heckman (1999)]{Calzetti99} Calzetti, D., \& Heckman, T. M.
        1999, ApJ, 519, 27
\bibitem[Chary \& Elbaz (2001)]{Chary01} Chary, R., \& Elbaz, D.
        2001, ApJ, 556, 562
\bibitem[Dale et al. (2005)]{Dale05} Dale, D. A., et al.
        2005, ApJ, in press (astro-ph/0507645)
\bibitem[Devriendt et al. (1999)]{Devriendt99} Devriendt, J. E. G., Guiderdoni, B., \& Sadat, R.
        1999, A\&A, 350, 381
\bibitem[Dole et al. (2004)]{Dole04} Dole, H., et al.
        2004, ApJS, 154, 93
\bibitem[Donley et al. (2005)]{Donley} Donley, J. L., Rieke, G. H., Rigby,
  J. R., P\'erez-Gonz\'alez P. G.
        2005, ApJ in press (astro-ph/0507676)
\bibitem[Faber et al. (2005)]{Faber05} Faber, S. M., et al.
        2005, ApJ, submitted (astro-ph/0506044)
\bibitem[Franceschini et al. (2003)]{Franceschini03} Franceschini, A., et al.
        2003, A\&A, 403, 501
\bibitem[Gordon et al. (2000)]{Gordon00} Gordon, K. D., Clayton, G. C., Witt, A. N., \& Misselt, K. A. 
        2000, ApJ, 533, 236
\bibitem[Gordon et al. (2005)]{Gordon} Gordon, K. D., et al.
        2005, PASP, 117, 503
\bibitem[Heavens et al. (2004)]{Heavens04} Heavens, A., Panter, B., Jimenez, R., \& Dunlop, J.
        2004, Nature, 428, 625
\bibitem[Hopkins (2004)]{Hopkins04} Hopkins, A. M.
        2004, ApJ, 615, 221
\bibitem[Hopkins et al. (2001)]{Hopkins01} Hopkins, A. M., Connolly, A. J., Haarsma, D. B., \& Cram, L. E. 
        2001, AJ, 122, 288
\bibitem[Ilbert et al. (2005)]{Ilbert} Ilbert, O., et al.
        2005, A\&A, in press (astro-ph/0409134)
\bibitem[Juneau et al. (2005)]{Juneau05} Juneau, S., et al.
        2005, ApJ, 619, L135
\bibitem[Kennicutt (1998a)]{Kennicutt98a} Kennicutt, Jr., R. C.
        1998a, ARA\&A, 36, 189
\bibitem[Kennicutt (1998b)]{Kennicutt98b} Kennicutt, Jr., R. C.
        1998b, ApJ, 498, 541
\bibitem[Le F$\acute{\rm e}$vre et al. (2004)]{LeFevre04} Le F$\acute{\rm e}$vre, O., et al.
        2004, A\&A, 428, 1043
\bibitem[Le Floc'h et al. (2005)]{LeFloc'h05} Le Floc'h E., et al.
        2005, ApJ, 632, 169 
\bibitem[Lehmer et al. (2005)]{Lehmer} Lehmer, B. D., et al.
        2005, ApJS, 161, 21
\bibitem[Madau et al. (1996)]{Madau96} Madau, P., Ferguson, H. C., Dickinson, M. E., Giavalisco, M., Steidel, C. C., \& Fruchter, A.
        1996, MNRAS, 283, 1388 
\bibitem[Martin et al. (2005)]{Martin05} Martin, D. C., et al.
        2005, ApJL, 619, L59
\bibitem[Mart\'inez-Sansigre et al. (2005)]{Martinez} Mart\'inez-Sansigre, A.,
  et al.
        2005, Nature, 436, 666
\bibitem[Meurer et al. (1999)]{Meurer99} Meurer, G. R., Heckman, T. M., \& Calzetti, D. 
        1999, ApJ, 521, 64
\bibitem[Mushotzky \& Loewenstein (1997)]{Mushotzky97} Mushotzky, R. F., \& Loewenstein, M.
        1997, ApJ, 481, L63
\bibitem[Papovich \& Bell (2002)]{Papovich02} Papovich, C., \& Bell, E. F.
        2002, ApJ, 579, L1
\bibitem[Papovich et al. (2004)]{Papovich04} Papovich, C., et al.
        2004, ApJS, 154, 70
\bibitem[P\'erez-Gonz\'alez et al. (2005)]{Perez05} P\'erez-Gonz\'alez, P. G., et al.
        2005, ApJ, 630, 82 
\bibitem[Rieke et al. (2004)]{Rieke04} Rieke, G. H., et al.
        2004, ApJS, 154, 25
\bibitem[Rix et al. (2004)]{Rix04} Rix, H.-W., et al.
        2004, ApJS, 152, 163
\bibitem[Schiminovich et al. (2005)]{Schiminovich05} Schiminovich, D., et al.
        2005, ApJ, 619, L47
\bibitem[Steidel et al. (1999)]{Steidel99} Steidel, C. C., Adelberger, K. L., Giavalisco, M., Dickinson, M., \& Pettini, M.
        1999, ApJ, 519, 1 
\bibitem[Sullivan et al. (2000)]{Sullivan00} Sullivan, M., Treyer, M. A., Ellis, R. S., Bridges, T. J., Milliard, B., \& Donas, J.
        2000, MNRAS, 312, 442
\bibitem[Takeuchi et al. (2003)]{Takeuchi03} Takeuchi, T., Yoshikawa, K., \& Ishii, T. T.
        2003, ApJ, 587, L39
\bibitem[Wang \& Heckman (1996)]{Wang96} Wang, B., \& Heckman, T. M.
        1996, ApJ, 457, 645
\bibitem[Willmer et al. (2005)]{Willmer05} Willmer, C. et al.
        2005, ApJ, submitted (astro-ph/0506041)
\bibitem[Wolf et al. (2003)]{Wolf03} Wolf, C., Meisenheimer, K., Rix, H.-W., Borch, A., Dye, S., \& Kleinheinrich, M.
        2003, A\&A, 401, 73
\bibitem[Wolf et al. (2004)]{Wolf04} Wolf, C., et al.
        2004, A\&A, 421, 913
\bibitem[Wolf et al. (2005)]{Wolf05} Wolf, C., et al.
        2005, ApJ, 630, 771 
\bibitem[Zheng et al. (2004)]{Zheng04} Zheng, X. Z., Hammer, F., Flores, H., Ass$\acute{\rm e}$mat, F., \& Pelat, D.
        2004, A\&A, 421, 847


\end{thebibliography}
